\documentclass[11pt,a4paper]{article}

\usepackage{microtype}
\usepackage{amsmath, amsfonts,amssymb,mathtools,nicefrac,nccmath,cases,physics}
\usepackage{bm}
\usepackage{soul}
\usepackage{dsfont}
\usepackage{booktabs,multirow}
\usepackage{tabularx}
\usepackage{makecell}
\usepackage[dvipsnames,table]{xcolor}
\usepackage{colortbl}
\usepackage{caption}
\usepackage{subcaption}
\usepackage{float}
\usepackage{hhline}
\usepackage{enumitem}
\usepackage{cite}
\numberwithin{equation}{section}
\numberwithin{table}{section}
\usepackage[linktocpage=true]{hyperref}
\hypersetup{colorlinks=true,linkcolor=colorloc3,citecolor=colorloc3,urlcolor=colorloc3}
\usepackage{afterpage}
\usepackage{tikz}

\def\hybrid{\topmargin -20pt    \oddsidemargin 0pt
	\headheight 0pt \headsep 0pt
	\textwidth 6.5in        
	\textheight 9in         
	\textwidth 6.25in       
	\textheight 9 in       
	\marginparwidth .875in
	\parskip 5pt plus 1pt 
	\jot = 1.5ex
}
\hybrid

\usepackage[framemethod=default]{mdframed}

\newmdenv[skipabove=10pt,
skipbelow=7pt,
rightline=false,
leftline=true,
topline=false,
bottomline=false,
linecolor=colorloc2,
backgroundcolor=colorloc1!5,
innerleftmargin=4pt,
innerrightmargin=0pt,
innertopmargin=0pt,
leftmargin=2pt,
rightmargin=0pt,
linewidth=2pt,
innerbottommargin=0pt,
frametitlebackgroundcolor=colorloc2]{lbBox}

\newmdenv[skipabove=10pt,
skipbelow=7pt,
rightline=false,
leftline=true,
topline=false,
bottomline=false,
linecolor=colorloc2,
backgroundcolor=colorloc2!5,
innerleftmargin=4pt,
innerrightmargin=4pt,
innertopmargin=0pt,
leftmargin=0pt,
rightmargin=0pt,
linewidth=2pt,
innerbottommargin=0pt,
frametitlebackgroundcolor=colorloc2]{lcBox}

\usepackage{fancyhdr}
\definecolor{colorloc1}{RGB}{164,42,46} 
\definecolor{colorloc2}{RGB}{100,100,100} 
\definecolor{colorloc3}{RGB}{204,119,34}  
\definecolor{colorloc4}{RGB}{25,25,112}  
\definecolor{colorloc5}{RGB}{0,128,128}  

\usepackage[linktocpage=true]{hyperref}
\hypersetup{colorlinks=true,linkcolor=colorloc5,citecolor=colorloc4,urlcolor=colorloc4}

\newcommand{\mysetminusD}{\hbox{\tikz{\draw[line width=0.6pt,line cap=round] (3pt,0) -- (0,6pt);}}}
\newcommand{\mysetminusT}{\mysetminusD}
\newcommand{\mysetminusS}{\hbox{\tikz{\draw[line width=0.45pt,line cap=round] (2pt,0) -- (0,4pt);}}}
\newcommand{\mysetminusSS}{\hbox{\tikz{\draw[line width=0.4pt,line cap=round] (1.5pt,0) -- (0,3pt);}}}
\newcommand{\mysetminus}{\mathbin{\mathchoice{\mysetminusD}{\mysetminusT}{\mysetminusS}{\mysetminusSS}}}

\begin{document}


	\baselineskip=14pt
	\parskip 5pt plus 1pt

	\vspace*{-1.5cm}
	\begin{flushright}    
		{\small ZMP - 24 - 28
				
			}
	\end{flushright}
	
	\vspace*{2.5cm}
	\begin{center}        
		{\bf \huge Asymptotics of 5d Supergravity Theories \\\vspace{2mm} and the Emergent String Conjecture}   
	\end{center}
	
	\vspace{0.5cm}
	\begin{center}        
		{\large \bf Lukas Kaufmann}\footnote{e-mail:
    lukas.kaufmann@desy.de}, {\large \bf Stefano Lanza}\footnote{e-mail:
    stefano.lanza@desy.de}, {\large \bf Timo Weigand}\footnote{e-mail:
    timo.weigand@desy.de}
        
	\vspace{0.2cm}

	\emph{II. Institut f\"ur Theoretische Physik, Universit\"at Hamburg,\\
	Luruper Chaussee 149, 22607 Hamburg, Germany}

    \emph{Zentrum f\"ur Mathematische Physik, Universit\"at Hamburg, \\Bundesstrasse 55,  20146 Hamburg, Germany  }

	\end{center}
	
	\vspace{2.5cm}
	
	
	\begin{abstract}
		\noindent 
        We invoke probe brane arguments to classify the asymptotic behavior of general five-dimensional supergravity theories with eight supercharges near infinite distance boundaries of the vector multiplet moduli space. Imposing consistency of supergravity strings we derive several constraints on the Chern-Simons couplings entering the prepotential, including their non-negativity. This establishes a classification of infinite distance limits analogous to those for theories obtained as Calabi-Yau compactifications, but without having to assume a geometric or string theoretic origin. All infinite distance limits are found to be either vector or tensor limits, depending on the nature of the gauge potential becoming weakly coupled at the fastest rate. In particular, we prove uniqueness results for the asymptotically leading gauge fields. The asymptotic physics along these limits is in perfect agreement with the predictions of the Emergent String Conjecture and hence serves as bottom-up evidence for the latter. Our findings imply that every consistent five-dimensional ${\cal N}=1$ supergravity with a non-compact vector multiplet moduli space either descends from six dimensions or contains a stringy subsector.

	\end{abstract}

	\thispagestyle{empty}
	\clearpage
	
	\setcounter{page}{1}
	
	
\newpage

\tableofcontents

\newpage

\section{Introduction and summary}
\label{sec:Introduction}

Although the landscape of  effective field theories that can arise from  quantum gravity  is believed to be
vast, it has been suggested that all such consistent theories obey some distinctive principles.
Understanding these general criteria for consistency with quantum gravity is the goal of the \emph{Swampland  Program} as initiated in \cite{Vafa:2005ui}, and reviewed, for instance, in \cite{Palti:2019pca,vanBeest:2021lhn,Grana:2021zvf,Agmon:2022thq}.

Among the most important, and best tested, of the proposed Swampland principles is the Distance Conjecture \cite{Ooguri:2006in}.
It asserts that 
whenever the moduli space of an effective theory in the quantum gravity landscape exhibits boundaries located at infinite geodesic distance,
infinite towers of states become
light
in the near-boundary regions, and in particular lighter than the effective theory cutoff. This leads to the breakdown of the original effective description.
 While the asymptotically light states have a distinguished quantum gravity origin, their precise  nature is a priori unspecified.

In this regard, the Emergent String Conjecture,  proposed in \cite{Lee:2019wij}, offers a more refined interpretation of the Distance Conjecture. 
 It posits that, in Einstein gravity\footnote{In this article we will always be working in flat spacetime. For recent progress on the Distance and Emergent String Conjecture in AdS spacetime and its holographic dual, see \cite{Baume:2020dqd,Perlmutter:2020buo,Baume:2023msm,Ooguri:2024ofs,Calderon-Infante:2024oed}.} in four or more dimensions, the leading infinite towers of states that realize the Distance Conjecture can only be of two kinds: They can be either Kaluza-Klein towers, associated to one or several internal dimensions that  decompactify in the limit,  or they are associated to the modes of a critical, asymptotically weakly coupled string.
If correct, this would have far-reaching consequences: 
It provides a lower bound on the exponential scale factor governing the rate at which towers of states become light asymptotically \cite{Etheredge:2022opl,Etheredge:2023odp,Etheredge:2024tok}.  Relatedly, it  identifies the species scale \cite{Castellano:2022bvr,vandeHeisteeg:2023ubh,Blumenhagen:2023yws,vandeHeisteeg:2023dlw,Herraez:2024kux}, first introduced in \cite{Dvali:2007hz,Dvali:2007wp}, beyond which a weakly coupled or local description of gravity is no longer possible, universally as either the string scale \cite{Dvali:2009ks,Dvali:2010vm} associated with a tower of asymptotically light string excitations or as the Planck scale \cite{Montero:2022prj} of an underlying higher-dimensional theory.
The Emergent String Conjecture furthermore
implies, on quite general grounds \cite{FierroCota:2023bsp}, the asymptotic version of the Tower Weak Gravity Conjecture \cite{Heidenreich:2015nta,Montero:2016tif,Andriolo:2018lvp}
 (as demonstrated in concrete setups e.g. in \cite{Lee:2018urn,Lee:2019wij,FierroCota:2023bsp}), and serves as one of the motivations behind the Dark Dimension Scenario of \cite{Montero:2022prj}.

The conjecture has been tested in large classes of effective field theories obtained from string or M-theory. This includes 
effective theories originating from F-theory compactifications in various dimensions \cite{Lee:2018urn, Lee:2018spm,Lee:2019tst,Klaewer:2020lfg, Lee:2021qkx, Lee:2021usk, Alvarez-Garcia:2023gdd,Alvarez-Garcia:2023qqj}
and their dual descriptions \cite{Collazuol:2022jiy,Collazuol:2022oey,Etheredge:2023odp,Etheredge:2024tok}, 
five- \cite{Lee:2019wij,Alvarez-Garcia:2021pxo,Rudelius:2023odg} and four-dimensional \cite{Xu:2020nlh} $\mathcal{N}=1$ M-theory effective theories, four-dimensional $\mathcal{N} = 2$ Type II effective theories \cite{Lee:2019wij, Baume:2019sry,Blumenhagen:2023yws}, with the Emergent String Conjecture offering a physics interpretation of the infinite distance limits, studied generally beginning with \cite{Grimm:2018ohb,Grimm:2018cpv,Corvilain:2018lgw,Grimm:2019bey}, as well as non-supersymmetric  \cite{Basile:2022zee} and even non-geometric \cite{Aoufia:2024awo} setups.

 Such tests of the Emergent String Conjecture within a given effective theory are typically grounded in the geometry of the underlying compactification manifold: Indeed, a key ingredient for the realization of the Emergent String Conjecture is the identification of the geometrical structures that could deliver either the Kaluza-Klein towers, or the emergent strings in the effective field theory, so as to have clear control over the associated mass scales in any infinite distance limit within the moduli space.
  In particular, the emergent critical strings must be unique and always accompanied by a tower of Kaluza-Klein states at the same mass scale. This non-trivial prediction is found to be realized thanks due to rather non-trivial properties of the compcatification geometry, sometimes even including the structure of the quantum corrections \cite{Lee:2019wij, Baume:2019sry, Klaewer:2020lfg, Alvarez-Garcia:2021pxo} of the theory. 

However, explaining whether, and how, the Emergent String Conjecture is realized genuinely from the bottom-up, without prior knowledge of the underlying compactification and with intrinsic effective field theory ingredients only, is a hard task. 
Remarkably, \cite{Basile:2023blg,Basile:2024dqq,Bedroya:2024ubj,Herraez:2024kux} have provided general arguments suggesting that consistent light towers in quantum gravity are either of Kaluza-Klein type or 
share key features of a stringy spectrum. 
 The arguments are rooted in various ideas on black hole thermodynamics including, in the case of \cite{Basile:2023blg,Basile:2024dqq,Herraez:2024kux}, the notion of species entropy introduced in
\cite{Cribiori:2023ffn} and, in the case of 
\cite{Bedroya:2024ubj}, properties of gravitational scattering amplitudes and the 
particle-black hole transition. 

In this work we pursue a different route to presenting evidence of the Emergent String Conjecture from a bottom-up perspective.
 We focus on five-dimensional $\mathcal{N}=1$ supergravity theories, endowed with eight supercharges, without assuming a realization of such a theory via string or M-theory.
Within these theories, we shall further restrict our attention to the vector multiplet sector, which accommodates real scalar fields, and an equal number of gauge one-forms.
The interactions between this sector and the gravity multiplet are fully governed by a prepotential. At the two-derivative level, it is a homogeneous function of degree three of the scalars residing in the vector multiplets. The scalar field metric, as well as the one-form gauge kinetic function depend on the second derivatives of the prepotential, while the Chern-Simons couplings are most readily identified with the triple derivatives of the prepotential.

Assuming positivity of the Chern-Simons terms, \cite{Heidenreich:2020ptx} has shown that all infinite distance limits in the vector multiplet moduli space of five-dimensonal supergravity are weak coupling limits (see \cite{Corvilain:2018lgw, Lee:2019wij} for related arguments). Furthermore, for special classes of limits, \cite{Etheredge:2022opl,Rudelius:2023odg} have observed that the asymptotic scaling behaviour of weakly coupled one- and two-forms matches expectations for Kaluza-Klein and for weakly coupled string limits, respectively.

To strengthen the case for the Emergent String Conjecture, additional information is required: In particular, the two-forms becoming weakly coupled at the fastest rate must be unique and always accompanied by at least one one-form field. 
 This is necessary, at the supergravity level, to guarantee uniqueness of the emergent string, charged under this two-form, and its criticality.

The main goal of this article is to show this in generality and under the minimal possible assumptions. 
 The key to achieving this are the \emph{supergravity strings} first introduced in \cite{Katz:2020ewz}.
 These are special BPS strings coupling to the two-forms in 
five-dimensional $\mathcal{N}=1$ supergravity theories.
 Their defining property is that the dual BPS particles display positive electric charges, in a convenient charge basis, with respect to the gauge field under which the supergravity string carries minimal magnetic charge.
 Whenever they are included within the effective field theory, the worldsheet anomalies induced by inflow from the bulk theory are encoded in a  nontrivial 't Hooft anomaly matrix defined on the string worldsheet. Consistency of the theory requires the inflow anomaly to be cancelled by an equal and opposite contribution form the worldsheet modes.
Furthermore, the signature of the 't Hooft anomaly matrix is crucially constrained: The 't Hooft anomaly matrix must display a single positive eigenvalue, while admitting some negative eigenvalues, related to the number of gauge two-forms under which the string is charged \cite{Katz:2020ewz}. This condition will have far-reaching consequences for us.

Analyzing the constraints which the supergravity strings must satisfy will give us crucial insights on the bulk supergravity theory that hosts them. 
 Indeed, their four-dimensional analogues, dubbed  
effective field theory strings in \cite{Lanza:2020qmt},
encode important information on the structure of infinite distance limits \cite{Lanza:2020qmt,Lanza:2021udy,Marchesano:2022axe,Grimm:2022sbl,Martucci:2024trp}, including even possible strongly coupled subsectors \cite{Marchesano:2022avb,Wiesner:2022qys}, as studied further in \cite{Marchesano:2023thx,Marchesano:2024tod,Castellano:2024gwi}.
 More generally, using consistency of probe strings has turned out to be an extremely fruitful avenue to constrain viable quantum gravity theories even in the bulk of the moduli space \cite{Kim:2019vuc,Lee:2019skh,Kim:2019ths,Katz:2020ewz,Tarazi:2021duw,Martucci:2022krl,Lee:2022swr,Kim:2024tdh,Kim:2024hxe}.

\noindent\textbf{Summary of main results.} Motivated by these successes, the main idea of this work is to employ supergravity strings as probes for the five-dimensional supergravity theory.
To this aim, we shall assume that the BPS spectrum of supergravity strings is complete, at least at the level of generators of the charge lattice, so as to exploit the associated supergravity strings  as probes for the theory. 
Then, enforcing the aforementioned constraints on the signature of the 't Hooft anomaly matrices associated to the strings, we will show that the prepotential distinguishing the five-dimensional theory hosting them is greatly constrained.

Specifically, we will show that in a suitable, so-called K\"ahler basis, the Chern-Simons couplings are non-negative, extending previous results in \cite{Kim:2024hxe}.
 Furthermore, exploiting the signature theorem for the 't Hooft matrix of supergravity strings \cite{Katz:2020ewz}, we derive 
a number of consistency relations for the Chern-Simons couplings that are summarized in Tables \ref{tab:N2_Sugra_JA}, \ref{tab:N2_Sugra_JB} and \ref{tab:N2_Sugra_JB2}. With this input we will be able to classify all possible infinite distance limits in the vector multiplet moduli space
 of five-dimensional $\mathcal{N}=1$ supergravity theories.
As in \cite{Lee:2019wij}, we will introduce a classification of limits -- to which we will refer as \emph{Class A} and \emph{Class B} -- which depends on the specific form of the prepotential, and along which the scalar fields display a distinguished asymptotic behavior. 
 What is remarkable is that a 
 formalism very similar to the one governing the asymptotics of compactifications of M-theory on Calabi-Yau threefolds \cite{Lee:2019wij} is also at work for general five-dimensional supergravities - without any assumption about their UV description or reference to an underlying Calabi-Yau geometry. The non-trivial input behind this is entirely the above-mentioned weak BPS completeness hypothesis and the resulting consistency conditions induced by the probe supergravity strings.

These constraints will furthermore allow us to show that the 
above classes of limits can only be of two kinds: they can either be \emph{vector limits}, or \emph{tensor limits}.
Along vector limits, the most weakly coupled gauge field is a one-form gauge potential; furthermore, we can prove that there is only a single such one-form becoming weakly coupled at the fastest rate. The latter is a special property of limits in the vector, as opposed to hypermultiplet, moduli space.
  Conversely, along tensor limits, 
  the most weakly coupled gauge field is a {\it unique} tensor field, {\it always} accompanied by one or several one-form fields becoming weakly coupled at the same rate. 
  Showing uniqueness constitutes, in fact, the hardest part of the analysis and makes full use of the aforementioned probe consistency conditions.

Furthermore, the rate at which the (unique) one-form potential becomes weakly coupled in vector limits will be shown to match exactly the behavior of the Kaluza-Klein gauge field in a circle reduction from six to five dimensions. Similarly, the unique two-form potential becoming weakly coupled in tensor limits does so at the rate of the Kalb-Ramond B-field in a five-dimensional string theory in the limit of weak string coupling.
We interpret this as strong evidence for the Emergent String Conjecture and conjecture that the vector limits are indeed decompactification limits to six dimensions, while the tensor limits describe emergent string limits.
  As we will explain, the latter statement makes further predictions for the structure of the higher-derivative, gravitational Chern-Simons couplings of the theory.

\noindent\textbf{Structure of the paper.} 
In Section~\ref{sec:5D} we review some of the salient features of five-dimensional $\mathcal{N}=1$ supergravities with an arbitrary number of vector multiplets. We recall the notion of supergravity strings introduced in \cite{Katz:2020ewz} and derive the positivity constraints on the Chern-Simons couplings that the theories must obey for consistency of the supergravity strings.
In Section~\ref{sec:Limits} we introduce a general classification of the infinite field distance limits that the five-dimensional supergravity theories can host, and we highlight how this classification relates to the emergent asymptotic physics.
In the following Sections~\ref{sec:N2_Sugra_JA} and~\ref{sec:N2_Sugra_JB} we particularize the discussion of Section~\ref{sec:Limits} to the two distinct classes of infinite field distance limits, namely Class A and Class B, by studying how key physical quantities behave asymptotically. 
Section~\ref{sec:EmString} is devoted to connecting the findings of Sections~\ref{sec:N2_Sugra_JA} and~\ref{sec:N2_Sugra_JB} to the predictions of the Emergent String Conjecture, by offering an interpretation of the light towers emerging along the aforementioned limits.
 Section \ref{sec:Conclusions} contains our conclusions and directions for future work.
Several technical details are contained in the appendices: Appendix~\ref{sec:Pos_FIJK} contains the detailed proof of the non-negativity of the Chern-Simons couplings of consistent five-dimensional supergravities, Appendix~\ref{app:BU_constraints} collects the proofs of the several, additional constraints that supergravity strings impose on the bulk theory, and Appendix~\ref{app:ClassB} includes some aspects of the asymptotic physics towards Class B limits.

\section{The geometrical structure of five-dimensional \texorpdfstring{$\mathcal{N}=1$}{N=1} supergravity}
\label{sec:5D}

In order to set the ground for the forthcoming sections, here we review some of the salient features of five-dimensional $\mathcal{N}=1$  supergravity.
Our main focus is the vector multiplet sector: After a brief overview of the interactions governing this sector, we introduce the key objects that couple, either electrically or magnetically, to the vector multiplets.
Among these are the \emph{supergravity strings} introduced in \cite{Katz:2020ewz}. After recalling their definition in Section~\ref{sec:5D_Cones}, we highlight the constraints that the five-dimensional supergravity ought to satisfy in order to accommodate them in its spectrum, including the important positivity constraints discussed in Section \ref{sec:5D_Constraints}.

\subsection{An overview of five-dimensional \texorpdfstring{$\mathcal{N}=1$}{N=1} supergravity}
\label{sec:5D_N2_Sugra}

We consider five-dimensional $\mathcal{N}=1$  supergravities which, besides the gravity multiplet, may host an arbitrary number of vector multiplets.
 Such supergravity theories are characterized by the following bosonic fields: the graviton and the graviphoton, being part of the supergravity multiplet; $n_V$ real scalar fields $\varphi^i$ and an equal number of gauge one-forms, which are contained in $n_V$ vector multiplets.
The two-derivative action governing the interactions among the bosonic components of the supergravity and vector multiplets is \cite{Bergshoeff:2004kh, Lauria:2020rhc}
\begin{equation}
	\label{eq:N2_S}
	\begin{aligned}
		S &= \int_{\mathbb{R}^{1,4}} \Bigg( \frac{M_{\rm P}^3}{2} \star R - \frac{M_{\rm P}^3}{4} g_{ij} {\rm d} \varphi^i \wedge \star {\rm d} \varphi^j - \frac{M_{\rm P}}{4} f_{IJ} F^I \wedge \star F^J - \frac{1}{12} \mathcal{F}_{IJK} A^I \wedge F^J \wedge F^K \Bigg)\,.
	\end{aligned}
\end{equation}
Here we have denoted with $M_{\rm P}$ the five-dimensional Planck mass, and $R$ is the Ricci scalar; additionally, we have packaged together the graviphoton and gauge fields within $A^I$, with $I = 0,1, \ldots, n_V$. We have not displayed the hypermultiplet sector, which will play no role in our discussion.

The interactions appearing in the action \eqref{eq:N2_S} are solely dictated by the \emph{prepotential}
\begin{equation}
	\label{eq:N2_F}
	\mathcal{F}[{\bf X}] = \frac{1}{3!} \mathcal{F}_{IJK} X^I X^J X^K \,, 
\end{equation}
with $\mathcal{F}_{IJK}\in\mathbb{Z}$ some constant parameters, and $X^I$ real coordinates.
Indeed, the field space, spanned by the scalar fields $\varphi^i$, can be understood as the $n_V$-dimensional hypersurface, within the $(n_V+1)$-dimensional space spanned by the coordinates $X^I$, obtained by setting, for instance,
\begin{equation}
	\label{eq:N2_Fconstr}
	\mathcal{F}[{\bf X}] = 1 \,.
\end{equation}
Furthermore, the gauge kinetic matrix entering \eqref{eq:N2_S} is expressed, in terms of the prepotential, as
\begin{equation}
	\label{eq:N2_a}
	f_{IJ} = \mathcal{F}_I \mathcal{F}_J - \mathcal{F}_{IJ}\,,
\end{equation}
where we have set $\mathcal{F}_I := \frac{\partial \mathcal{F}}{\partial X^I} = \frac12 \mathcal{F}_{IJK} X^J X^K$ and $\mathcal{F}_{IJ} := \frac{\partial^2 \mathcal{F}}{\partial X^I \partial X^J} = \mathcal{F}_{IJK} X^K$.
The inverse of \eqref{eq:N2_a} is given by
\begin{equation}
	\label{eq:N2_ainv_b}
	f^{IJ} = g^{ij} \partial_i X^I \partial_j X^J +  \frac13 X^I X^J \,,
\end{equation}
with $\partial_i = \frac{\partial}{\partial \varphi^i}$. 
Moreover, the field space metric can be understood as the pull-back of the gauge kinetic matrix onto the hypersurface specified by the constraint \eqref{eq:N2_Fconstr} as
\begin{equation}
	\label{eq:N2_g}
	g_{ij} = f_{IJ} \partial_i X^I \partial_j X^J \,.
\end{equation}
Finally, the Chern-Simons couplings appearing in \eqref{eq:N2_S} are most readily identified with the triple derivatives of the prepotential \eqref{eq:N2_F}.

It is sometimes useful to encode the information contained in the scalars $X^I$ compactly in the vector object\footnote{In the following, as a shorthand notation, we will also denote with ${\bf X}$ the column vector $(X^I)^T$, denoting the components of the vector ${\bf X}$, expressed in the canonical basis ${\bf e}_1 = (1,0,\ldots, 0)$, ${\bf e}_2 = (0,1,0,\ldots, 0)$, \ldots, ${\bf e}_{n+1} = (0,\ldots, 0,1)$.} 
\begin{equation} 
    \label{eq:Xvec}
    {\bf X} = X^I  \, K_I \,.
\end{equation}
The Chern-Simons couplings can then be viewed as a triple product 
\begin{equation}
    {\cal F}_{IJK} =: K_I \cdot K_J \cdot K_K
\end{equation}
and in this notation the prepotential at the two-derivative level takes the compact form
\begin{equation}
    {\cal F} [{\bf X}]= \frac{1}{3! }{\bf X} \cdot {\bf X} \cdot {\bf X} \,.
\end{equation}

The supergravity action \eqref{eq:N2_S} further admits couplings to fundamental BPS objects.
Indeed, BPS particles, electrically charged under the gauge one-forms $A^I$, can be coupled to \eqref{eq:N2_S} via 
\begin{equation}
	\label{eq:N2_Spart}
	S_{\rm part} = - \int_{\mathcal{L}} {\rm d} \tau \sqrt{-\gamma}\, M_{\rm part} + q_I \int_{\mathcal{L}} A^I\,.
\end{equation}
Here, the second, Wess-Zumino term expresses the minimal coupling of the particle to the gauge fields $A^I$, with $q_I$ being the particle electric charges.
In the first, Nambu-Goto term, $\tau$ denotes the coordinate spanning the particle worldline $\mathcal{L}$, and $\gamma$ is the determinant of the metric over the particle worldline. Moreover, we have introduced the particle mass $M_{\rm part}$, which can be expressed in terms of the ungauged scalar fields $X^I$ as 
\begin{equation}
    \label{eq:N2_Mpart}
    M_{\rm part} = M_{\rm P} q_I X^I \,.
\end{equation}
It is also convenient to define the \emph{particle physical coupling} as 
\begin{equation}
	\label{eq:N2_Qpart}
	\mathfrak{q}^2_{\rm part} = \frac{2}{M_{\rm P}} q_I f^{IJ} q_J\,.
\end{equation}
 It quantifies the strength of the gauge coupling associated to the linear combination of the $\mathrm{U}(1)$'s to which the particle is coupled.
However, the physical charge \eqref{eq:N2_Qpart} and mass \eqref{eq:N2_Mpart} are not independent quantities: in fact, employing \eqref{eq:N2_ainv_b}, it is immediate to show that they are related via the following relation:
\begin{equation}
	\label{eq:N2_BPSpart}
	M_{\rm P}^3 \mathfrak{q}^2_{\rm part} = \frac23 M_{\rm part}^2 + 2 g^{ij} \partial_i M_{\rm part} \partial_j M_{\rm part} \,.
\end{equation}

Additionally, the supergravity action \eqref{eq:N2_S} can be coupled to fundamental BPS strings, minimally interacting with gauge two-forms $B_I$, which are the electric-magnetic duals of the gauge one-forms $A^I$.
The effective action describing a BPS fundamental string is
\begin{equation}
	\label{eq:N2_Sstr}
	S_{\rm str} = -  \int_{\mathcal{W}} {\rm d}^2 \xi  \sqrt{-h}\,  \mathcal{T}_{\rm str}  + p^I \int_{\mathcal{W}} B_I\,.
\end{equation}
The second term contains the string elementary charges $p^I$ and expresses the strings electric coupling to the gauge two-forms $B_I$.
In the first term, $\xi^{\iota}$, $\iota = 1,2$, are the coordinates spanning the string worldsheet $\mathcal{W}$, with $h$ the determinant of the worldsheet metric. The string tension $\mathcal{T}_{\rm str}$ can be further expressed in terms of the prepotential as
\begin{equation}
    \label{eq:N2_Tstr}
    \mathcal{T}_{\rm str} = \frac{M_{\rm P}^2}{2}  p^I \mathcal{F}_I \,.
\end{equation}
As for the BPS particles, we can introduce the \emph{string physical coupling}
\begin{equation}
	\label{eq:N2_Qstr}
	\mathcal{Q}^2_{\rm str} = \frac{M_{\rm P}}2 p^I f_{IJ} p^J\,,
\end{equation}
which entails the strength of the gauge coupling associated to the linear combination of the $\mathrm{U}(1)$'s to which the BPS string is coupled.
Furthermore, it can be shown that the string physical charge \eqref{eq:N2_Qstr} is related to the string tension \eqref{eq:N2_Tstr} via the relation
\begin{equation}
	\label{eq:N2_BPSstr}
	M_{\rm P}^3 \mathcal{Q}^2_{\rm str} = \frac23 \mathcal{T}^2_{\rm str} + 2 g^{ij} \partial_i \mathcal{T}_{\rm str} \partial_j \mathcal{T}_{\rm str} \,.
\end{equation}

For ease of exposition, unless stated otherwise, we will from now on set 
\begin{equation}
M_{\rm P} \equiv 1 \,.
\end{equation}

\subsection{Conical structures and supergravity strings}
\label{sec:5D_Cones}

The key objects of our investigation are the \emph{supergravity strings} first introduced in \cite{Katz:2020ewz}. 
While these may still be regarded as BPS strings, being effectively described by the action \eqref{eq:N2_Sstr}, their non-trivial definition rests on the conical structures induced by the BPS spectrum of the five-dimensional $\mathcal{N}=1$ supergravity, which we shall now review. 

First, let us focus on the spectrum of fundamental BPS particles, preserving half of the bulk supersymmetry on their worldline.
 Their central charge is given by the mass \eqref{eq:N2_Mpart}. Since the coordinates $X^I$ are real, whether the associated particle is BPS (as opposed to anti-BPS) only depends on the sign of \eqref{eq:N2_Mpart}.
In this work we shall follow the convention that fundamental BPS particles are those described by the action \eqref{eq:N2_Spart} with $M_{\rm part} \geq 0$.
 The set of mutually BPS charges with this property forms a cone 
which we denote by 
\begin{equation}
    \label{eq:Cones_Cpart-def}
    \mathcal{C}_{\rm part} = \{ {\bf q}: \mathbf{q} \,\, \text{supports mutually compatible BPS particles}
    \} \,,
\end{equation}
with ${\bf q} = (q_I)^T$.

Given the BPS cone ${\cal C}_{\rm part}$, the inhomogeneous coordinates $X^I$ 
of the vector multiplet moduli space lie in the dual cone\footnote{The statement that the $X^I$ lie in a cone is of course only correct prior to imposing the prepotential constraint \eqref{eq:N2_Fconstr}.} such that $M_{\rm part} \geq 0$ for BPS particles, i.e. in the cone
\begin{equation}
     \label{eq:Cones_CXI-def}
     \Delta = \{ {\bf X} \in \mathbb{R}^{n+1}: q_I X^I \geq 0  \quad \forall \mathbf{q} \in {\cal C}_{\rm part} \} \,.
 \end{equation}
This cone is oftentimes referred to as the K\"ahler cone of the supergravity theory.
 Conversely, the BPS cone is the dual of the K\"ahler cone, 
\begin{equation}
    \label{eq:Cones_Cpart_dual}
    \mathcal{C}_{\rm part} = \Delta^{\vee} \cap N_{\mathbb{Z}} \,,
\end{equation}
with $N_{\mathbb{Z}}$ being the trivial, full-rank lattice generated by the canonical basis of $\mathbb{R}^{n+1}$, ${\bf e}^{1} = (1,0,\ldots, 0)^T$, ${\bf e}^{2} = (0,1,0,\ldots, 0)^T$, \ldots, ${\bf e}^{n+1} = (0,\ldots, 0,1)^T$.\footnote{In $\mathbb{R}^{n+1}$, a lattice $N_{\mathbb{Z}}$ generated by some linearly-independent vectors ${\bf v}^{(i)}$, where $i = 1, \ldots, m + 1$ with $m \geq 1$, is the set $N_{\mathbb{Z}} = \{ \alpha_{i} {\bf v}^{(i)} :\alpha_i \in \mathbb{Z}\}$. If $m = n$, the lattice is said to be of full rank.}

For example, if the BPS cone is simply given by the non-negative charges, i.e. if
 \begin{equation} \label{eq:Cones_Cpart}
 \mathcal{C}_{\rm part} =
 \{ {\bf q} \in \mathbb{N}^{n+1} \} \,,
 \end{equation}
 then the dual K\"ahler cone is 
 \begin{equation} \label{eq:Cones_CXI}
 \Delta = 
 \{ {\bf X} \in \mathbb{R}^{n+1}_{\geq 0} \}\,. 
 \end{equation}

These two examples of cones are trivially simplicial and convex.
More generally, the structure of the moduli space is more involved.
In particular, a given patch of the  moduli space may be composed by distinct, simplicial, convex components, so that the local description may not be simplicial, or convex. 
However, each of these simplicial, convex components can be described by a distinct set of ungauged coordinates $X^I$ such that, after an appropriate rescaling, the coordinates $X^I$ describing this patch reside in the trivial cone \eqref{eq:Cones_CXI}. 
The full K\"ahler cone can be obtained by patching together these local simplicial K\"ahler subcones.

Now, let us consider the spectrum of BPS strings.
Once the mutually supersymmetric BPS particles have been fixed as above, the fundamental BPS strings are those whose central charge, given by the tension \eqref{eq:N2_Tstr}, is non-negative.
Such BPS strings are also naturally associated with conical structures.
 Let us introduce the cone generated by the first derivatives of the prepotential
\begin{equation}
    \label{eq:Cones_CFI}
    \Gamma = \left\{ \bm{\mathcal{F}} \in \mathbb{R}^{n+1} : \mathcal{F}_I = \frac{\partial \mathcal{F}}{\partial X^I} \Big|_{{\bf X} \in \Delta } \right\} \,,
\end{equation}
where $\bm{\mathcal{F}} = (\mathcal{F}_I)^T$. Then, BPS strings are those whose elementary charges ${\bf p} = (p_I)^T$ reside in the dual cone
\begin{equation}
    \label{eq:Cones_Cstr}
    \mathcal{C}_{\rm str} = \Gamma^{\vee} \cap N_{\mathbb{Z}}\,,
\end{equation}
where the intersection with the trivial lattice is necessary to enforce the integrality of the BPS string charges.

As originally conceived in \cite{Katz:2020ewz}, \emph{supergravity strings} are BPS strings with the following additional property: all BPS particles carry non-negative charge 
under the gauge field under which the string carries minimal magnetic charge.
In other words, the elementary charges of supergravity strings reside in the cone
\begin{equation}
    \label{eq:Cones_Csugra}
    \mathcal{C}_{\text{\tiny{SUGRA}}} = \Delta \cap N_{\mathbb{Z}} \,.
\end{equation}
Thus, given the definition \eqref{eq:Cones_CXI}, supergravity strings are characterized by non-negative elementary charges.
However, it is worth stressing that \eqref{eq:Cones_Csugra} is independent of the BPS condition. Indeed, in order to guarantee that supergravity strings are BPS, we additionally need to require that  
\begin{equation}
    \label{eq:Cones_Csugra_BPS}
    \mathcal{C}_{\text{\tiny{SUGRA}}} \subseteq \mathcal{C}_{\rm str} \,,
\end{equation}
or, equivalently, that $\Delta \subseteq \Gamma^{\vee}$.

The importance of supergravity strings lies in the fact that they do not decouple from gravity. This leads to some distinguishing features exhibited by their worldsheet theories.
 In general, the worldsheet of a BPS string hosts an $\mathcal{N}=(0,4)$ theory that flows to an SCFT in the infra-red. The gauge and gravitational anomalies developed by the string worldsheet degrees of freedom cancel the corresponding anomalies induced by anomaly inflow from the bulk supergravity  \eqref{eq:N2_S} \cite{Ferrara:1996hh,Mizoguchi:1998wv,Boyarsky:2002ck,Katz:2020ewz}.
 The worldsheet anomalies are encoded in the `t Hooft anomaly matrix. Given any BPS string with elementary charges $\mathbf{p} = (p^I)^T$, the `t Hooft anomaly matrix can be understood as a bilinear form
\begin{equation}
    \label{eq:Pos_tHmat_def_gen}
    k^{(\bf p)}:\mathcal{A}\times\mathcal{A}\longrightarrow\mathbb{R}
\end{equation}
defined on the vector space $\mathcal{A}$ that is spanned by the generators  $K_I$
introduced in \eqref{eq:Xvec}. From the worldsheet anomaly polynomial one  reads off the form
\begin{equation}
	\label{eq:Pos_tHmat_def}
	k^{({\bf p})}_{IJ} := k(K_I,K_J) = \mathcal{F}_{IJK} p^K\,.
\end{equation}
Oftentimes we will drop the superscript in $k^{({\bf p})}_{IJ}$ if it is clear from the context
which supergravity string it refers to.

The positive (negative) eigenvalues of the `t Hooft anomaly matrix are related to the number of Abelian currents in the right-(left-)moving sector of the worldsheet theory.
 The key property of a supergravity string is that 
on its worldsheet, charged under $r$ gauge fields, only a single right-moving Abelian current algebra is present, associated to the coupling with the graviphoton, while the remaining Abelian currents are left-moving \cite{Katz:2020ewz}. 
Thus, the signature of the `t Hooft anomaly matrix of a supergravity string is  
\begin{equation}
	\label{eq:Pos_tHmat_sig}
	\boxed{{\rm sgn}(k^{(\bf p)}_{IJ}) = (1,r-1) \quad \forall {\bf p} \in {\cal C}_{\text{\tiny{SUGRA}}}\qquad \text{with} \quad r \leq n := n_V +1 }\,.
\end{equation}
In particular, there exists a (possibly non integral) basis of generators $\{\tilde{K}_I\}$ in which
\begin{equation}
	\label{eq:Pos_tHmat_diag}
    \tilde{k}^{(\bf p)}_{IJ} := k(\tilde{K}_I, \tilde{K}_J) = {\rm diag}(1,-1,\ldots,-1,0,\ldots,0)\,.
\end{equation}
For the following discussion, it is convenient to split the indices of this basis as $\{\tilde{K}_I\} = \{\tilde{K}_0, \tilde{K}_\alpha, \tilde{K}_\zeta\}$, with the $0$-index associated to the positive eigenvalue, the indices $\alpha = 1, \ldots, r-1$ related to the eigenspace with negative eigenvalue, and the indices $\zeta = 1, \ldots, n - r$ associated to the null eigenspace.

\subsection{Positivity constraints on the  prepotential}
\label{sec:5D_Constraints}

In this work we are going to impose consistency of supergravity string probes
to derive constraints on the prepotential of the five-dimensional supergravity theory. 
To this end, we are making the key assumption that along each ray of elementary\footnote{An
elementary supergravity string charge is one that cannot be written as the sum of two or more
other supergravity string charges.} supergravity string charges there exists a physical string
in the spectrum of the theory for at least one charge on the ray.
This mild version of the BPS completeness hypothesis \cite{Polchinski:2003bq,Banks:2010zn} is imposed on the theory as a Swampland constraint. 
 Note that the mild version of BPS completeness imposed here, which is to be satisfied only for some multiples of the generators of the BPS cone, is in particular consistent with the counter-examples for BPS completeness found in theories with sixteen supercharges \cite{Montero:2022vva}.

\begin{figure}[t]
	\centering
	\includegraphics[width=11cm]{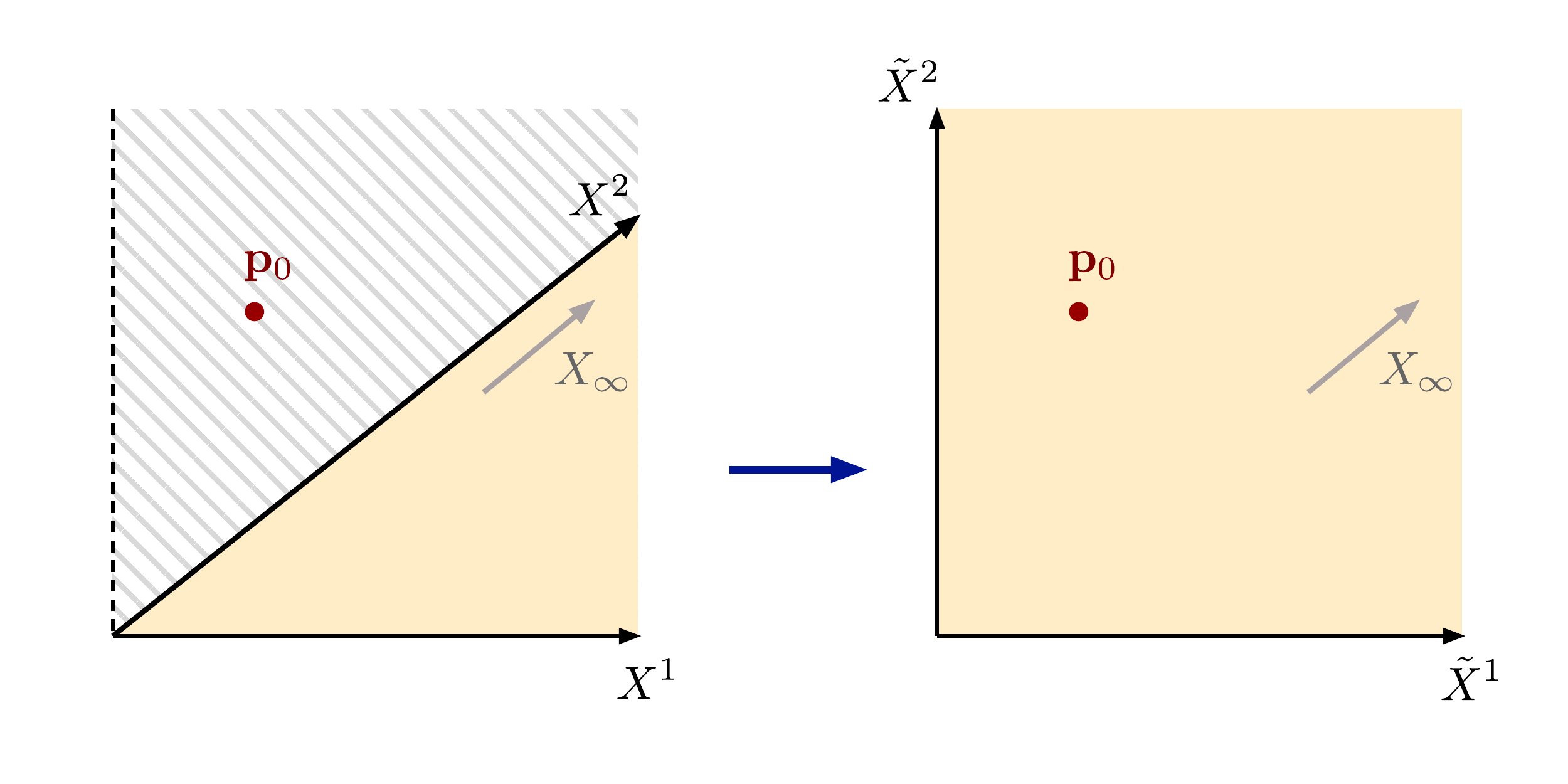}
	\caption{\footnotesize 
    On the left we illustrate, in yellow, a subcone of the full K\"ahler cone (given by the union of the yellow and the hashed region) spanned by $X^i \geq 0$  to analyze an infinite distance point $X_\infty$. To describe a supergravity string charge ${\bf p}_0$ outside this subcone, we can pass to a new subcone that contains ${\bf p}_0$ along with 
    $X_\infty$.
    		\label{Fig:Cones_5D}}
\end{figure}

Specifically, we will cover the K\"ahler moduli space $\Delta$ with open patches, each described
by a convex and simplicial (sub-)cone with coordinates that can be taken, w.l.o.g., as 
$X^I \geq 0$. Then in this coordinate system, charges $p^I \geq 0$ are associated with 
supergravity strings. There may exist, in addition, supergravity string charges for 
which some of the charges are negative with respect to the chosen coordinates $X^I$. As illustrated in Figure \ref{Fig:Cones_5D}, to analyze the constraints from these, 
we can pass to a different K\"ahler subcone (with new generators $\tilde X^I \geq 0$) in which this particular string has non-negative charges.

Quite generally, positivity will play a pivotal role in our analysis.
Apart from the positivity of the supergravity string charges in the K\"ahler basis, this includes the Chern-Simons couplings $\mathcal{F}_{IJK}$ entering the bulk action
\eqref{eq:N2_S}. We will argue\footnote{We thank Kai Xu for important discussions
on this.} in this section that in the simplicial (sub-)cones described above 
\begin{equation}
    \label{eq:Pos_III}
    \boxed{\mathcal{F}_{IJK} \geq 0 \qquad \forall I, J,K\in\mathcal{J}} \,.
\end{equation}
For the special case in which two or more indices are identical, this has already been shown
in \cite{Kim:2024tdh}, whose arguments we will review in the following. The key idea
is that, apart from the rank condition \eqref{eq:Pos_tHmat_sig}, extra constraints arise by 
imposing compatibility of the definition of supergravity strings
\eqref{eq:Cones_Csugra_BPS} with the BPS condition \eqref{eq:Cones_Cstr}.

First, consider a supergravity string with elementary charge ${\bf p}^{(I)} = {\bm \delta}^I$
(in a given patch, as explained above), where ${\bm \delta}^I$ denotes the $(n_V+1)$-dimensional
unit vector whose only non-zero component is in the $I$-th position. 
Now, fix a limit for which all the coordinates $X^J \to 0$, with $J \neq I$.
Along this limit, the tension of the supergravity string with elementary charge 
${\bf p}^{(I)} = {\bm \delta}^I$ is
\begin{equation}
	\label{eq:Pos_I_T}
	\mathcal{T}_{{\bf p}^{(I)}} \longrightarrow \frac14 \mathcal{F}_{III} (X^I)^2\,.
\end{equation}
Thus, in order to guarantee that such a string remains BPS, while still being regarded as a
supergravity string, we must demand that \cite{Kim:2024tdh}
\begin{equation}
    \label{eq:Pos_I}
   {\mathcal{F}_{III} \geq 0} \,,
\end{equation}
which ought to hold for all the indices $I\in\mathcal{J}$.

By similar reasoning, the condition \eqref{eq:Pos_I} can be extended to include couplings
of the form $\mathcal{F}_{IIJ}$ for $J\neq I$. To this end, we consider the limit along which
all the scalar fields $X^K \to 0$ (including $X^I)$, except for $X^J$.
In this limit, the tension of the supergravity string ${\bf p}^{(I)} = {\bm \delta}^I$ behaves
as
\begin{equation}
	\label{eq:Pos_II_T}
	\mathcal{T}_{{\bf p}^{(I)}} \longrightarrow \frac14 \mathcal{F}_{IJJ} (X^J)^2 \,.
\end{equation}
Requiring such a tension to remain non-negative along the limit implies that \cite{Kim:2024tdh}
\begin{equation}
    \label{eq:Pos_II}
    {\mathcal{F}_{IJJ} \geq 0} \,.
\end{equation}

To deduce the condition \eqref{eq:Pos_III} for distinct indices $I,J,K\in\mathcal{J}$, we
make use of the rank condition \eqref{eq:Pos_tHmat_sig} as follows: Pick three generators 
$K_I$, $K_J$, $K_K$ and consider the supergravity string with charge 
${\bf p}^{(I)} = {\bm \delta}^I$. The inner product defined by its 't Hooft matrix 
$k^{({\bm\delta}^I)}(K_J,K_K) = {\cal F}_{IJK}$ defines a Lorentzian metric on the vector
space $\mathcal{A}$ as a consequence of \eqref{eq:Pos_tHmat_sig}. This means in particular
that we can define future- and past-directed lightcones $\mathcal{C}_+$ and $\mathcal{C}_-$.

Writing $k=k^{({\bm\delta}^I)}$ from now on to avoid clutter of notation, we know from 
\eqref{eq:Pos_I_T} and \eqref{eq:Pos_II} that 
\begin{equation} \label{eq:ineq-k-1}
    k(K_I,K_I) \geq 0 \,, \quad 
    k(K_J,K_J) \geq 0 \,, \quad k(K_K,K_K) \geq 0 \,
\end{equation}
as well as 
\begin{equation} \label{eq:ineq-k-2}
    k(K_I,K_J) \geq 0 \,, \quad  k(K_I,K_K) \geq 0 \,.
\end{equation}
Assume first that all inequalities hold strictly. 
Since $k(K_I,K_I)>0$, the generator $K_I$ is a time-like vector with respect to the Lorentzian
metric $k$ and therefore an element of either ${\cal C}_+$ or ${\cal C}_-$. Assume without loss 
of generality that $K_I \in {\cal C}_+$. Any two vectors $v_1, v_2 \in {\cal C}_+$ have 
positive inner-product, $k(v_1,v_2) > 0$, and likewise for $v_1, v_2 \in {\cal C}_-$, while for 
$v_1 \in {\cal C}_+$ and $v_2\in {\cal C}_-$ one finds $k(v_1,v_2) <0$. From $k(K_J,K_J)>0$ it
follows that $K_J \in {\cal C}_+ \cup {\cal C}_-$, and since 
$k(K_I,K_J)>0$ we can in fact deduce that $K_J \in {\cal C}_+$. The same argument applied to 
$K_K$ leads to $K_K \in {\cal C}_+$. But this implies $k(K_J,K_K)=\mathcal{F}_{IJK} >0$, which
is precisely the sought-after condition \eqref{eq:Pos_III}.

If some of the inequalities in \eqref{eq:ineq-k-1} or \eqref{eq:ineq-k-2} are saturated, more
care is required. We relegate this discussion to Appendix \ref{sec:Pos_FIJK}, where we derive
\eqref{eq:Pos_III} for all possible cases allowed by \eqref{eq:ineq-k-1} and \eqref{eq:ineq-k-2}.

 Note that the positivity condition \eqref{eq:Pos_III} is indeed realized in concrete, geometric constructions of five-dimensional $\mathcal{N}=1$ supergravities stemming from M-theory compactifications on a Calabi-Yau three-fold because the K\"ahler cone generators are semi-ample divisors.
 It was also imposed in the bottom-up analysis of five-dimensional supergravities presented in \cite{Heidenreich:2020ptx,Heidenreich:2021yda}.

\section{Systematics of infinite distance limits in supergravity}
\label{sec:Limits}

With this preparation we can now launch our investigation of infinite distance limits from a genuine 
bottom-up perspective, without relying on the geometric origin of the ingredients constituting
the effective field theory. 

\subsection{Vector versus tensor limits}

To begin with, we introduce an appropriate characterization of
infinite distance limits that is motivated by the Emergent String Conjecture \cite{Lee:2019wij} and that will allow us
to systematically identify the emergent asymptotic physics.

According to the Emergent String Conjecture, every infinite distance limit is either a 
decompactification limit or a limit along which a unique critical string becomes asymptotically
weakly coupled such that the tower of string excitations is accompanied by a Kaluza-Klein tower
at the same scale. If the Kaluza-Klein tower is charged under a Kaluza-Klein $\mathrm{U}(1)$,
the latter becomes weakly coupled in the limit. The same applies, under an analogous assumption, to the two-form to which an
emergent string couples. The statement of the Emergent String Conjecture is then that in an 
emergent string limit, the two-form becoming weakly coupled at the fastest rate 
is unique and does so at a comparable rate as one (or several) $\mathrm{U}(1)_{\text{\tiny KK}}$
gauge groups.\footnote{The assumption that both the Kaluza-Klein tower and the emergent string are charged under one- or two-form gauge fields, respectively, is indeed appropriate for limits in the vector multiplet moduli sector of five-dimensional supergravity theories. More generally, it suffices that the emergent string is unique and accompanied by a Kaluza-Klein tower at the same mass scale.}

To turn this into a useful criterion to distinguish limits in supergravity, we must first
identify the condition for a one-form and a two-form to become weakly coupled at the same rate: 
We say that a one-form gauge field with physical coupling $\mathfrak{q}$ \textit{becomes weakly
coupled at the same rate} as a two-form gauge field with physical coupling $\mathcal{Q}$ if\footnote{We write $A \sim B$ (or $A \prec B$) for two quantities $A$ and $B$ if asymptotically, in the infinite distance limit under consideration, $A/B \to 1$ (or $A/B \to 0$). }
\begin{equation}
    \mathfrak{q}^2\sim\mathcal{Q}\longrightarrow0
    \label{eq:qQcompare}
\end{equation}
in the infinite distance limit. 
The motivation behind this definition is that, invoking the Asymptotic Tower Weak Gravity
Conjecture as in \cite{Cota:2022yjw,Cota:2022maf}, this amounts to comparing the WGC scales associated
with the one-form and the two-form, as follows by comparing \eqref{eq:N2_BPSpart} and
\eqref{eq:N2_BPSstr}. 

An infinite distance limit is then said to be a
\begin{itemize}
    \item \textit{vector limit} if the fastest vanishing one-form gauge coupling
    $\mathfrak{q}_{\mathrm{min}}$ satisfies that 
    $\mathfrak{q}_{\mathrm{min}}^2\prec\mathcal{Q}_{\mathbf{p}}$ for all
    elementary supergravity string charge vectors $\mathbf{p} 
    \in {\cal C}_{\text{\tiny{SUGRA}}}$, or
    
    \item \textit{tensor limit} if there exists a unique ray ${\cal R}_0$ in the supergravity string lattice
    ${\cal C}_{\text{\tiny{SUGRA}}}$ such that, for the minimal charge ${\bf p}_0 \in  {\cal R}_0$, it holds that $\mathcal{Q}_{{\bf p}_0} 
    \prec \mathcal{Q}_{{\bf p}} \, \forall \mathbf{p}\in{\cal C}_{\text{\tiny{SUGRA}}}$
    with $\mathbf{p}\not\in {\cal R}_0 $ and furthermore $\mathcal{Q}_{{\bf p}_0}\sim 
    \mathfrak{q}_{\rm min}^2$, with $\mathfrak{q}_{\rm min}$ the fastest vanishing one-form 
    coupling.
\end{itemize}
In both cases, $\mathfrak{q}_{\rm min}$ can in principle be associated with several one-forms
becoming weakly coupled at the same rate.
 However, we will see that in vector limits in the vector multiplet moduli sector of five-dimensional supergravity, only a single such field occurs.

\subsection{Class A versus Class B limits}

The characterization of infinite distance limits given above is driven by the asymptotic
physics.  For classification purposes, another, complementary way to characterize infinite
distance limits will turn out to be useful which closely mimics the one developed in 
\cite{Lee:2019wij} for supergravity theories obtained by compactifying M-theory on Calabi-Yau
threefolds.
 
The starting point is to cover the full, in general non-simplicial, K\"ahler cone by local
patches, each described by a simplicial sub-cone with coordinates $X^I \geq 0$.
We are assuming that the path towards the limiting locus in moduli space can be described within
one such subcone of the form \eqref{eq:Cones_CXI}.\footnote{This amounts to the assumption that
no switching between cones at infinitum occurs, as guaranteed in particular for geodesic paths,
which are the relevant ones for our analysis. See also \cite{Heidenreich:2020ptx} for a similar
discussion.} The limit is then classified according to the behavior that the coordinates $X^I$
display asymptotically.

First, recall how geodesic distances may be computed in the scalar field space of the vector
multiplet sector of five-dimensional $\mathcal{N}=1$ supergravities. Consider a path
within the physical field space, spanned by the coordinates $\varphi^i$ appearing in the action
\eqref{eq:N2_S}. Such a path can be thought of as a function $\varphi^i(\lambda)$, with 
$\lambda$ being the affine parameter describing the path. Conventionally, we shall assume that
the affine parameter can be chosen in the domain $[1, \infty[$, so that the path is threaded
between a point described by the local coordinates $\varphi^i_0 = \varphi^i (1)$, and the point
$\varphi^i_\infty = \varphi^i (\infty)$, reached as $\lambda \to \infty$. The geodesic length of
a path so chosen can be computed as
\begin{equation}
	\label{eq:InfDist_d}
	\mathrm{d}(\varphi^i_0, \varphi^i_\infty) 
    = \int_1^{\infty} {\rm d} \lambda \sqrt{\frac12 g_{ij}(\varphi) 
    \frac{{\rm d} \varphi^i}{{\rm d} \lambda} \frac{{\rm d} \varphi^j}{{\rm d} \lambda}}\,.
\end{equation}
Thus, a point $\varphi^i_\infty$ is said to be located at infinite field distance whenever
$\mathrm{d}(\varphi^i_0, \varphi^i_\infty) \to \infty$.

According to the remarks above, we will describe such paths as threaded in a subcone of the form
\eqref{eq:Cones_CXI}, spanned by the coordinates $X^I$. Denoting $X^I_0 = X^I(1) :=
X^I(\varphi^i_0)$, $X^I_\infty = X^I (\infty) = X^I(\varphi^i_\infty)$, the path described above
can be seen as a path lying in \eqref{eq:Cones_CXI} as
\begin{equation}
	\label{eq:InfDist_Pb}
	\mathcal{P} = \{\, {\bf X}(\lambda) \in \Delta : \mathcal{F}[{\bf X}] = 1,\, \lambda \in [1, \infty[ \, \} \,.
\end{equation}
Hence, upon employing \eqref{eq:N2_g}, the geodesic distance \eqref{eq:InfDist_d} can be written
in the more convenient form
\begin{equation}
	\label{eq:InfDist_db}
	\mathrm{d}(X^I_0, X^I_\infty) 
    = \int_1^\infty {\rm d} \lambda \sqrt{ \frac12 f_{IJ} \frac{{\rm d} X^I}{{\rm d} \lambda}
    \frac{{\rm d} X^J}{{\rm d} \lambda}} \Bigg|_{\mathcal{F}[{\bf X}] = 1} \,.
\end{equation}

In order for the point $X^I_\infty$ to be located at infinite field distance, 
\eqref{eq:InfDist_db} needs to diverge. This requires some of the coordinates $X^I$ to diverge
as $\lambda\to\infty$. Correspondingly, given the non-negativity of all Chern-Simons couplings
\eqref{eq:Pos_III}, in order to consistently preserve the hypersurface constraint $\mathcal{F}
[{\bf X}]=1$, some of the remaining coordinates need to appropriately fall-off along the limit. 
We further parametrize these paths in such a way that the fastest growing coordinate, say $X^0$,
scales like $\lambda$, with the remaining scaling at most as $\lambda$; namely
\begin{equation}
X^0\sim\lambda \,, \qquad X^I\precsim\lambda \quad 
\forall \, I\in\mathcal{J}\setminus\{0\} \,.
\end{equation}
We denote by $\mathcal{J}_\lambda$ those indices $I\in\mathcal{J}$
for which $X^I\sim\lambda$ holds, i.e.
\begin{equation}
    \label{eq:InfDist_Jl}
    \mathcal{J}_\lambda=\{I\in\mathcal{J}\,\vert\,X^I\sim\lambda\}.
\end{equation}
The realization of the hypersurface constraint \eqref{eq:N2_Fconstr} along the limits requires
\begin{equation}
    \label{eq:InfDist_Fabc}
    \mathcal{F}_{abc}=0  \quad \forall a,b,c\in\mathcal{J}_\lambda \,,
\end{equation}
and in particular $\mathcal{F}_{000}=0$.

From now on, we fix some coordinate $X^0$ with $0\in\mathcal{J}_\lambda$.
Then, the full index set $\mathcal{J}$ can be split with respect to $X^0$ as
\begin{equation}
    \mathcal{J}=\bigsqcup_{k=0}^3\mathcal{J}_k
\end{equation}
with $\mathcal{J}_0=\{0\}$ and 
\begin{subequations}
\label{eq:N2_Sugra_J_split}
    \begin{align}
        \label{eq:N2_Sugra_J_split_1}
        &\mathcal{J}_1=\{i\in\mathcal{J}\,:\,\mathcal{F}_{00i}\neq0\},
        \\
        \label{eq:N2_Sugra_J_split_2}
        &\mathcal{J}_2=\{\mu\in\mathcal{J}\,:\,\mathcal{F}_{00\mu}=0\,\,\mathrm{and}\,\,\exists
                \,\nu\in\mathcal{J}_2:\,\mathcal{F}_{0\mu\nu}\neq0\},
        \\
        \label{eq:N2_Sugra_J_split_3}
        &\mathcal{J}_3=\{r\in\mathcal{J}\,:\,\mathcal{F}_{00r}=0\,\,\mathrm{and}\,\,\,\mathcal{F}_{0rs}=0
                \,\,\forall
                s\in\mathcal{J}_2\sqcup\mathcal{J}_3\} \,.
    \end{align}
\end{subequations}
Analogously to the geometric infinite distance limits in \cite{Lee:2019wij}, we can then split
all infinite distance limits in two disjoint classes depending on the set $\mathcal{J}_1$.
Concretely, the infinite distance limit $\lambda\to\infty$ is said to be of 
\begin{itemize}
    \item \textit{Class A} if $\mathcal{J}_1\neq\varnothing$, namely ${\cal F}_{00i} \neq 0$ for
    the generators labeled by $i \in {\cal I}_1$, or 
    \item \textit{Class B} if $\mathcal{J}_1=\varnothing$. 
\end{itemize}
Notice that each infinite distance limit is either of Class A or B and that a single
supergravity can allow for both Class A \textit{and} Class B limits with respect to a different 
coordinate $X^0$, singled out from the set $\mathcal{J}_\lambda$ defined in 
\eqref{eq:InfDist_Jl}.

\subsection{Summary of main results}

The upcoming sections are devoted to analyzing the infinite distance behavior 
of both the classes of limits, and relating them to the above characterization of vector and
tensor limits. The following pattern will be established:
\begin{itemize}
    \item Class A limits are always vector limits, with a {\it single} one-form field becoming 
    asymptotically weakly coupled at the fastest rate.
    Concretely, the one-form field in question is given by the linear combination 
    \begin{equation}
    A_{\rm min}=\sum_{i\in\mathcal{J}_1}
    c_i A^i \,,
    \label{eq:A-U(1)KKa}
\end{equation}
    with  $c_i = {\cal F}_{00i}$ and coupling scaling as $\mathfrak{q}^2_{\mathrm{min}}\sim\lambda^{-4}$. This will be shown in Section \ref{sec:N2_Sugra_JA}.

    \item For Class B limits we have to differentiate between the cases of 
    $\mathcal{J}_\lambda=\{0\}$ and $|\mathcal{J}_\lambda|>1$. While the latter always give rise
    to vector limits in the above sense, the former can give rise to both vector and tensor 
    limits, depending on the scaling of the coordinates. In tensor limits, the minimal tensor coupling scales as 
$\mathcal{Q}_{\mathrm{min}}^2 = \mathcal{Q}_{{\bm\delta}^0}^2\sim\lambda^{-2}$.
The vector limits exhibit only a single asymptotically weakly coupled
    one-form field, see \eqref{Amin-B1}.\\
    More details and proofs can be found in Section \ref{sec:N2_Sugra_JB}.

\end{itemize}

The results summarized above form the basis for the interpretation of the vector and tensor limits in light of the Emergent String Conjecture, which we will give in Section \ref{sec:EmString}. The reader mainly interested in the physics interpretation rather than detailed proofs can jump directly there.

\section{Class A limits as vector limits}
\label{sec:N2_Sugra_JA}

We begin with the analysis of infinite distance limits of Class A as introduced in the previous section (see \cite{Lee:2019wij} for the analogous discussion in the context of M-theory compactified on a Calabi-Yau threefold).
 A key idea is to employ supergravity strings as probes of the bulk theory.
As explained in Section \ref{sec:5D_Constraints}, this amounts to imposing the rank condition 
\eqref{eq:Pos_tHmat_sig} for the 't Hooft matrix of supergravity strings along with the positivity conditions \eqref{eq:Pos_III} in a given simplicial K\"ahler subcone.

A Class A limit can only exist if there is a coordinate $X^0$ 
(singled out by its maximal scaling as $X^0 \sim \lambda$, with $\lambda \to \infty$, in the limit)
 with ${\cal F}_{000} =0$ and ${\cal F}_{00i} \neq 0$  for some other coordinates $X^i$. In Appendix~\ref{app:BU_constraints} we  derive from the rank condition \eqref{eq:Pos_tHmat_sig} for supergravity strings a number of constraints on the Chern-Simons couplings along directions associated with the sets \eqref{eq:N2_Sugra_J_split}. We collect these in Table~\ref{tab:N2_Sugra_JA}.

As a result in particular of ${\cal F}_{0rs} = 0 = {\cal F}_{rst}$ for $r,s,t \in {\cal J}_3$, the prepotential of a supergravity theory allowing for an infinite distance limit of Class A (with respect to the coordinate $X^0$) takes the form
\begin{equation}
    \label{eq:N2_Sugra_JA_F}
    \begin{aligned}
        \mathcal{F}&=\frac{1}{2}\mathcal{F}_{00i}(X^0)^2X^i+\frac{1}{2}\mathcal{F}_{0ij}X^0X^iX^j+\frac{1}{6}\mathcal{F}_{0ir}X^0X^iX^r
        \\
        &\quad\, +\frac{1}{6}\mathcal{F}_{ijk}X^iX^jX^k+\frac{1}{2}\mathcal{F}_{ijr}X^iX^jX^r+\frac{1}{2}\mathcal{F}_{irs}X^iX^rX^s,
    \end{aligned}    
\end{equation}
where $i,j,k\in\mathcal{J}_1$ and $r,s\in\mathcal{J}_3$.

\begin{table}[H]
	\begin{center}
		\begin{tabular}{ | c | c  | } 
			\hline
			\rowcolor{colorloc4!20} Supergravity string & Constraints
			\\[.05cm]
			\Xhline{1pt} 
            {Positivity constraints}& ${\cal F}_{IJK} \geq 0$ \quad $\forall I, J, K
            \in\mathcal{J}$
			\\[.05cm]
            \hline
			\hline

            ${\bf p}^{(0)} = {\bm \delta}^0$ & $\mathcal{J}_2 = \varnothing$
			\\[.05cm]
            \hline
			\hline
			${\bf p}^{(i)} = {\bm \delta}^i$ & $\mathcal{F}_{0ir} \neq 0$ \quad $\forall\, i \in \mathcal{J}_1$ and $\forall\, r \in \mathcal{J}_3$
			\\[.05cm]
            \hline
			\hline
			${\bf p}^{(r)} = {\bm \delta}^r$ & $\mathcal{F}_{rst} = 0$ \quad $\forall\, r,s,t \in \mathcal{J}_3$
			\\[.05cm]
            \hline
			\hline
			${\bf p}^{(0)} = {\bm \delta}^0$ \,, ${\bf p}^{(r)} = {\bm \delta}^r$ & $c_i\mathcal{F}_{jab}=c_j\mathcal{F}_{iab}  \quad \forall i,j \in \mathcal{J}_1$ \, and $ \forall  a,b\in\{0\}\sqcup\mathcal{J}_3$ \\
            \hline   
		\end{tabular}
		\caption{\footnotesize{Consistency constraints in Class A limits. Using the supergravity string probe with 
        elementary charge vector $\mathbf{p}$ as specified in the left column and enforcing the correct
        signature of the 't Hooft anomaly matrix on its worldsheet yields the corresponding constraint
        in the right column.}\label{tab:N2_Sugra_JA}}
	\end{center}
\end{table}

To determine the infinite distance behavior of Class A limits 
$\lambda\to\infty$, first recall that $X^0\sim\lambda$, while the coordinates $X^r$, 
$r\in\mathcal{J}_3$, obey the general bound $X^r\precsim\lambda$.
The prepotential constraint \eqref{eq:N2_Fconstr} implies that $X^i\precsim\lambda^{-2}$ for each
$i\in\mathcal{J}_1$. In fact, to realize \eqref{eq:N2_Fconstr}, there must
exist at least one $i_0\in\mathcal{J}_1$ with $X^{i_0}\sim\lambda^{-2}$ because
at least one term in the prepotential must be parametrically constant while all others 
must scale like $\precsim1$. 
This can be combined with the  constraint from Table \ref{tab:N2_Sugra_JA} to compute the
scalings of the elementary tensions to be
\begin{equation}  \label{eq:stringtensions-A}
    \mathcal{F}_0\supset \mathcal{F}_{00i} X^0 X^i\sim\lambda^{-1},\qquad
    \mathcal{F}_i\supset \mathcal{F}_{00i}(X^0)^2\sim\lambda^{2},\qquad
    \mathcal{F}_r\supset \mathcal{F}_{0ir}X^0X^i\sim\lambda^{-1},
\end{equation}
where $i\in\mathcal{J}_1$ and $r\in\mathcal{J}_3$. 

A priori, these scalings do not automatically carry over to those of the gauge kinetic matrix
\eqref{eq:N2_a} due to possible cancellations among the positive and negative contributions
therein. However, for all $i,j\in\mathcal{J}_1$, one immediately finds that
\begin{equation}
    f_{ij}=\mathcal{F}_i\mathcal{F}_j-\mathcal{F}_{ij}\sim\lambda^4\,,
\end{equation}
and there are no other elements of $(f_{IJ})$ that can scale faster with $\lambda$. 
In other words, $\mathcal{Q}_{\mathrm{max}}^2\sim\lambda^4$, which by electric-magnetic duality
means that there must exist one or several one-form gauge fields becoming weakly coupled at the
rate
\begin{equation} \label{eq:qminClassA}
\mathfrak{q}^2_{\mathrm{min}}\sim\lambda^{-4} \,.
\end{equation}
To identify the limit as either a vector or a tensor limit, we must compare this minimal 
asymptotic one-form coupling to the two-form couplings along supergravity string tensor
directions. This requires us to establish lower bounds on the scaling of the gauge kinetic
matrix. For a generic elementary supergravity string charge vector $\mathbf{p}\in
\mathcal{C}_{{\rm\scriptscriptstyle SUGRA}}$, the corresponding two-form coupling reads
\begin{equation}
    \mathcal{Q}_{\mathbf{p}}^2=\frac{1}{2}(p^0)^2f_{00}+p^0f_{0i}p^i+p^0f_{0r}p^r
    +\frac{1}{2}p^if_{ij}p^j+p^if_{ir}p^r+\frac{1}{2}p^rf_{rs}p^s.
\end{equation}

Consider first the supergravity string charges with $p^I\geq0$.
We can assume $p^i=0$ for all $i\in\mathcal{J}_1$ as otherwise
$\mathcal{Q}_{\mathbf{p}}^2\sim\lambda^4$ as noted before. To arrive at a lower bound
for $\mathcal{Q}_{\mathbf{p}}^2$ it is therefore enough to find bounds on $f_{00}$ and
$f_{rr}$ for each $r\in\mathcal{J}_3$. 
In fact, we will show momentarily that 
\begin{equation} \label{eq:f00frr}
f_{00} \sim \lambda^{-2} \,, \qquad f_{rr} \sim \lambda^{-2} \,,
\end{equation}
and hence $\mathcal{Q}^2_{\bf p} \succsim \lambda^{-2}$ for $p^I \geq 0$.
If the K\"ahler cone and its dual are simplicial, this covers all supergravity string charges.
More generally, some of the supergravity string charges can be negative in the coordinates
corresponding to the K\"ahler subcone with $X^I \geq 0$, as chosen for the description of the
limit. In that case, pick any one such supergravity string charge ${\bf p}_0$. 
We can then switch to a different K\"ahler subcone containing the same limiting point such that
this supergravity string has non-negative charge in the new basis and repeat the analysis, see Figure \ref{Fig:Cones_5D}.\footnote{We thank Damian van de Heisteeg for discussions on this point.} This
procedure allows us to assume from now on, and without loss of generality, that $p^I \geq 0$ for
all supergravity string charges. In particular, together with \eqref{eq:f00frr}, we conclude
that 
 \begin{equation}
 \mathcal{Q}^2_{\bf p} \succsim \lambda^{-2} \,  \quad \forall {\bf p} \in {\cal C}_{\text{\tiny{SUGRA}}} \,.
 \end{equation}
By comparison with (\ref{eq:qminClassA}), this identifies the Class A limits as vector limits. 

To show \eqref{eq:f00frr}, consider first $f_{00}$, whose leading part at 
$\mathcal{O}(\lambda^{-2})$ can be written as
\begin{equation}
    f_{00}\vert_{\mathcal{O}(\lambda^{-2})}
    =\mathcal{F}_{00i}X^i\left( \mathcal{F}_{00j}(X^0)^2X^j+2\mathcal{F}_{0jr}X^0X^jX^r-1\right)
   +\mathcal{F}_{00i}\beta_s(i)\mathcal{F}_{0jr}X^iX^jX^rX^s,
   \label{eq:f00-A-lead}
\end{equation}
where the implicit sums run only over those $i\in\mathcal{J}_1$ for which $X^i\sim\lambda^{-2}$
and those $r\in\mathcal{J}_3$ with $X^r\sim\lambda$. We have also introduced the ratio 
\begin{equation}
\beta_r(i)=\frac{\mathcal{F}_{0ir}}{\mathcal{F}_{00i}} \,.
\end{equation}
Importantly, this ratio is independent of $i$, as follows from the relation \eqref{eq:unique1A}
derived from certain supergravity string constraints. Using this, we can factor out the sum 
$\mathcal{F}_{00i}X^i$ and make use of the asymptotic hypersurface constraint (with the same
summation conventions as in \eqref{eq:f00-A-lead}),
\begin{equation}
    \lim_{\lambda\rightarrow\infty}\mathcal{F}[{\bf X}]
    =\frac{1}{2}\mathcal{F}_{00i}(X^0)^2X^i+\mathcal{F}_{0ir}X^0X^iX^r
    +\frac{1}{2}\mathcal{F}_{irs}X^iX^rX^s=1\,,
\end{equation}
to write
\begin{equation}
    \frac{f_{00}\vert_{\mathcal{O}(\lambda^{-2})}}{\mathcal{F}_{00i}X^i}=
    \frac{1}{2}\mathcal{F}_{00j}(X^0)^2X^j+\mathcal{F}_{0jr}X^0X^jX^r+\left(\mathcal{F}_{0jr}\beta_s-\frac{1}{2}\mathcal{F}_{jrs}\right)X^jX^rX^s.
\end{equation}
If we can show that the coefficient of the last term is always non-negative, then
$f_{00}\sim\lambda^{-2}$ in Class A limits. To do so, we consider a supergravity string with
elementary charge vector $\mathbf{p}^{(j)}={\bm\delta}^j$ to write the coefficient as
\begin{equation}
    \frac{k(K_0,K_r)k(K_0,K_s)}{k(K_0,K_0)}-\frac{1}{2}k(K_r,K_s) \geq 0  \,,
\end{equation}
where $k=k^{({\bm\delta}^j)}$ denotes the 't Hooft anomaly matrix on the supergravity string
worldsheet. The inequality is the content of \eqref{eq:ineq-f00-nocancel}, which we derive in
Appendix~\ref{sec:BU_constraints_sugra_JA} as yet another important consequence of the
supergravity consistency conditions. 

We move on to the discussion of possible cancellations in $f_{rr}$. The analysis is similar to
the one of $f_{00}$ so we will be brief. By invoking \eqref{eq:unique1A} again, we can write
\begin{align}
    \begin{split}
        \frac{f_{rr}\vert_{\mathcal{O}(\lambda^{-2})}}{\mathcal{F}_{irr}X^i}=&
        \left(\beta^2\mathcal{F}_{jrr}-\frac{1}{2}\mathcal{F}_{00j}\right)(X^0)^2X^j
        +\left(2\beta\gamma_s\mathcal{F}_{jrr}-\mathcal{F}_{0js}\right)X^0X^jX^s\\
        &+\left(\gamma_s\gamma_t\mathcal{F}_{jrr}-\frac{1}{2}\mathcal{F}_{jst}\right)X^jX^sX^t,
    \end{split}
\end{align}
where we have defined 
\begin{equation}
\beta=\frac{\mathcal{F}_{0jr}}{\mathcal{F}_{jrr}} \quad 
\text{and} \quad  
\gamma_s=\frac{\mathcal{F}_{jrs}}{\mathcal{F}_{jrr}} \,.
\end{equation}
Non-negativity of the three types of coefficients follows from the inequalities 
\eqref{eq:inequl-3}. In fact, the first inequality in \eqref{eq:inequl-3} always holds strictly
so that we conclude $f_{rr}\sim\lambda^{-2}$ for all $r\in\mathcal{J}_3$.

While our analysis shows that every Class A limit is a vector limit with minimal one-form
coupling \eqref{eq:qminClassA}, it remains to investigate how many gauge fields become weakly
coupled at the same time. Candidates for such fields are linear combinations of the gauge fields
$A^i$ with $i\in\mathcal{J}_1$, but the precise number of weakly coupled fields in the limit is
given by the rank of the leading part of the gauge kinetic submatrix
$(f_{ij}\vert_{\mathcal{O}(\lambda^4)})$. The discussion is analogous to that in 
\cite{Cota:2022maf} for weak coupling limits geometrically realized as M-theory compactified on
a Calabi-Yau three-fold. We write\footnote{Here we use the same summation conventions as in
\eqref{eq:f00-A-lead}. Namely, we sum only over those $i\in\mathcal{J}_1$ with
$X^i\sim\lambda^{-2}$ and those $r\in\mathcal{J}_3$ with $X^r\sim\lambda$.}
\begin{align}
    \begin{split}
        f_{ij}\vert_{\mathcal{O}(\lambda^4)}=&
        \left(\frac{1}{2}\mathcal{F}_{00i}(X^0)^2+\mathcal{F}_{0ir}X^0X^r 
        +\frac{1}{2}\mathcal{F}_{irs}X^rX^s\right)\\
        &\times\left(\frac{1}{2}\mathcal{F}_{00j}(X^0)^2+\mathcal{F}_{0jt}X^0X^t
        +\frac{1}{2}\mathcal{F}_{itu}X^tX^u\right)
    \end{split}
    \label{eq:fij-scaling}
\end{align}
and recall that a reduced rank of $(f_{ij}\vert_{\mathcal{O}(\lambda^4)})$ is equivalent to some
of its rows or columns being linearly dependent. 
In particular, $\mathrm{rk}(f_{ij}\vert_{\mathcal{O}(\lambda^4)})=1$ means that for all 
$i,j\in\mathcal{J}_1$ there exist coefficients $c_i,c_j\in\mathbb{Q}$ such that
\begin{equation}
    c_if_{jk}\vert_{\mathcal{O}(\lambda^4)}=c_jf_{ik}\vert_{\mathcal{O}(\lambda^4)}\qquad
    \forall k\in\mathcal{J}_1\,.
    \label{eq:fij-rank1}
\end{equation}
By combining consistency conditions for the supergravity strings along 
$\mathbf{p}^{(i)} = {\bm \delta}^i$ and $\mathbf{p}^{(r)} = {\bm \delta}^r$ we can derive the
important relation
\begin{equation}
    c_i\mathcal{F}_{jab}=c_j\mathcal{F}_{iab}  \quad 
    \text{for all $a,b\in\{0\}\sqcup\mathcal{J}_3$} \,,
\end{equation}
see \eqref{eq:unique1A}, which in turn implies that the relation 
\eqref{eq:fij-rank1} is in fact solved by $c_i= \mathcal{F}_{00i}$.

In total, we have thus shown that there is a unique one-form gauge field associated with the
minimal coupling \eqref{eq:qminClassA}. The corresponding $\mathrm{U}(1)$ gauge potential can be
written in terms of the Chern-Simon couplings $c_i = {\cal F}_{00i}$ as
\begin{equation}
    A_{\rm min}=\sum_{i\in\mathcal{J}_1}
    c_i A^i.
    \label{eq:A-U(1)KK}
\end{equation}
To conclude, infinite distance limits of Class A are always vector limits. While the two-form
couplings of supergravity strings are bounded from below by
$\mathcal{Q}_{\mathrm{min}}^2\sim\lambda^{-2}$, there exists a unique one-form associated with
$\mathrm{U}(1)_{\rm min}$, see \eqref{eq:A-U(1)KK}, whose coupling scales as 
$\mathfrak{q}_{\mathrm{min}}^2\sim\lambda^{-4}$. We will come back to the interpretation of this
weakly coupled one-form gauge field in light of the Emergent String Conjecture in Section
\ref{sec:EmString}.


\section{Class B limits as vector or tensor limits}
\label{sec:N2_Sugra_JB}

According to the definition given in Section \ref{sec:Limits}, Class B limits are those 
infinite distance limits for which 
\begin{equation}
    {\cal F}_{00I} = 0 \quad \forall I\in\mathcal{J} \,,
\end{equation}
or $\mathcal{J}_1=\varnothing$ in short. Using the same philosophy as for Class A limits, we can
show that for every index $\mu$ (for which by the above ${\cal F}_{00\mu} =0$) one finds
another index $\nu$, possibly equal to $\mu$, so that ${\cal F}_{0 \mu \nu} \neq 0$.
In the notation of \eqref{eq:N2_Sugra_J_split}, this means that $\mathcal{J}_3=\varnothing$.
The proof can be found in Appendix \ref{app:BU_constraints}, see also Table
\ref{tab:N2_Sugra_JB}.
\begin{table}[H]
	\begin{center}
		\begin{tabular}{ | c | c  | } 
			\hline
			\rowcolor{colorloc4!20} Supergravity string & Constraints
			\\[.05cm]
			\Xhline{1pt} 
            {Positivity constraints}& ${\cal F}_{IJK} \geq 0$ \quad $\forall I, J, K$
			\\[.05cm]
			\hline
			\hline
			${\bf p} = (1,1,\ldots, 1)$ & $\forall \mu \, \, \exists \nu: {\cal F}_{0\mu \nu} \neq 0$ 
			\\[.05cm]
			\hline
		\end{tabular}
		\caption{\footnotesize{Consistency constraints in Class B limits. Using the supergravity string probe with 
        elementary charge vector $\mathbf{p}$ as specified in the left column and enforcing the correct
        signature of the 't Hooft anomaly matrix on its worldsheet yields the corresponding constraint
        in the right column.}\label{tab:N2_Sugra_JB}}
	\end{center}
\end{table}
\noindent Using this, together with the splitting of the index set $\mathcal{J}$ as in 
\eqref{eq:N2_Sugra_J_split}, allows us to write the most general prepotential of a theory
allowing for a Class B infinite distance limit as
\begin{equation}
    \label{eq:N2_Sugra_JB_F}
    \begin{aligned}
        \mathcal{F}&= \frac12 \mathcal{F}_{0\mu\nu} X^0 X^\mu X^\nu 
        + \frac16  \mathcal{F}_{\mu\nu\rho} X^\mu X^\nu X^\rho,
    \end{aligned}    
\end{equation}
where $X^0\sim\lambda$ and $X^\mu\precsim\lambda$ for all $\mu\in\mathcal{J}_2$.
It turns out that in order to determine the infinite distance behavior along these limits it is
useful to make the further distinction of whether $X^0$ is the only coordinate scaling as
$\lambda$ or not, i.e. of whether the set $\mathcal{J}_\lambda$ of $X^I$ which scale exactly as
$\lambda$ is given by $\mathcal{J}_\lambda=\{0\}$ or $|\mathcal{J}_\lambda|>1$.

\subsection{Class B with \texorpdfstring{$\mathcal{J}_\lambda=\{0\}$}{Jl=0}: Tensor or vector limits}
\label{sec:ClassB_Jl=1}

Similar to the Class A discussion, we wish to establish bounds on the scalings of the elementary
tensions. Using various scaling arguments which are based on the hypersurface constraint as well
as consistency of supergravity probe strings, we derive in Appendix \ref{app:ClassB1} that in
this class of infinite distance limits the tensions scale like
\begin{equation}
    \mathcal{F}_0\sim\lambda^{-1},\qquad
    \mathcal{F}_\mu\succ\lambda^{-1}
    \label{eq:tensionsB}
\end{equation}
for all $\mu\in\mathcal{J}_2$. Importantly, the proof of the scaling
$\mathcal{F}_0\sim\lambda^{-1}$ shows that any order one contribution to the prepotential comes
from $X^0\mathcal{F}_0$. In other words, there always exist $\rho_0,\sigma_0\in\mathcal{J}_2$
(with $\rho_0=\sigma_0$ possible) such that $\mathcal{F}_{0\rho_0\sigma_0}\neq0$ and
$X^0X^{\rho_0}X^{\sigma_0}\sim 1$ so that it is always a term of the form
$\mathcal{F}_{0\mu\nu}$ which satisfies the hypersurface constraint. Any order one contribution
of the form $\mathcal{F}_{\mu\nu\rho}X^\mu X^\nu X^\rho$ would lead to inconsistencies in the
worldsheet theories of supergravity strings associated with 
$\mathbf{p}={\bm\delta}^{\mu,\nu,\rho}$.

Using these bounds, we can continue and discuss the behavior of the gauge kinetic matrix in
Class B limits with $\mathcal{J}_\lambda=\{0\}$. Notice first that since ${\cal F}_{00I} =0$ for
all $I\in\mathcal{J}$ in the present set of limits, the scaling $f_{00}\sim\lambda^{-2}$ can
never be subject to any cancellations. We will further distinguish two complementary cases in
this set of limits.\\

\noindent\textbf{Tensor limits from Class B.} Let us first assume that 
\begin{equation}
    X^\nu\precsim\lambda^{-1/2} \qquad \forall \nu \in \mathcal{J}_2 \,.
    \label{eq:B_tensorlim}
\end{equation}
Under this assumption we will prove that the most weakly coupled supergravity two-form in the limit is unique and associated with the direction $\mathbf{p}^{(0)}={\bm\delta}^0$ in the supergravity string charge lattice, i.e.
\begin{equation}
    \mathcal{Q}_{\mathrm{min}}^2=\mathcal{Q}_{{\bm\delta}^0}^2=\frac{1}{2}f_{00}\sim\lambda^{-2}.
\end{equation}
To do so, we are first concerned with the submatrix 
\begin{equation}
    f'=(f_{\mu\nu})\vert_{\mathcal{O}(\lambda)}=
    \left(\mathcal{F}_{0\mu\rho}\mathcal{F}_{0\nu\sigma}(X^0)^2X^\rho X^\sigma\vert_{\mathcal{O}(\lambda)}
    -\mathcal{F}_{0\mu\nu}X^0\right),
    \label{eq:changeofcoords}
\end{equation}
which, due to our additional assumption \eqref{eq:B_tensorlim}, is the leading part of $f$ on
the subspace $\mathcal{C}=\langle K_\mu\,\vert\,\mu\in\mathcal{J}_2\rangle\subset\mathcal{A}$. 
Using the signature of the 't Hooft matrix on the supergravity string with $\mathbf{p}^{(0)}=
{\bm\delta}^0$ we now show that $f'$ is always non-zero. Since
$k^{({\bm\delta}^0)}_{\mu\nu}=k_{\mu\nu}=\mathcal{F}_{0\mu\nu}$, $f'$ takes the form
\begin{equation}
    f'_{\mu\nu}=X^0\left(k_{\mu\rho}k_{\nu\sigma}X^0X^\rho X^\sigma-k_{\mu\nu}\right).
\end{equation}
Let $\Lambda:\mathcal{C}\longrightarrow\mathcal{C}'$ be the linear transformation from our
current basis to the eigenbasis $\{\Tilde{K}_0,\Tilde{K}_\alpha,\Tilde{K}_\zeta\}$\footnote{We
do not lose the positive eigenvalue when restricting to $\mathcal{C}'=\Lambda(\mathcal{C})$.} 
of $k$ and let $\Sigma$ denote the inverse. Under this transformation, $f'$ is mapped
to\footnote{More explicitly, under $\Lambda$, the gauge one-forms $A^I$ and the coordinates
$X^I$ transform as
\begin{equation*}
    A^I\longmapsto \Lambda^I_J A^J,\quad X^I\longmapsto \Lambda^I_J X^J\,,
\end{equation*}
while the 't Hooft matrix $k_{IJ}$, the Chern-Simons couplings $\mathcal{F}_{IJK}$, and the
gauge kinetic matrix $f_{IJ}$ transform as
\begin{equation*}
        k_{IJ}\longmapsto \Sigma^L_I\Sigma^M_J k_{LM},\quad 
        \mathcal{F}_{IJK}\longmapsto \Sigma^L_I\Sigma^M_J\Sigma^N_K \mathcal{F}_{LMN},\quad
        f_{IJ}\longmapsto \Sigma^L_I\Sigma^M_J f_{LM}\,,
\end{equation*}
with $\Sigma = \Lambda^{-1}$, which follows from $K_I\longmapsto \Sigma^J_IK_J$.}

\begin{equation}
    \Tilde{f}'_{\mu\nu}
    =X^0\left(\Tilde{k}_{\mu\rho}\Tilde{k}_{\nu\sigma}X^0\Tilde{X}^\rho\Tilde{X}^\sigma
    -\Tilde{k}_{\mu\nu}\right),
\end{equation}
or, in matrix notation,
\begin{equation}
    \Tilde{f}'=X^0\left(\begin{array}{ccc}
       X^0\left(\delta_{0\rho}\Tilde{X}^\rho\right)^2-1 & 0 & 0 \\
       0 & \left(X^0\delta_{\alpha\gamma}\Tilde{X}^\gamma\delta_{\beta\tau}\Tilde{X}^\tau
       +\delta_{\alpha\beta}\right) & 0\\
       0 & 0 & 0
    \end{array}\right),
    \label{eq:Hatf-matrix}
\end{equation}
where $\rho,\gamma,\tau$ run through all new coordinates 
$\Tilde{X}^0,\Tilde{X}^\alpha,\Tilde{X}^\zeta$, which in general can also be negative. 
Regardless, we can see right away that this matrix is non-zero. Thus, our original $f'$ is
non-zero as well. Moreover, it follows from the hypersurface constraint
in the new coordinates $\Tilde{X}^\alpha$ that also $\Tilde{f}'_{00}\neq0$. 

Now we come back to the original claim about the uniqueness of the scaling of 
$\mathcal{Q}^2_{{\bm\delta}^0}\sim\lambda^{-2}$. For this it has to be checked that any other
supergravity charge vector $\mathbf{p}\in\mathcal{C}_{\rm\scriptscriptstyle SUGRA}$ with some
$p^\mu>0$ leads to $\mathcal{Q}_{\mathbf{p}}^2\succ\lambda^{-2}$. This is clear if $f'$ has full
rank, $\mathrm{rk}(f')=n_V$. Hence, we focus on the case where $f'$ has reduced rank, i.e. there
exist some $\mu_0,\nu_0\in\mathcal{J}_2$ with $f'_{\mu_0\rho}=c f'_{\nu_0\rho}$ 
for all $\rho\in\mathcal{J}_2$, where the constant $c$ might vanish. Since the first subleading 
contribution to $(f_{\mu\nu})$ is at order $\mathcal{O}(\lambda^{-1/2})$, we have to exclude
that also this subleading part has reduced rank due to the same rows/\,columns, i.e.  
$f_{\mu_0\rho}\vert_{\mathcal{O}(\lambda^{-1/2})} 
=cf_{\nu_0\rho}\vert_{\mathcal{O}(\lambda^{-1/2})}$. 

To see that this can never happen, we show that the system of equations
\begin{equation}
    \begin{gathered}
        f'_{\mu_0\rho}=c f'_{\nu_0\rho},\qquad
        f_{\mu_0\rho}\vert_{\mathcal{O}(\lambda^{-1/2})} =cf_{\nu_0\rho}
        \vert_{\mathcal{O}(\lambda^{-1/2})}
    \end{gathered}
\end{equation}
cannot have a solution which respects the hypersurface constraint. After adding both equations
and some algebraic manipulations we arrive at the condition
\begin{equation}
    \left(\mathcal{F}_{\mu_0}-c\mathcal{F}_{\nu_0}\right)
    \left(1-\mathcal{F}_{0\rho\sigma}X^0X^\rho X^\sigma\right)
    =\frac{1}{2}\mathcal{F}_{\rho\gamma\delta}X^\rho X^\gamma X^\delta 
    \left(\mathcal{F}_{0\mu_0\nu}-c\mathcal{F}_{\nu_0\nu}\right)X^0X^\nu,
    \label{eq:reduced-rank-cancel}
\end{equation}
for which we consider two cases:
\begin{itemize}
    \item If there exist constants $c_{\mu_0},c_{\nu_0}\in\mathbb{Q}$ such that
    $c_{\nu_0} \mathcal{F}_{\mu_00\alpha} = c_{\mu_0}\mathcal{F}_{\nu_00\alpha}$ for all 
    $\alpha\in\mathcal{J}_2$\footnote{Notice that these coefficients are not directly related
    to those in \eqref{eq:unique1A} due to the different nature of indices appearing.}, then the
    previous condition simplifies to
    \begin{equation}
        \frac{1}{2}\left(\mathcal{F}_{\mu_0\alpha\beta}-
        c\mathcal{F}_{\nu_0\alpha\beta}\right)X^\alpha X^\beta
        \left(1-\mathcal{F}_{0\rho\sigma}X^0X^\rho X^\sigma\right)=0,
    \end{equation}
    in which the right factor approaches the constant value $-1$ asymptotically. Hence, it can
    only be satisfied if the left factor vanishes identically. However, this implies
    $\mathcal{F}_{\mu_0IJ}=c\mathcal{F}_{\nu_0IJ}$ for all $I,J\in\mathcal{J}$ from which we
    would conclude that $K_{\mu_0}$ and $K_{\nu_0}$ are linearly dependent. This contradicts the
    fact that the $K_I$ form a basis of $\mathcal{A}$.

    \item If no such $c_{\mu_0},c_{\nu_0}$ exist, then \eqref{eq:reduced-rank-cancel} cannot be
    satisfied either. Indeed, while the left hand side scales like $\sim\lambda^{1/2}$ (by
    invoking the asymptotic hypersurface constraint), the right hand side vanishes like
    $\sim\lambda^{-1}$.
\end{itemize}
Thus, as soon as $p^\mu>0$ for some $\mu\in\mathcal{J}_2$, we find
$\mathcal{Q}_{\mathbf{p}}^2\succsim\lambda^{-1/2}$ and conclude that only
$\mathcal{Q}_{{\bm\delta}^0}^2$ scales like $\sim\lambda^{-2}$.

Additionally, with \eqref{eq:Hatf-matrix} we have shown that the most weakly coupled one-forms
scale like 
\begin{equation}
    \mathfrak{q}_{\mathrm{min}}^2\sim\lambda^{-1} \,,
\end{equation}
while the number $N_{\rm min}$ of such weakly coupled gauge fields is given by 
\begin{equation}
    N_{\rm min} = {\rm rk}(f') \geq 1 \,.
\end{equation}
We can therefore conclude that Class B limits with $\mathcal{J}_\lambda=\{0\}$ and 
$X^\mu\precsim\lambda^{-1/2}$ are tensor limits as defined in Section \ref{sec:Limits}.

Before moving on, we note that all components $f_{0\mu}$ of the gauge kinetic matrix scale as
$\lambda^{-2}$ as well. However, this is not a problem since each linear combination of two-form
fields whose coupling involves $f_{0\mu}$ also includes $f_{\mu\mu}$ and hence scales slower
than $\lambda^{-2}$ in the limit. In this sense, there is no gauge kinetic mixing between $A^0$ 
and the $A^\mu$ in the limit.\\

\noindent\textbf{Vector limits from Class B.} Now, let us consider the complementary case with
respect to the one studied above: namely, let us assume that there exists at least one
$\mu_0\in\mathcal{J}_2$ such that 
\begin{equation}
    X^{\mu_0}\sim\lambda^{-1/2+x}   \,, 
    \qquad    \text{for some $x\in\left(0,\frac{3}{2}\right)$} \,.
\end{equation}
Without loss of generality we assume that $x$ is the largest such parameter in the limit.
As before, we are interested in bounds on the gauge kinetic matrix.

For an upper bound, notice that the hypersurface constraint implies
$\mathcal{F}_{0\mu_0\mu_0}=0$. By definition of the index set $\mathcal{J}_2$, see
\eqref{eq:N2_Sugra_J_split_2}, there exists some $\mu_0\neq\nu\in\mathcal{J}_2$ with
$\mathcal{F}_{0\mu_0\nu}\neq0$. We can therefore say
\begin{equation}
    \mathcal{Q}_{\mathrm{max}}^2\supset\frac{1}{2}f_{\nu\nu}\sim\lambda^{1+2x}
    \sim\mathfrak{q}_{\mathrm{min}}^{-2},
    \label{eq:B12-maxQ}
\end{equation}
where it follows readily that this in fact the upper bound on the scaling of the gauge kinetic
matrix. Furthermore, this shows that the scaling $f_{\mu\mu}\sim\lambda^{1+2x}$ holds for all 
$\mu\in\mathcal{J}_2$ with $\mathcal{F}_{0\mu_0\mu}\neq0$.

To arrive at a lower bound, we invoke the BPS nature of supergravity strings. Since the
Yukawa-term in the no-force condition \eqref{eq:N2_BPSstr} is non-negative, we can estimate
\begin{equation}
    \mathcal{Q}_{\mathrm{min}}^2\geq\frac{2}{3}\mathcal{T}_{\mathrm{min}}^2\succ\lambda^{-2},
\end{equation}
where the second step follows via \eqref{eq:tensionsB}. We also note that parametrically
different scales are allowed in the first inequality, i.e. the case $\mathcal{Q}_{\mathrm{min}}
\succ\mathcal{T}_{\mathrm{min}}$ is included.\footnote{Since we are considering only supergravity
strings which do not decouple from gravity in the limit, we actually expect 
$\mathcal{Q}_{\mathrm{min}}\sim\mathcal{T}_{\mathrm{min}}$.} This implies
\begin{equation}
    \mathcal{Q}^2_{\mathrm{min}}\succ\lambda^{-2}
    \label{eq:B12-minQ}
\end{equation}
and together with \eqref{eq:B12-maxQ} we have shown that limits of Class B with 
$\mathcal{J}_\lambda=\{0\}$ and at least one $X^{\mu_0}\succ\lambda^{-1/2}$ are always vector
limits.

Having established the vector nature of the current set of limits, we wish to determine the
number of one-form gauge fields becoming weakly coupled at the rate \eqref{eq:B12-maxQ}.
Collecting indices $\nu\in\mathcal{J}_2$ with $\mathcal{F}_{0\mu_0\nu}\neq0$ in the subset 
$\varnothing\neq\mathcal{J}_2^{\mu_0}\subseteq\mathcal{J}_2$ and comparing with the Class A
analysis in Section \ref{sec:N2_Sugra_JA}, we know that the number of leading one-form gauge
fields is given by the rank of the gauge kinetic submatrix 
$(f_{\mu\nu}\vert_{\mathcal{O}(\lambda^{1+2x})})$, where $\mu,\nu\in\mathcal{J}_2^{\mu_0}$.
Without loss of generality, we can assume that there is only one such
$\mu_0\in\mathcal{J}_2$.\footnote{If $X^{\mu_0}\sim\lambda^{-1/2+x}\sim X^{\nu_0}$, then 
$\mathcal{F}_{0\mu_0\mu_0}=\mathcal{F}_{0\mu_0\nu_0}=\mathcal{F}_{0\nu_0\nu_0}=0$ by the
hypersurface constraint. Thus, on the supergravity string with 
$\mathbf{p}^{(0)}={\bm\delta}^0$ we have $K_{\mu_0}\vert_{\mathcal{B}^0}\propto
K_{\nu_0}\vert_{\mathcal{B}^0}$.} Then the rank of the relevant submatrix is one if and
only if there exist $c_{\mu},c_{\nu}\in\mathbb{Q}$ for all 
$\mu,\nu\in\mathcal{J}_2^{\mu_0}$ such that
\begin{equation}
    c_\mu \mathcal{F}_{0\mu_0\nu}=c_\nu \mathcal{F}_{0\mu_0\mu},
\end{equation}
which is satisfied for $c_\mu=\mathcal{F}_{0\mu_0\mu}$. Thus, also in vector limits which are
realized as limits of Class B with $\mathcal{J}_\lambda=\{0\}$ we find a unique leading one-form
gauge field, given by
\begin{equation} \label{Amin-B1}
A_{\rm min} = \sum_{\mu\in\mathcal{J}_2^{\mu_0}}c_{\mu} A^\mu \,.
\end{equation}

\subsection{Class B with \texorpdfstring{$|\mathcal{J}_\lambda|>1$: Vector limits}{|Jl|>1}}
\label{sec:ClassB_Jl>1}

We will first define a new splitting of the index set $\mathcal{J}$ which is better suited for
the present class of limits. Denoting indices in $\mathcal{J}_\lambda$ by $a,b,\ldots$, we define
the sets
\begin{equation}
    \begin{gathered}
        \mathcal{J}_2'=\{\mu'\in\mathcal{J}_2\,:\,
        \mathcal{F}_{ab\mu'}=0\,\,\,\,\forall\,a,b\in\mathcal{J}_\lambda\},\\
        \mathcal{J}_2''=\{\mu''\in\mathcal{J}_2\,:\,
        \mathcal{F}_{ab\mu''}\neq0\,\,\,\,\forall a\neq b\in\mathcal{J}_\lambda\}.
    \end{gathered}
    \label{eq:JB2-split}
\end{equation}
Then it holds that $\mathcal{J}=\mathcal{J}_\lambda\sqcup\mathcal{J}_2'\sqcup\mathcal{J}_2''$
as we show in Appendix \ref{app:B-Jl>1}. In particular, 
\begin{equation}
    X^{\mu''}\precsim\lambda^{-2} \quad \forall \, \mu''\in\mathcal{J}_2'', 
    \qquad X^{\mu'}\prec\lambda \quad \forall \, \mu'\in\mathcal{J}_2' \,.
\end{equation}
In Appendix \ref{app:ClassB1} we furthermore derive prepotential constraints shown in Table 
\ref{tab:N2_Sugra_JB2} from the consistency conditions for supergravity strings.

\begin{table}[H]
	\begin{center}
		\begin{tabular}{ | c | c  | } 
			\hline
			\rowcolor{colorloc4!20} Supergravity string & Constraints
			\\[.05cm]
			\Xhline{1pt} 
            {Positivity constraints}& ${\cal F}_{IJK} \geq 0$ \quad $\forall I, J, K$
			\\[.05cm]
            \hline
			\hline
            \multirow{2}{*}{${\bf p}^{(a)} = {\bm\delta}^a$} & $\mathcal{F}_{a\mu'\nu'}=0$ \\
            & $\mathcal{F}_{a\mu'\nu''}\neq0$
			\\[.05cm]
			\hline
            \hline
            ${\bf p}^{(\mu')} = {\bm\delta}^{\mu'}$ & $\mathcal{F}_{\mu'\nu'\rho'}=0$
            \\[.05cm]
            \hline
		\end{tabular}
		\caption{\footnotesize{Consistency constraints in Class B limits with $|\mathcal{J}_\lambda|>1$.
        Here we use the notation set in \eqref{eq:JB2-split}.}\label{tab:N2_Sugra_JB2}}
	\end{center}
\end{table}

Comparing the constraints in Table \ref{tab:N2_Sugra_JB2} to those for Class A limits in Table
\ref{tab:N2_Sugra_JA}, we see that the sets $\mathcal{J}_2'$ and $\mathcal{J}_3$ behave in very
similar ways. A crucial difference, however, is that in the present case $\mathcal{F}_{aaI}=0$
for all $I\in\mathcal{J}$, otherwise the limit would be of Class A to start with. This seems to
make a direct comparison to the discussion in Section \ref{sec:N2_Sugra_JA} difficult at first
sight.

However, taking inspiration from the geometric analysis in \cite{Lee:2019wij}, we can define a
new set of vector multiplet variables in which these complications can be avoided. 
Consider an arbitrary Class B limit with $|{\cal J}_\lambda|\geq 2$ and split this set as 
${\cal J}_\lambda = \{ 0  \}   \cup {\cal J}'_\lambda  $. Then, by definition, the coordinates
$X^a$ for $a \in {\cal J}_\lambda$ scale to infinity as
\begin{equation}
    X^a = \lambda \, \hat X^a \qquad a = 0, \, \alpha,\qquad \alpha\in {\cal J}'_\lambda \,,
\end{equation}
for finite values of $\hat X^a$, cf.~\eqref{eq:vecSDC-Xhat}. Given 
such a limit, we define the linear combination
\begin{equation}
    K_D :=  \sum_{a\in\mathcal{J}_\lambda} \hat X^a K_a  \, 
\end{equation}
of generators $K_I$ defined in \eqref{eq:Xvec} and take $\{K_D, K_\alpha, K_{\mu'}, K_{\mu''}\}$
as the new generating set in which we describe the limiting point in question as 
 \begin{equation}
    {\bf X} = \lambda \, K_D + X^{\mu'} K_{\mu'} + X^{\mu''} K_{\mu''} \,,
    \qquad X^{\mu'},X^{\mu''}  \prec \lambda \,.
\end{equation}
The key observation is that the Chern-Simons terms involving the new generator $K_D$ obey the
relations
\begin{equation}
    {\cal F}_{D D D} = 0 \,, \qquad {\cal F}_{D D \alpha}  = 0 \,,
    \qquad  {\cal F}_{D D \mu'}  = 0  \,, \qquad   {\cal F}_{D D \mu''}   \neq 0  \,.
\end{equation}
In view of \eqref{eq:N2_Sugra_J_split_1} and \eqref{eq:N2_Sugra_J_split_3}, the generator sets
of this Class B limit can hence be identified with the generator sets in a Class A limit as
follows:
\begin{eqnarray}
    \{ D \} \longleftrightarrow \{ 0 \} \,, \,\,  \qquad 
    {\cal J}_\lambda' \cup {\cal J}_2' \longleftrightarrow  {\cal J}_3 \,,  \, \,  \qquad 
    {\cal J}_2''    \longleftrightarrow  {\cal J}_1 \,.
\end{eqnarray}

This allows us to adopt the notation and results from the discussion of Class A limits.
In particular, there exists at least one $\mu''\in\mathcal{J}_2''$ which satisfies $X^{\mu''}\sim\lambda^{-2}$ and
it also holds that
\begin{equation}
    \mathfrak{q}_{\mathrm{min}}^{-2} \sim\mathcal{Q}_{\mathrm{max}}^2
    \sim\lambda^4\sim f_{\mu''\nu''}
\end{equation}
for all $\mu'',\nu''\in\mathcal{J}_2''$. Furthermore, translating the discussion around
(\ref{eq:fij-scaling}) to the present context, the rank of the leading part of this gauge
kinetic submatrix $(f_{\mu''\nu''}|_{{\cal O}(\lambda^4)})$ is found to be
\begin{equation}
    {\rm rk}(f_{\mu''\nu''}|_{{\cal O}(\lambda^4)}) = 1 \,.
\end{equation}
This establishes the existence of exactly one linear combination of one-form gauge fields
becoming weakly coupled at the fastest scale $\mathfrak{q}_{\rm min}^2 \sim \lambda^{-4}$.

Similarly, the arguments spelled out for Class A limits show that all supergravity tensor gauge
couplings scale more slowly, more precisely
\begin{equation}
    \mathcal{Q}_{\mathrm{min}}^2\sim\lambda^{-2},\qquad
    \mathcal{Q}_{\mathrm{max}}^2\sim\mathfrak{q}_{\mathrm{min}}^{-2}\sim\lambda^4 \,.
    \label{eq:scaling-B2}
\end{equation}
In showing the relation $\mathcal{Q}_{\mathrm{min}}^2\sim\lambda^{-2}$ one has to scan over all
supergravity string charges. Due to the change of generator sets, some of these charges with
$p^I \geq 0$ in the original coordinates may have some negative components along the directions
associated with the generators $K_a$. In this case, one can always adjust the generators by 
switching to a new K\"ahler subcone in which the charge in question has non-negative coordinates.

This establishes that Class B limits with $|\mathcal{J}_\lambda|>1$ behave exactly like Class A
limits: They are characterized as vector limits with a unique weakly coupled one-form vector.

\section{Relation to the Emergent String Conjecture}
\label{sec:EmString}

We now interpret our classification of infinite distance limits as vector or tensor limits in
light of the Emergent String Conjecture \cite{Lee:2019wij}.  

\subsection{Vector limits as decompactification limits to six dimensions}
Consider first the vector limits, in which, by definition, the most weakly coupled gauge field
is a one-form potential. For the vector multiplet moduli space of five-dimensional
$\mathcal{N}=1$ supergravity we have been able to show that the one-form gauge field 
$A_{\rm min}$ becoming weakly coupled at the fastest rate is unique and given by 
\eqref{eq:A-U(1)KK} and \eqref{Amin-B1} in Class A and Class B limits, respectively. The
uniqueness of $A_{\rm min}$ is a non-trivial consequence of the consistency conditions imposed
by supergravity string probes.

To characterize the nature of this weakly coupled gauge field further, we can calculate the 
scaling of the physical coupling with the geodesic distance \eqref{eq:InfDist_db} in moduli space. Let us illustrate this for vector limits realized as Class A limits,
with $A_{\rm min}$ corresponding to the linear combination \eqref{eq:A-U(1)KK} of gauge fields.

Focusing on those $i\in\mathcal{J}_1$ with $X^i\sim\lambda^{-2}$ and those $r\in\mathcal{J}_3$
with $X^r\sim\lambda$, we can write at leading order
\begin{equation}
    \frac{\mathrm{d}X^0}{\mathrm{d}\lambda}=\Hat{X}^0,\quad
    \frac{\mathrm{d}X^i}{\mathrm{d}\lambda}=-2\lambda^{-3}\Hat{X}^i,\quad
    \frac{\mathrm{d}X^0}{\mathrm{d}\lambda}=\Hat{X}^r,
    \label{eq:vecSDC-Xhat}
\end{equation}
where by $\Hat{X}^I$ we denote the finite $\mathcal{O}(1)$-part of $X^I$ after scaling
out the leading behavior with $\lambda$. Using the asymptotic hypersurface constraint
\begin{equation}
    \lim_{\lambda\rightarrow\infty}\mathcal{F}[{\bf X}]=
    \frac{1}{2}\mathcal{F}_{00i}(X^0)^2X^i
    +\mathcal{F}_{0ir}X^0X^iX^r+\frac{1}{2}\mathcal{F}_{irs}X^iX^rX^s=1,
\end{equation}
we find
\begin{equation}
    f_{IJ}\frac{\mathrm{d}X^I}{\mathrm{d}\lambda}\frac{\mathrm{d}X^J}{\mathrm{d}\lambda}
    =6\lambda^{-2}+\mathcal{O}(\lambda^{-3}),
\end{equation}
 which, together with \eqref{eq:InfDist_db}, establishes the scaling 
\begin{equation}
    \mathfrak{q}_{\mathrm{min}}\sim\lambda^{-2}
    =
    \exp\left(- \alpha\,
    \mathrm{d}(X^I_0,X^I_\lambda)\right) \,, \qquad  \alpha = \frac{2}{\sqrt{3}}    =  \sqrt{\frac{d-1}{d-2}} \Big|_{d=5} \,.
    \label{eq:vector-rate1}
\end{equation}
 As stressed in 
\cite{Etheredge:2022opl,Agmon:2022thq,Etheredge:2023odp,Etheredge:2024tok,Rudelius:2023odg, Rudelius:2023mjy}, the 
exponential decay rate $\alpha  = \sqrt{\frac{d-1}{d-2}} \Big|_{d=5} $ coincides with the 
scaling of a Kaluza-Klein gauge field in the decompactification limit from five to six dimensions. 
 We view this as strong evidence in favor of the Emergent String Conjecture, which instructs us to interpret the weakly coupled one-form $A_{\rm min}$ - in absence of a competing tensor field  -
as a Kaluza-Klein vector.

Correspondingly,
a BPS particle charged under this gauge symmetry therefore has a mass decaying in this infinite
distance limit as
\begin{equation} \label{eq:scaling-vector}
    \frac{m_{\scriptscriptstyle\mathrm{KK}}}{M_{\mathrm{Pl}}}\sim
    \exp\left(- \alpha\,
    \mathrm{d}(X^I_0,X^I_\lambda)\right) \,,
\end{equation}
where we have reinstated the five-dimensional Planck mass $M_{\rm P}$ for clarity.
In fact, the Asymptotic Tower Weak Gravity Conjecture \cite{Heidenreich:2015nta,Montero:2016tif, Andriolo:2018lvp,Cota:2022maf,Cota:2022yjw,FierroCota:2023bsp} predicts a tower of (at best
marginally super-) extremal charged states at this mass scale, which in the present case must play the role of the Kaluza-Klein tower.

Of course, the existence of a tower of at most marginally super-extremal states charged under
the weakly coupled gauge field $A_{\rm min}$, i.e. the Asymptotic Weak Gravity Conjecture, is 
a non-trivial assumption going beyond our analysis of the limits in supergravity, and so is, a
priori, the interpretation of the tower as a Kaluza-Klein tower. For the latter, independent
bottom-up arguments have been provided in 
\cite{Basile:2023blg,Basile:2024dqq,Bedroya:2024ubj,Herraez:2024kux}.
The fact that the exponential rate is given by $\alpha = \sqrt{\frac{d-1}{d-2}} \Big|_{d=5} $ 
strongly supports this claim in the present context. 
In fact, already \cite{Etheredge:2022opl,Rudelius:2023odg}  have observed this scaling for a special type of limits in the vector 
multiplet moduli space of five-dimensional supergravity. 
Our analysis rigorously excludes the possibility of asymptoting towards a theory in more than
$d=6$ dimensions by taking a limit in vector multiplet moduli space: Apart from our result of
uniqueness of the leading weakly coupled gauge field for all possible limits in vector multiplet
moduli space, such a higher dimensional origin would modify the scaling of the exponent in 
\eqref{eq:scaling-vector}. Our reasoning can hence be summarized as the statement that every 
five-dimensional ${\cal N}=1$ supergravity admitting a vector limit in its vector multiplet
moduli space has a six-dimensional origin.

This is in agreement with the geometric situation of M-theory on Calabi-Yau threefolds
\cite{Lee:2019wij}. Indeed, the only possible vector limits are limits of Type $T^2$, in which 
the Calabi-Yau threefold is a genus-one fibration and the limit is nothing but the F-theory 
limit of M-theory.

\subsection{Tensor limits as emergent string limits}
Tensor limits arise only in infinite distance limits of Class B with a single coordinate $X^0$
diverging at the fastest rate as $X^0 \sim \lambda$, i.e. $\mathcal{J}_\lambda=\{0\}$, and all
other coordinates $X^\mu$, $\mu\in\mathcal{J}_2$, satisfying the bound 
$X^\mu\precsim\lambda^{-1/2}$. 

It is again instructive to compute how the couplings of the unique weakly coupled two-form and
its accompanying weakly coupled one-forms,
\begin{equation}
    \mathcal{Q}_{\mathrm{min}}^2\sim\lambda^{-2}\sim\mathfrak{q}_{\mathrm{min}}^4,
\end{equation}
scale with the geodesic distance. The asymptotic hypersurface constraint in this case reads
\begin{equation}
    \lim_{\lambda\rightarrow\infty}\mathcal{F}[{\bf X}]
    =\frac{1}{2}\mathcal{F}_{0\mu\nu}X^0X^\mu X^\nu=1,
\end{equation}
where we sum only over those $\mu,\nu\in\mathcal{J}_2$ which satisfy $X^\mu\sim X^\nu\sim 
\lambda^{-1/2}$. At leading order, this also allows us to write
\begin{equation}
    \frac{\mathrm{d}X^0}{\mathrm{d}\lambda}=\Hat{X}^0,\qquad 
    \frac{\mathrm{d}X^\mu}{\mathrm{d}\lambda}=-\frac{1}{2}\lambda^{-3/2}\Hat{X}^\mu.
\end{equation}
The integrand of \eqref{eq:InfDist_db} is then readily computed to be
\begin{equation}
    f_{IJ}\frac{\mathrm{d}X^I}{\mathrm{d}\lambda}\frac{\mathrm{d}X^J}{\mathrm{d}\lambda}
    =\frac{3}{2}\lambda^{-2}+\mathcal{O}(\lambda^{-3}),
\end{equation}
 and correspondingly
\begin{equation}
    \label{eq:rate-tensor}\mathcal{Q}^{1/2}_{\mathrm{min}}\sim\mathfrak{q}_{\mathrm{min}}\sim
    \exp\left(-   \alpha \mathrm{d}(X_0,X_\lambda)\right) \,, \qquad \alpha = \frac{1}{\sqrt{3}} = \frac{1}{\sqrt{d-2}}\Big|_{d=5} \,.
\end{equation}
 This agrees with the scaling of the gauge coupling for the Kalb-Ramond two-form of a critical string theory in $d=5$ dimensions in the limit of small string coupling \cite{Lee:2019wij,
Etheredge:2022opl,Agmon:2022thq,Rudelius:2023mjy,Rudelius:2023odg}.
  The tension of a BPS string charged under this two-form  scales like
$\mathcal{Q}_{\mathrm{min}}$. The string scale associated with this BPS string is therefore set
by $\mathcal{Q}_{\mathrm{min}}^{1/2}$ and we find
\begin{equation} \label{eq:Mstr-scaling}
    \frac{M_{\rm str}}{M_{\mathrm{P}}}\sim
    \exp\left(-\alpha \, \mathrm{d}(X_0,X_\lambda)\right) \,,
\end{equation}
as expected for an emergent string limit in five dimensions. 

A crucial characteristic of emergent string limits is that the string becoming tensionless
is in fact a \textit{critical} string.
The appearance of the correct scaling behavior \eqref{eq:rate-tensor} is very suggestive 
that this is the case, as stressed already in \cite{Etheredge:2022opl,Rudelius:2023odg}.
Another necessary condition for criticality are uniqueness of the string, and
hence of the weakly coupled two-form, and the appearance of at least one accompanying 
one-form gauge field becoming weakly coupled at the same rate. Both requirements have been
established for the tensor limits in the vector multiplet moduli space with the help of the
supergravity string consistency conditions.

Further information on the nature of the asymptotically tensionless string is encoded, among 
other things, in the anomaly polynomial associated with the gravitational anomaly on the 
worldsheet. This part of the anomaly polynomial is related, by anomaly inflow, to the pullback
of higher-derivative gravitational Chern-Simons couplings in the five-dimensional bulk, which
are of the form \cite{Antoniadis:1997eg,Grimm:2017okk}
\begin{equation} \label{ARRcoupling}
    S_{ARR}= \frac{1}{192}\int_{\mathbb{R}^{1,4}}C_I A^I\wedge
    \mathrm{tr}\left({\cal R}\wedge {\cal R}\right)
\end{equation}
with $C_I\in\mathbb{Z}$ suitable couplings. 
 The gravitational anomaly on the string is in turn expressed in terms of the central charges of the SCFT to which the worldsheet theory flows in the infra-red.
  Applied to the tensionless strings  in the tensor limits, which
carry charge $\mathbf{p}^{(0)}={\bm\delta}^0$, this reasoning expresses the left- and 
right-moving central charges of the string as
\cite{Katz:2020ewz}
\begin{equation}
    c_L=C_0,\qquad c_R=\frac{1}{2}C_0.
\end{equation}
Unitarity of the worldsheet theory therefore requires $C_0\in\mathbb{N}_0$. The right-moving
central charge $c_R$ is furthermore related to the $\mathrm{SU}(2)_R$ anomaly coefficient
$k_R\in\mathbb{Z}$ via $6k_R=c_R$ and hence $c_R\in6\mathbb{N}_0$. This yields the 
constraint
\begin{equation}
    C_0\in12\mathbb{N}_0 \,.
\end{equation}

However, demanding consistency with the Emergent String Conjecture allows us to go beyond this
result: For the string to be a critical string, it has to correspond either to a Type II string,
with $c_L = c_R = 0$, or to a heterotic string, with $c_L =0$, $c_R = 12$. Consequently, the
Emergent String Conjecture predicts the two possible values 
\begin{equation}
    \begin{aligned}
        &C_0 = 0  \quad &\text{(Type II emergent string limit)\,,} \\
        &C_0 = 24  \quad &\text{(heterotic emergent string limit)}
    \end{aligned}
\end{equation}
for the higher-derivative coupling \eqref{ARRcoupling} of a supergravity theory admitting a 
tensor limit with asymptotically tensionless string charge ${\bf p}^{(0)} = {\bm \delta}^0$.
More compactly, this can be phrased as the statement that for a basis of K\"ahler cone 
generators 
\begin{equation} \label{eq:constraintsESC}
    {\cal F}_{00I} =0 \qquad \forall I\in\mathcal{J} \qquad   
    \stackrel{\text{ESC}}{\Longrightarrow} \qquad  C_0 \in \{0,24\}\,.
\end{equation}
 Proving this general prediction independently, e.g. by more refined probe brane arguments, would therefore constitute substantial further evidence for the Emergent String Conjecture.

\section{Conclusions and outlook}
\label{sec:Conclusions}

In this work we have provided a systematic classification of infinite distance limits in the vector multiplet moduli space of five-dimensional ${\cal N}=1$ supergravity theories, irrespective of a possible UV completion of the theory by string or M-theory. 
The key assumption entering our analysis is a weak version of the BPS completeness hypothesis. Specifically, we assume that 
for each generator of 
the lattice of supergravity string charges ${\cal C}_{\text{\tiny{SUGRA}}}$,  in the sense of \cite{Katz:2020ewz}, a physical BPS string exists in the spectrum for at least one multiple of the minimal charge. This relatively mild hypothesis, combined with consistency of the supergravity strings which it predicts, has surprisingly far-reaching consequences.

First, as explained in Section \ref{sec:5D_Constraints}, BPS completeness together with the signature condition (\ref{eq:Pos_tHmat_sig})  obtained in \cite{Katz:2020ewz} imply that the Chern-Simons couplings entering the prepotential must be non-negative. 
  The signature condition (\ref{eq:Pos_tHmat_sig}) allows us to derive a number of further constraints on the Chern-Simons couplings, which are summarised in Tables \ref{tab:N2_Sugra_JA}, \ref{tab:N2_Sugra_JB} and \ref{tab:N2_Sugra_JB2}.
 With these results at hand, we have provided a classification of infinite distance limits in vector multiplet moduli space which closely resembles the systematics developed in \cite{Lee:2019wij} for theories obtained as M-theory compactifications on Calabi-Yau threefolds. 
  That essentially the same pattern applies even more generally, without having to assume 
such a geometric origin, is quite remarkable and a rather non-trivial consequence of the signature constraint of \cite{Katz:2020ewz} for supergravity probe strings.

 The main finding of our analysis is then that all vector multiplet moduli space limits are of one of the following two types:
 In a vector limit, the gauge field becoming weakly coupled at the parametrically fastest rate is a one-form potential, and in fact there is only a unique such weakly coupled gauge field. 
 The rate at which its gauge coupling asymptotes to zero, given by \eqref{eq:vector-rate1}, precisely matches the behavior of a Kaluza-Klein gauge field for decompactification from five to six dimensions \cite{Etheredge:2022opl,Agmon:2022thq,Etheredge:2023odp,Rudelius:2023odg,Etheredge:2024tok,Rudelius:2023mjy}. 
  This strongly suggests that vector limits in the vector multiplet moduli space are decompactification limits from five to six dimensions.

  The second possible limit is a tensor limit, in which a unique two-form gauge field becomes weakly coupled faster than any other two-form coupling to a supergravity string, and is always accompanied by at least one equally weakly coupled one-form gauge field.
   The scaling rate \eqref{eq:rate-tensor} 
   matches exactly the expectations for the Kalb-Ramond two-form that becomes weakly coupled in heterotic or Type II string theory as the dilaton is sent to infinity. We view this as strong evidence for the interpretation of such a tensor limit as an emergent string limit in the sense of \cite{Lee:2019wij}, i.e. as a limit in which a critical fundamental string becomes asymptotically weakly coupled and tensionless with respect to the Planck scale. 

In combination, these results provide intriguing support for the Emergent String Conjecture without having to rely on a UV completion of the five-dimensional supergravity via string or M-theory. 
 They complement the bottom-up ideas developed in \cite{Basile:2023blg,Basile:2024dqq,Bedroya:2024ubj,Herraez:2024kux} arguing for a dichotomy of possible light towers in quantum gravity as Kaluza-Klein towers or as towers with string-like exponential degeneracy.

Under the assumptions stated above,   we have provided evidence for the fact that all minimally supersymmetric  gravity theories in five dimensions with a non-compact vector multiplet moduli space either descend from a theory in six dimensions or contain a potentially weakly coupled subsector corresponding to a heterotic or Type II string theory. In view of the recent bounds of  \cite{Kim:2024hxe} on six-dimensional minimal supergravity theories, this is a welcome result as far as a potential classification of quantum consistent five-dimensional supergravity theories is concerned.

 Clearly, it would be exciting to provide more evidence for this general picture. To lend additional support to the
  interpretation of tensor limits as emergent string limits,
 it would be important to further constrain the worldsheet theory of the candidate critical strings from first principles. 
 For these weakly coupled strings to correspond to critical strings, the higher-derivative gravitational Chern-Simons couplings of the bulk theory must follow the pattern \eqref{eq:constraintsESC}. It would be remarkable if this prediction could be derived entirely by relying on probe brane consistency arguments.
 Furthermore, while the found constraints on the two-derivative Chern-Simons terms generally agrees in its structure with expectations for compactification limits and weakly coupled string limits, a detailed analysis of this match is likewise left for future work. 

 Finally it will be interesting to see to what extent the program of constraining the asymptotics of supergravity theories by probe brane arguments can be extended to theories in other dimensions and with less supersymmetry. We hope to return to these questions in the near future.

 \vspace{3mm}

\noindent{\textbf{Acknowledgments.}} We are grateful to Ivano Basile, Vicente Cort\'es, Naomi Gendler, Damian van de Heisteeg, Seung-Joo Lee, Luca Melotti, Miguel Montero, Cumrun Vafa, Max Wiesner, and Kai Xu for useful discussions.
TW thanks the Harvard theory group for hospitality.
This work is supported in part by Deutsche Forschungsgemeinschaft under Germany’s Excellence Strategy EXC 2121 Quantum Universe 390833306, by Deutsche Forschungsgemeinschaft through a German-Israeli Project Cooperation (DIP) grant “Holography and the Swampland” and by Deutsche Forschungsgemeinschaft through the Collaborative Research Center 1624 “Higher Structures, Moduli Spaces and Integrability”.

\appendix

	
\section{Positivity constraints on the Chern-Simons couplings}
\label{sec:Pos_FIJK}
As advertised in Section \ref{sec:5D_Constraints} and as explicitly seen in Sections
\ref{sec:BU_constraints_sugra_JA} and \ref{sec:BU_constraints_sugra_JB}, non-negativity
of the Chern-Simons couplings entering the action \eqref{eq:N2_S} is crucial for our 
analysis of infinite distance limits in the vector multiplet moduli space of
five-dimensional $\mathcal{N}=1$ supergravities. In this appendix we derive the
positivity constraint
\begin{equation}
    \mathcal{F}_{IJK}\geq0\qquad \forall I,J,K\in\mathcal{J}
    \label{eq:App_FIJK_1}
\end{equation}
in full generality, thereby completing the discussion after \eqref{eq:Pos_III}.
Throughout this section we fix some distinct $I,J,K\in\mathcal{J}$ and denote the
't Hooft anomaly matrix on the supergravity string with elementary charge $\mathbf{p}^{(I)}=
{\bm\delta^I}$ by $k:=k^{({\bm\delta}^I)}$. From \eqref{eq:Pos_tHmat_sig} we know that $k$ is a 
mostly-negative Lorentzian metric on the vector space $\mathcal{A}$. Therefore, we can define
future- and past-directed lightcones $\mathcal{C}_+$ and $\mathcal{C}_-$, the boundaries of
which are populated by light-like vectors whose $k$-norm vanishes. We know
two time-like vectors with positive inner product to lie in the same lightcone.

To arrive at \eqref{eq:App_FIJK_1}, we will start by classifying all cases of the
second generator $K_J$ that need to be considered using the constraints \eqref{eq:Pos_I}
and \eqref{eq:Pos_II}.

First, we assume $k(K_I,K_I)=\mathcal{F}_{III}>0$. As discussed in the main text, we
can choose $K_I$ to be future-directed, i.e. $K_I\in\mathcal{C}_+$. Next, for $I\neq J\in\mathcal{J}$, according to \eqref{eq:Pos_II} we have to distinguish two cases:
\begin{itemize}
    \item[(a)] If $K_J$ is time-like, meaning $k(K_J,K_J)>0$, we know it to lie in the
    union $K_J\in\mathcal{C}_+\cup\mathcal{C}_-$. Since by \eqref{eq:Pos_II} we know 
    that $k(K_I,K_J)\geq0$, $K_J$ cannot lie in $\mathcal{C}_-$ and hence $K_J\in\mathcal{C}_+$.
    \item[(b)] If, on the other hand, $k(K_J,K_J)=0$ and hence $K_J$ is light-like, we conclude
    $K_J\in\partial\mathcal{C}_+\cup\partial\mathcal{C}_-$. To decide which boundary component
    $K_J$ resides in, we recall that  $k(K_I,K_J)\geq0$. If $k(K_I,K_J)>0$, it follows that
    $K_J\in\partial\mathcal{C}_+$. The second possibility, $k(K_I,K_J)=0$, cannot occur, as we now show. To do
    so, assume the contrary so that we have
    \begin{equation}
        k(K_I,K_J)=\mathcal{F}_{IIJ}=0,\qquad k(K_J,K_J)=\mathcal{F}_{IJJ}=0.
        \label{eq:App_FIJK_2}
    \end{equation}
    Expanding $K_I$ and $K_J$ in the eigenbasis of $k$, see \eqref{eq:Pos_tHmat_diag},
    \begin{equation}
        \begin{gathered}
            K_I=a_0\Tilde{K}_0+a_\alpha\Tilde{K}_\alpha+a_\zeta\Tilde{K}_\zeta,\\
            K_J=b_0\Tilde{K}_0+b_\alpha\Tilde{K}_\alpha+b_\zeta\Tilde{K}_\zeta,
        \end{gathered}
        \label{eq:App_FIJK_3}
    \end{equation}
    we can write \eqref{eq:App_FIJK_2} as
    \begin{equation}
        a_0b_0=a_\alpha b_\alpha,\qquad b_0^2=b_\alpha b_\alpha,
    \end{equation}
    where the sum over $\alpha=1,\ldots,r-1$ is left implicit. By the Cauchy-Schwarz inequality
    we can deduce
    \begin{equation}
        a_0^2=\frac{(a_\alpha b_\alpha)^2}{b_0^2}=\frac{(a_\alpha b_\alpha)^2}{b_\alpha b_\alpha}
        \leq a_\alpha a_\alpha,
    \end{equation}
    which in our original notation means $k(K_I,K_I)=\mathcal{F}_{III}\leq0$. This is
    in contradiction to our assumption that $\mathcal{F}_{III}>0$.
\end{itemize}

According to \eqref{eq:Pos_I}, the only other possibility to consider is
$k(K_I,K_I)=\mathcal{F}_{III}=0$, i.e. $K_I\in\partial\mathcal{C}_+\cup\partial\mathcal{C}_-$
is light-like. We again pick $J\neq I$ and distinguish whether $K_J$ is time-like or 
light-like:
\begin{itemize}
    \item[(c)] If $K_J$ is time-like, $k(K_J,K_J)>0$, we may choose $K_J\in\mathcal{C}_+$.
    In case $k(K_I,K_J)>0$, it also follows that $K_I\in\partial\mathcal{C}_+$. The
    other possibility, $k(K_I,K_J)=0$, cannot happen. By employing the expansion 
    \eqref{eq:App_FIJK_3}, we may write the current setup as
    \begin{equation}
        b_0^2>b_\alpha b_\alpha,\qquad a_0b_0=a_\alpha b_\alpha.
    \end{equation}
    Combining these relations with the Cauchy-Schwarz inequality yields
    \begin{equation}
        a_0^2=\frac{(a_\alpha b_\alpha)^2}{b_0^2}<\frac{(a_\alpha b_\alpha)^2}{b_\alpha b_\alpha}
        \leq a_\alpha a_\alpha,
    \end{equation}
    which means $k(K_I,K_I)=\mathcal{F}_{III}<0$, a contradiction.
    \item[(d)] Assume now that also $K_J$ is light-like, $k(K_J,K_J)=0$. If
    $k(K_I,K_J)>0$, both $K_I$ and $K_J$ live in the same boundary component which we may choose
    to be $\partial\mathcal{C}_+$. The case of $k(K_I,K_J)=0$ is more involved and will be
    dealt with in point 3.\,below.
\end{itemize}
With this preparation we can now turn to the proof of \eqref{eq:App_FIJK_1}.
\begin{enumerate}
    \item If $k(K_I,K_I)>0$ as well as $k(K_I,K_J),k(K_I,K_K)>0$, all three generators
    live in the future-directed lightcone $\mathcal{C}_+$ and we deduce 
    $k(K_J,K_K)=\mathcal{F}_{IJK}>0$. By our reasoning in case (a) above, it also follows
    that $\mathcal{F}_{IJK}\geq0$ if $k(K_I,K_J),k(K_I,K_K)\geq0$. Using the arguments from (b)
    this holds in case the generators $K_J$ and/\,or $K_K$ are light-like, as well.

    \item Consider now a light-like $K_I$. If both generators $K_J$ and $K_K$ are time-like, 
    (c) shows that we can deduce $\mathcal{F}_{IJK}\geq0$. Even if one or both are light-like 
    with $k(K_I,K_J),k(K_I,K_K)>0$ this conclusion holds, see (d).

    \item The only case left to consider is when $K_I$ is light-like and at least one of the
    remaining generators, say $K_J$, is also light-like with $k(K_I,K_J)=0$. This means that ${\cal F}_{III} =0$, ${\cal F}_{IJJ} =0$ and ${\cal F}_{IIJ} =0$.
    
    Note first that ${\cal F}_{IJJ}=0$ and ${\cal F}_{IIJ}=0$ imply that also ${\cal F}_{JJJ} =0$, by applying the Cauchy-Schwarz inequality to the Lorentzian metric induced by the 't Hooft anomaly matrix 
    $k^{({\bm \delta}^J)}$.

       Furthermore, assume from now on that ${\cal F}_{IJK} \neq 0$ (otherwise the inequality (\ref{eq:App_FIJK_1}) is trivially satisfied).  
      This is only possible if ${\cal F}_{IIK} \neq 0$ and also ${\cal F}_{JJK} \neq 0$.
     To see this, note that 
     from $\mathcal{F}_{IIK}=0=\mathcal{F}_{III}$ we would conclude $\beta K_I\vert_{\mathcal{B}^I}
    =K_K\vert_{\mathcal{B}^I}$ for some $\beta\in\mathbb{Q}$, where 
     $\mathcal{B}^I\subset\mathcal{A}$ denotes the subspace on which $k$ is non-degenerate,
     see \eqref{eq:Pos_tHmat_diag}.\footnote{See the discussion around \eqref{eq:Pos_JA_I_d} for details of this argument in a similar context.} It follows that $\mathcal{F}_{IJK}=\beta\mathcal{F}_{IIJ}=0$.
     A similar argument shows that $\mathcal{F}_{JJK}\neq0$.

    With this preparation we shall now show that a negative value for ${\cal F}_{IJK}$ leads to an inconsistency. To this end, consider the (finite distance) limit in which
    all $X^L\to0$ for $L\in\mathcal{J}\mysetminus\{I,J,K\}$, leading to the prepotential
    \begin{align}
        \begin{split}
            \mathcal{F}[X]&=\frac{1}{2}\mathcal{F}_{IIK}(X^I)^2X^K
            +\mathcal{F}_{IJK}X^IX^JX^K+\frac{1}{2}\mathcal{F}_{IKK}X^I(X^K)^2\\
            &+\frac{1}{2}\mathcal{F}_{JJK}(X^J)^2X^K+\frac{1}{2}\mathcal{F}_{JKK}X^J(X^K)^2
            +\frac{1}{6}\mathcal{F}_{KKK}(X^K)^3 \,.
        \end{split}
        \label{eq:App_FIJK_4}
    \end{align}
    Let us furthermore scale the remaining coordinates as 
    $X^I\sim X^J\sim\lambda$ and $X^K\sim\lambda^{-2}$ for $\lambda \gg 1$.

The physical charge of a BPS string, characterized by elementary charges ${\bf p}^{(I)} = {\bm \delta}^I$, becomes, to leading order in $1/\lambda$, 
\begin{equation}
{\mathcal Q}_{{\bm \delta}^I}^2 = {\cal F}^2_{I}  - {\cal F}_{II} = \left({\cal F}_{IIK} X^I X^K + {\cal F}_{IJK}  X^J X^K + {\cal O}\left(\frac1{\lambda^4}\right)\right)^2 - {\cal F}_{IIK} X^K \,.
\end{equation}
We see that if ${\cal F}_{IJK} <0$, the terms in bracket can vanish, to leading order, at $X^I = - \frac{{\cal F}_{IJK}}{{\cal F}_{IIK}} X^J$ and the whole expression becomes
negative for $\lambda\gg 1$, but finite.  This is a physical inconsistency and hence we conclude that  ${\cal F}_{IJK} >0$.


\end{enumerate}

To summarize, we have shown that in an appropriate simplicial subcone of the Kähler cone, all
cubic Chern-Simons couplings appearing in the bulk action \eqref{eq:N2_S} must be 
non-negative,
\begin{equation}
    \boxed{
    \mathcal{F}_{IJK}\geq0\quad\forall I,J,K\in\mathcal{J}
    }\,.
\end{equation}

\section{Prepotential constraints from bottom-up limits}
\label{app:BU_constraints}

We shall now examine additional constraints on the Chern-Simons couplings that arise from consistency of
supergravity string probes. These constraints are, in particular, relevant for our analysis of infinite distance limits, but hold more generally. 
As in Appendix \ref{sec:Pos_FIJK}, a pivotal ingredient of the analysis is the 't
Hooft anomaly matrix \eqref{eq:Pos_tHmat_def}, defined on the supergravity string worldsheet,
which must have the signature \eqref{eq:Pos_tHmat_sig}.
For the sake of clarity, we split the discussion of prepotential constraints into those
entering our discussion of Class A and  of Class B limits.

\subsection{Prepotential constraints from Class A limits}
\label{sec:BU_constraints_sugra_JA}

In this section we derive the prepotential constraints that five-dimensional $\mathcal{N}=1$ supergravity satisfy whenever ${\cal F}_{000} =0$, but 
${\cal F}_{00i}  \neq 0$ for some $i$.
 These are also the conditions 
 to host the Class A limits introduced in Section~\ref{sec:N2_Sugra_JA}, to which we refer for notation.
The constraints, succinctly listed in Table~\ref{tab:N2_Sugra_JA}, originate from assuming the coupling with some elementary BPS supergravity strings, and enforcing consistency of their worldsheet theory.

\noindent\textbf{Constraints from the supergravity string with charges ${\bf p}^{(0)} = {\bm \delta}^0$.} 
 We first show that the 
set of indices $\mathcal{J}_2$, defined in \eqref{eq:N2_Sugra_J_split_2}, is empty under the assumption that $\mathcal{J}_1$ is non-empty (Class A). 
 This claim can be proved by contradiction. 
 Indeed, let us assume there exist some indices $\mu\in\mathcal{J}_2$. 
Then, for some indices $\mu,\nu \in\mathcal{J}_2$, there is a non-null Chern-Simons coupling $\mathcal{F}_{0\mu\nu}\neq 0$. 
Notice that, in particular, if $\mathcal{J}_2$ hosts only a single index $\mu\in\mathcal{J}_2$, we need $\mathcal{F}_{0\mu\mu}\neq 0$ for that index. 
Moreover, recall that $\mathcal{F}_{000}=\mathcal{F}_{00\mu} = 0$ due to the definition of $\mathcal{J}_2$ in \eqref{eq:N2_Sugra_J_split_2}.
    
Consider now 
a BPS supergravity string with only non-null, unitary charge in the $0$-th position.
The abelian 't Hooft matrix defined on its  worldsheet  is $k^{({\bm\delta}^0)}_{IJ}=k_{IJ}=\mathcal{F}_{0IJ}$. 
Then, there exists a basis of generators $\Tilde{K}_I$ where the 't Hooft matrix takes the form \eqref{eq:Pos_tHmat_diag}.
In particular, we can expand $K_0$ and $K_\mu$ in this basis as
\begin{equation}
    \label{eq:Pos_JA_I_Aexp}
    K_0=a_0\Tilde{K}_0+a_\alpha\Tilde{K}_\alpha+a_\zeta\Tilde{K}_\zeta,\quad 
    K_\mu=b^\mu_0\Tilde{K}_0+b^\mu_\alpha\Tilde{K}_\alpha+b^\mu_\zeta\Tilde{K}_\zeta\,,
\end{equation}
where we have employed the same notation as in the decomposition of \eqref{eq:Pos_tHmat_diag}. In particular, notice that $a_0\neq0$ since $K_0$ is non-zero.

The coefficients appearing in the expansions \eqref{eq:Pos_JA_I_Aexp} can be related among each other by using  $\mathcal{F}_{000} \neq 0$ and $\mathcal{F}_{00\mu} \neq 0$.
In fact, employing the expansions \eqref{eq:Pos_JA_I_Aexp} yields
\begin{equation}
    \label{eq:Pos_JA_I_a}
    0 = \mathcal{F}_{000}= k(K_0,K_0) = a_0^2-a_\alpha a_\alpha, \qquad 0 = \mathcal{F}_{00\mu}=k(K_0,K_\mu)=a_0b^\mu_0-a_\alpha b^\mu_\alpha \,,
\end{equation}
where we sum over indices $\alpha=1,\ldots,n-r$; hence, the relations \eqref{eq:Pos_JA_I_a} lead to
\begin{equation}
    \label{eq:Pos_JA_I_b}
    a_0^2 = a_\alpha a_\alpha, \qquad b^\mu_0 = \frac{a_\alpha b^\mu_\alpha}{a_0}.
\end{equation}

Now, let us consider the Chern-Simons coupling
\begin{equation}
    \label{eq:Pos_JA_I_c}
    \mathcal{F}_{0\mu\mu} = k(K_\mu,K_\mu)= (b^\mu_0)^2-b^\mu_\alpha b^\mu_\alpha=
    \frac{(a_\alpha b^\mu_\alpha)^2}{a_0^2}-b^\mu_\alpha b^\mu_\alpha \,,
\end{equation}
which must be non-negative, due to \eqref{eq:Pos_II}.
Notice that in \eqref{eq:Pos_JA_I_c} no sum over the index $\mu$ is understood.
Then, applying the Cauchy-Schwarz inequality, \eqref{eq:Pos_JA_I_c} leads to
\begin{equation}
    \label{eq:Pos_JA_I_d}
    \mathcal{F}_{0\mu\mu} \leq\frac{(a_\alpha a_\alpha)(b^\mu_\beta b^\mu_\beta)}{a_0^2}-b^\mu_\alpha b^\mu_\alpha = 0\,,
\end{equation}
whose consistency with \eqref{eq:Pos_II} implies that $\mathcal{F}_{0\mu\mu}=0$.
Thus, the Cauchy-Schwarz inequality in \eqref{eq:Pos_JA_I_d} is necessarily saturated.
In turn, this is only possible if the vectors $K_0$ and $K_\mu$ are linearly dependent on the subspace $\mathcal{B}^0\subseteq\mathcal{A}$ where the 't Hooft matrix $k_{IJ}$ is non-degenerate, namely $K_0\vert_{\mathcal{B}^0}=\beta_{(\mu)} K_\mu\vert_{\mathcal{B}^0}$, where no sum over the index $\mu$ is understood.
Clearly, such a relation holds for every index $\mu \in \mathcal{J}_2$. 

This shows in particular that $k$ cannot have full rank and hence $\mathcal{B}^0$ is a proper
subspace of $\mathcal{A}$. Consequently, we can rewrite 
\begin{equation}
    \label{eq:Pos_JA_I_e}
    \begin{aligned}
        0 &= \mathcal{F}_{00\mu}=k(K_0,K_\mu)= a_0 b_0^\mu - a_\alpha b_\alpha^\mu
        \\
        &= \beta_{(\nu)} \left(b_0^\nu b_0^\mu - b^\nu_\alpha b_\alpha^\mu \right) = \beta_{(\nu)} k(K_\nu,K_\mu)= \beta_{(\nu)}  \mathcal{F}_{0\mu\nu}\,,
    \end{aligned}
\end{equation}
where we have employed the relations \eqref{eq:Pos_JA_I_b} and used the fact that, since $K_0\vert_{\mathcal{B}^0}=\beta_{(\mu)} K_\mu\vert_{\mathcal{B}^0}$, we have $a_0 = \beta_{(\mu)} b_0^\mu$, $a_\alpha = \beta_{(\mu)} b_\alpha^\mu$ for some non-null $\beta_{(\mu)}$.
However, \eqref{eq:Pos_JA_I_e} contradicts the definition \eqref{eq:N2_Sugra_J_split_2} of $\mathcal{J}_2$, unless $\beta_{(\mu)} = 0$; but $\beta_{(\mu)} = 0$ would imply $k(K_0,K_I)=0$ for all $I\in\mathcal{J}$, which in turn contradicts $\mathcal{J}_1\neq\varnothing$. 

Thus, in sum, in order to allow for Class A limits - or equivalently for there to exist some ${\cal F}_{000}=0$, but ${\cal F}_{00i} \neq 0$ for some $i \in {\cal J}_1$ -  the consistent coupling with the BPS supergravity string with fundamental charge ${\bf p}^{(0)} = {\bm \delta}^0$ leads to the condition
\begin{equation}
    \label{eq:Pos_JA_I}
    \boxed{\mathcal{J}_2 = \varnothing } \,.
\end{equation}

\noindent\textbf{Constraints from supergravity strings with charges ${\bf p}^{(i)} = {\bm \delta}^i$.}
The 't Hooft anomaly matrix on the worldsheet of a BPS superstring with elementary charges ${\bf p}^{(i)} = {\bm \delta}^i$ takes the form $k^{({\bm\delta}^i)}_{IJ}=k_{IJ}=\mathcal{F}_{iIJ}$. 
Then, there exists a basis of generators $\{\Tilde{K}_I\}$ where the 't Hooft anomaly matrix \eqref{eq:Pos_tHmat_def} reduces to the simple, diagonal matrix \eqref{eq:Pos_tHmat_diag}.
The original generators $K_0$ and $K_r$, with $r\in\mathcal{J}_3$, can be expanded in this basis as follows
\begin{equation}
    \label{eq:Pos_JA_II_a}
    K_0=a_0\Tilde{K}_0+a_\alpha\Tilde{K}_\alpha+a_\zeta\Tilde{K}_\zeta,\quad
    K_r=c_0\Tilde{K}_0+c_\alpha\Tilde{K}_\alpha+c_\zeta\Tilde{K}_\zeta.
\end{equation}

The coefficients appearing in the expansions \eqref{eq:Pos_JA_II_a} satisfy some non-negativity constraints.
In fact, consider the Chern-Simons couplings $\mathcal{F}_{00i}$ and $\mathcal{F}_{irr}$, for some $i \in \mathcal{J}_1$, $r \in \mathcal{J}_3$.
From \eqref{eq:Pos_II} and the definition \eqref{eq:N2_Sugra_J_split_1} of $\mathcal{J}_1$ we read off for the expansions \eqref{eq:Pos_JA_II_a} that
\begin{subequations}
    \begin{align}
        \label{eq:Pos_JA_II_ba}
        &\mathcal{F}_{00i}=k(K_0,K_0)=a_0^2-a_\alpha a_\alpha>0,
        \\
        \label{eq:Pos_JA_II_bb}
        &\mathcal{F}_{irr}=k(K_r,K_r)=c_0^2-c_\alpha c_\alpha\geq 0,
    \end{align}
\end{subequations}
from which we conclude $|a_0|>\sqrt{a_\alpha a_\alpha}$ and $|c_0|\geq\sqrt{c_\alpha c_\alpha}$.

Now, let us consider the Chern-Simons coupling $\mathcal{F}_{0ir}$, which can be expanded as
\begin{equation}
    \label{eq:Pos_JA_II_c}
    k(K_0,K_r)= \mathcal{F}_{0ir} = a_0c_0-a_\alpha c_\alpha.
\end{equation}
Such couplings can never be zero: Indeed, if $\mathcal{F}_{0ir}=0$, then
\begin{equation}
    \label{eq:Pos_JA_II_d}
    |c_0|=\frac{|a_\alpha c_\alpha|}{|a_0|}<\frac{|a_\alpha c_\alpha|}{\sqrt{a_\alpha a_\alpha}}
    \leq\sqrt{c_\alpha c_\alpha},
\end{equation}
where, in the last step, we have used the Cauchy-Schwarz inequality. However, \eqref{eq:Pos_JA_II_d} contradicts \eqref{eq:Pos_JA_II_bb}, implying that
\begin{equation}
    \label{eq:Pos_JA_II}
    \boxed{\mathcal{F}_{0ir}\neq 0 \quad \text{for all $i\in\mathcal{J}_1$ and $r\in\mathcal{J}_3$} } \,.
\end{equation}

\noindent\textbf{Constraints from supergravity strings with charges ${\bf p}^{(r)} = {\bm \delta}^r$.}
Consider next a BPS supergravity string with fundamental charges ${\bf p}^{(r)} = {\bm \delta}^r$.
 We show that consistency implies  $\mathcal{F}_{rst} = 0$, for all $r,s,t\in\mathcal{J}_3$, under the assumptions of a Class A limit.
In order to show this, consider its 't Hooft matrix $k^{({\bm\delta}^r)}_{IJ}=k_{IJ} = \mathcal{F}_{rIJ}$. 
In the diagonal basis  $\{\Tilde{K}_I\}$, see \eqref{eq:Pos_II_T},  
the generators $K_0$ and $K_s$, $s\in\mathcal{J}_3$, can be expanded  as
\begin{equation}
    \label{eq:Pos_JA_III_a}
    K_0=a_0\Tilde{K}_0+a_\alpha\Tilde{K}_\alpha+a_\zeta\Tilde{K}_\zeta,\qquad
    K_s=b^s_0\Tilde{K}_0+b^s_\alpha\Tilde{K}_\alpha+b^s_\zeta\Tilde{K}_\zeta.
\end{equation}
By definition of $\mathcal{J}_3$, we have $k(K_0,K_0)= \mathcal{F}_{r00} = 0$ and $k(K_0,K_s)= \mathcal{F}_{rs0} = 0$, which leads to the following relations for the coefficients of the expansions in \eqref{eq:Pos_JA_III_a}:
\begin{equation}
    \label{eq:Pos_JA_III_b}
    a_0^2-a_\alpha a_\alpha=0,\quad a_0 b^s_0-a_\alpha b^s_\alpha=0.
\end{equation}
Since $k(K_s,K_s) = \mathcal{F}_{rss} \geq 0$ due to the universal constraint \eqref{eq:Pos_II}, we can combine these equalities to get
\begin{equation}
    \label{eq:Pos_JA_III_c}
    |b_0^s|=\frac{|a_\alpha b_\alpha^s|}{|a_0|}=\frac{|a_\alpha b^s_\alpha|}{\sqrt{a_\alpha a_\alpha}},\qquad
    |b_0^s|\geq\sqrt{b_\alpha^s b_\alpha^s}.
\end{equation}
Additionally, using the Cauchy-Schwarz inequality on the first relation yields $|b_0^s|\leq\sqrt{b_\alpha^s b_\alpha^s}$, and its consistency with the second relation in \eqref{eq:Pos_JA_III_b} implies that $|b_0^s|=\sqrt{b_\alpha^s b_\alpha^s}$. 
In particular, since the Cauchy-Schwarz equality is saturated, there exists some $\beta_{(s)}\in\mathbb{R}$ such that
$a_\alpha=\beta_{(s)} b_\alpha^s$; in turn, the second equality in \eqref{eq:Pos_JA_III_b} further yields $a_0=\beta_{(s)} b_0$.
Therefore, $K_0\vert_{\mathcal{B}^r}= \beta_{(s)} K_s\vert_{\mathcal{B}^r}$, where $\mathcal{B}^r\subset\mathcal{A}$ denotes the
subspace of $\mathcal{A}$ on which the 't Hooft anomaly pairing $k$ is non-degenerate. Notice that $\mathcal{B}^r$ must be a proper subspace since $K_0$ and $K_s$ are part of a basis of $\mathcal{A}$, and therefore are linearly independent.

Clearly, the above discussion can be repeated for any index $s \in \mathcal{J}_3$.
Thus, picking a different index $t\in\mathcal{J}_3$, we similarly find 
$K_0\vert_{\mathcal{B}}=\beta_{(t)} K_t\vert_{\mathcal{B}^r}$. 
Furthermore, $\beta_{(s)}$, $\beta_{(t)}$ need to be non-null.
In fact, if $\beta_{(s)}=0$ or $\beta_{(t)}=0$, then also $a_0 = 0$ and 
$k(K_I,K_0) = \mathcal{F}_{r I 0} =0$ for all $r \in \mathcal{J}_3$, $I\in\mathcal{J}$, contradicting our previous finding \eqref{eq:Pos_JA_II}. 
Thus, we can write
\begin{equation}
    \mathcal{F}_{rst}=k(K_s,K_t)=\frac{1}{\beta_{(s)}\beta_{(t)}}k(K_0,K_0)=\frac{1}{\beta_{(s)}\beta_{(t)}} \mathcal{F}_{00r}=0 \,,
\end{equation}
and we deduce that
\begin{equation}
    \label{eq:Pos_JA_III}
    \boxed{\mathcal{F}_{rst} = 0 \quad \text{for all $r,s,t\in\mathcal{J}_3$}} \,.
\end{equation}

\noindent\textbf{Ratios of Chern-Simons couplings.}
Consider the supergravity string with $\mathbf{p}^{(0)}={\bm\delta}^0$. Since $\mathcal{F}_{000}=\mathcal{F}_{0rr}=\mathcal{F}_{00r}=0$
for all $r\in\mathcal{J}_3$, it follows that $\beta_{(r)} K_0\vert_{\mathcal{B}^0}=K_r\vert_{\mathcal{B}^0}$ with $\beta_{(r)} \neq 0$ by
using the Cauchy-Schwarz inequality for the 't Hooft matrix $k=k^{({\bm\delta}^0)}$
on this supergravity string. This in turn implies
\begin{equation}
    \mathcal{F}_{0ir}=k(K_i,K_r)=\beta_{(r)}k(K_0,K_i)=\frac{k(K_0,K_i)}{k(K_0,K_j)}k(K_j,K_r),
\end{equation}
or, equivalently,
\begin{equation} \label{eq:cCequation1}
   \boxed{ c_i\mathcal{F}_{0jr}=c_j\mathcal{F}_{0ir} \quad 
 \text{for all $i,j\in\mathcal{J}_1$ and $r\in\mathcal{J}_3$}}\,, 
\end{equation}
where
\begin{equation}
c_i=\mathcal{F}_{00i} \,.
\end{equation}
Using the same 
logic on the supergravity string with $\mathbf{p}^{(r)}={\bm\delta}^r$ to relate $\beta_{(s)}'K_0\vert_{\mathcal{B}^r}
=K_s\vert_{\mathcal{B}^r}$ we can actually conclude
\begin{equation} 
    \boxed{c_i\mathcal{F}_{jab}=c_j\mathcal{F}_{iab}  \quad \text{for all $a,b\in\{0\}\sqcup\mathcal{J}_3$}}\,.
    \label{eq:unique1A}
\end{equation}

\noindent\textbf{Important inequalities.}
Consider the supergravity string with charge $\mathbf{p}^{(j)}={\bm\delta}^j$ and consider
the two vectors $A=K_s$ and $B=k(K_r,A)K_0-k(K_0,A)K_r$ of $\mathcal{A}$,
where $k=k^{({\bm\delta}^j)}$. In the eigenbasis $\{\Tilde{K}_0,\Tilde{K}_\alpha,
\Tilde{K}_\zeta\}$ of $k$ we can expand $K_s$ and $B$ as
\begin{equation}
    K_s=a_0\Tilde{K}_0+a_\alpha\Tilde{K}_\alpha+a_\zeta\Tilde{K}_\zeta,\quad
    B=b_0\Tilde{K}_0+b_\alpha\Tilde{K}_\alpha+b_\zeta\Tilde{K}_\zeta,
\end{equation}
and by construction we have
\begin{equation}
    k(K_s,K_s)=a_0^2-a_\alpha a_\alpha\geq0,\qquad
    k(K_s,B)=a_0b_0-a_\alpha b_\alpha=0.
\end{equation}
From the Cauchy-Schwarz inequality it follows that $k(B,B)\leq0$. In other words,
\begin{equation}
    k(K_r,K_s)^2k(K_0,K_0)+k(K_0,K_s)^2k(K_r,K_r)-2k(K_r,K_s)k(K_0,K_s)k(K_r,K_0)\leq0.
\end{equation}
Using $k(K_r,K_r)\geq0$ and $k(K_r,K_s)>0$ (as otherwise there is no issue to start with), this yields
\begin{equation}
    \boxed{k(K_0,K_r)k(K_0,K_s)\geq\frac{k(K_0,K_0)k(K_r,K_s)}{2} }\,.
    \label{eq:ineq-f00-nocancel}
\end{equation}
Similar reasoning allows us to deduce the inequalities
\begin{equation} \label{eq:inequl-3}
    \begin{gathered}
        k(K_0,K_r)^2\geq\frac{k(K_0,K_0)k(K_r,K_r)}{2},\quad
        k(K_0,K_r)k(K_r,K_s)\geq\frac{k(K_0,K_s)k(K_r,K_r)}{2},\\
        k(K_r,K_s)k(K_r,K_t)\geq\frac{k(K_r,K_r)k(K_s,K_t)}{2} .
    \end{gathered}
\end{equation}

\subsection{Prepotential constraints from Class B limits}
\label{sec:BU_constraints_sugra_JB}

We now turn to the prepotential constraints which arise when ${\cal F}_{00I} =0$ for all $I$. This is also the condition for Class B limits to be possible.

\noindent\textbf{Constraints from supergravity strings with charges ${\bf p} = (1,1,\ldots, 1)^T$.} 
Consider a BPS supergravity string coupled with all the vector multiplet $\mathrm{U}(1)$ gauge fields that the supegravity theory supports. 
For instance, we choose a BPS supergravity string with elementary charges ${\bf p} = (1,1,\ldots, 1)^T$. 
As we will now show, under the assumption that ${\cal J}_1$ defined in \eqref{eq:N2_Sugra_J_split_1} is empty (Class B), such a BPS supergravity string can be coupled to the theory only if the set of indices $\mathcal{J}_3$, defined in \eqref{eq:N2_Sugra_J_split_3}, is empty as well.

To this end, let us assume that there exist some indices $r \in \mathcal{J}_3$, and consider the 't Hooft anomaly matrix defined on
the worldsheet of such a supergravity string, whose elements are
\begin{equation}
    k^{(\mathbf{p})}_{IJ} = k_{IJ} = \sum\limits_{L} \mathcal{F}_{LIJ}.
\end{equation}
In its eigenbasis $\{\Tilde{K}_I\}$, this matrix acquires the simple, diagonal form \eqref{eq:Pos_tHmat_diag}, with no null eigenvalues, for the string is coupled to all the gauge fields.
We can then expand the original generators $K_0$ and $K_r$ in this basis as
\begin{equation}
    \label{eq:Pos_JB_I_a}
    K_0=a_0\Tilde{K}_0+a_\alpha\Tilde{K}_\alpha,\qquad K_r=b_0^r\Tilde{K}_0+b_\alpha^r\Tilde{K}_\alpha.
\end{equation}
Then, since $\mathcal{F}_{00I} = 0$ along any Class B limit, and $\mathcal{F}_{rrI} \geq 0$ due to \eqref{eq:Pos_II}, we have the following constraints for the coefficients appearing in the expansions \eqref{eq:Pos_JB_I_a}:
\begin{subequations}
    \label{eq:Pos_JB_I_b}
    \begin{align}
        \label{eq:Pos_JB_I_ba}
        &k(K_0,K_0) = \sum\limits_{L} k_{L00} = a_0a_0-a_\alpha a_\alpha=0,
        \\
        \label{eq:Pos_JB_I_bc}
        &k(K_r,K_r)= \sum\limits_{L} k_{Lrr}  = b_0^r b_0^r - b_\alpha^r b_\alpha^r \geq 0,
    \end{align}
\end{subequations}
with the latter most readily leading to $|b_0^r| \geq\sqrt{b_\alpha^r b_\alpha^r}$. 

Now, consider
\begin{equation}
    \label{eq:Pos_JB_I_bb}
    k(K_0,K_r)= \sum\limits_{L} k_{L0r}  = a_0b_0^r-a_\alpha b_\alpha^r ,
\end{equation}
which has to be non-zero. Indeed, assume that $k(K_0,K_r) = 0$; then, applying the Cauchy-Schwarz inequality on \eqref{eq:Pos_JB_I_bb}, and using \eqref{eq:Pos_JB_I_ba}, we get
\begin{equation}
    \label{eq:Pos_JB_I_c}
    |b_0^r|=\frac{|a_\alpha b^r_\alpha |}{\sqrt{a_\alpha a_\alpha}}\leq\sqrt{b^r_\alpha b^r_\alpha}.
\end{equation}
By consistency with \eqref{eq:Pos_JB_I_bc}, we need to have $|b_0^r| = \sqrt{b^r_\alpha b^r_\alpha}$, and the generators $K_0$ and $K_r$ are linearly dependent. 
Hence, we need to have $k(K_0,K_r) \neq 0$.

However, since $k(K_0,K_r) = \sum_{L} k_{L0r} = \sum_{i} k_{I0r}$, there exists some index $I \in\mathcal{J}$, with $\mathcal{F}_{0Ir} \neq 0$. 
Furthermore, since $\mathcal{J}_1= \varnothing$ along Class B limits, and $\mathcal{F}_{00r} = 0$, it follows that $I\in\mathcal{J}_2\sqcup\mathcal{J}_3$, in contradiction with the fact that $r\in\mathcal{J}_3$. 
Therefore, we conclude that, along Class B limits
\begin{equation}
    \label{eq:Pos_JB_I}
    \boxed{\mathcal{J}_3= \varnothing} \,.
\end{equation}

The constraints that we derive in the reminder of this section assume that $|\mathcal{J}_\lambda|>1$. Hence, for convenience, we adopt the splitting of indices introduced in \eqref{eq:JB2-split}.

\noindent\textbf{Constraints from supergravity strings with charges $\mathbf{p}^{(a)}={\bm\delta}^a$.}
Consider a supergravity string with charge vector $\mathbf{p}^{(a)}={\bm\delta}^a$ with
$a\in\mathcal{J}_\lambda$. Since $\mu'\in\mathcal{J}_2'$, it holds that
\begin{equation}
    k^{({\bm\delta}^a)}(K_b,K_{\mu'})=k(K_b,K_{\mu'})=\mathcal{F}_{ab\mu'}=0
\end{equation}
for all $b\in\mathcal{J}_\lambda$, see \eqref{eq:JB2-split}. Since also $k(K_b,K_b)=0$, applying
the Cauchy-Schwarz inequality, similarly as done in \eqref{eq:Pos_JB_I_c}, shows that $k(K_{\mu'},K_{\mu'})=0$, as
well as $K_{\mu'}\vert_{\mathcal{B}^a}\propto K_b\vert_{\mathcal{B}^a}$. 
But the choice of index $\mu'\in\mathcal{J}_2'$ is arbitrary, and we conclude
\begin{equation}
    \mathcal{F}_{a\mu'\nu'}=k(K_{\mu'},K_{\nu'})\propto k(K_b,K_{\nu'})=\mathcal{F}_{ab\nu'}
    \label{eq:app-Camu'nu'}
\end{equation}
for all $\mu',\nu'\in\mathcal{J}_2'$. Invoking \eqref{eq:JB2-split}, we see
\begin{equation}
    \boxed{
        \mathcal{F}_{a\mu'\nu'}=0
    }\,.
\end{equation}

As a further consistency condition on the supergravity string with $\mathbf{p}^{(a)}={\bm\delta}^a$, we
can derive $\mathcal{F}_{a\mu'\nu''}\neq0$ for all $\mu'\in\mathcal{J}_2'$ and $\nu''\in\mathcal{J}_2''$.
To this end, we pick any $a\neq b\in\mathcal{J}_\lambda$\footnote{Recall that here, since 
$|\mathcal{J}_\lambda|>1$, we can always find some $b\neq a$.} and recall that
\begin{equation}
    k(K_b,K_b)=k(K_b,K_{\mu'})=0,
\end{equation}
by \eqref{eq:JB2-split}, where also $\mu'\in\mathcal{J}_2'$ is arbitrary. By the Cauchy-Schwarz inequality
we deduce that $K_b\vert_{\mathcal{B}^a}=\beta K_{\mu'}\vert_{\mathcal{B}^a}$, where $\beta>0$. Thus,
\begin{equation}
    \mathcal{F}_{ab\nu''}=k(K_b,K_{\nu''})=\beta k(K_{\mu'},K_{\nu''})=\mathcal{F}_{a\mu'\nu''},
\end{equation}
which holds for all $\nu''\in\mathcal{J}_2''$. 
Then, by the definition of $\mathcal{J}_2''$ in
\eqref{eq:JB2-split} it follows that
\begin{equation}
    \boxed{
        \mathcal{F}_{a\mu'\nu''}\neq0
    }\,.
\end{equation}

\noindent\textbf{Constraints from supergravity strings with charges $\mathbf{p}^{(\mu')}={\bm\delta}^{\mu'}$.}
The 't Hooft anomaly matrix on a supergravity string with elementary charge vector 
$\mathbf{p}^{(\mu')}={\bm\delta}^{\mu'}$ reads $k^{({\bm\delta}^{\mu'})}_{IJ}=k_{IJ}=\mathcal{F}_{\mu'IJ}$. 
For $a\in\mathcal{J}_\lambda$ and $\nu'\in\mathcal{J}_2'$, we can expand $K_a$ and $K_{\nu'}$ in the eigenbasis of $k$ as
\begin{equation}
    K_a=a_0\Tilde{K}_0+a_\alpha\Tilde{K}_\alpha+a_\zeta\Tilde{K}_\zeta,\quad
    K_{\nu'}=b_0\Tilde{K}_0+b_\alpha\Tilde{K}_\alpha+b_\zeta\Tilde{K}_\zeta.
\end{equation}
As in \eqref{eq:app-Camu'nu'}, from $k(K_a,K_a)=0$ and $k(K_a,K_{\nu'})=0$, and applying the
Cauchy-Schwarz inequality, we deduce that
\begin{equation}
    \label{eq:Pos_JB_III_Arel}
    K_a\vert_{\mathcal{B}^{\mu'}}=\beta^a_{\nu'}K_{\nu'}\vert_{\mathcal{B}^{\mu'}},
\end{equation}
where there is no sum over $\nu'$ on the right hand side. Since $\mathcal{F}_{a\mu'\nu''}\neq0$, as derived
above, we know that $\beta^a_{\nu'}>0$. 
Thus, employing the relation \eqref{eq:Pos_JB_III_Arel} twice yields
\begin{equation}
    \mathcal{F}_{\mu'\nu'\rho'}=\beta^a_{\nu'}\mathcal{F}_{a\mu'\rho'}
    =\beta^a_{\nu'}\beta^b_{\rho'}\mathcal{F}_{ab\mu'},
\end{equation}
which vanishes by the defining relation of $\mathcal{J}_2'$, \eqref{eq:JB2-split}. 
Thus, we finally arrive at
\begin{equation}
    \boxed{
        \mathcal{F}_{\mu'\nu'\rho}=0
    }\,.
\end{equation}


\section{Asymptotic features of Class B limits}
\label{app:ClassB}

In this appendix we collect some additional asymptotic features exhibited by Class B limits.
In particular, we show the general asymptotic behavior of the tensions of supergravity strings for Class B limits, characterized by the index set $\mathcal{J}_\lambda$ defined in \eqref{eq:InfDist_Jl} with $|\mathcal{J}_\lambda| = 1$; furthermore, for Class B limits with $|\mathcal{J}_\lambda| > 1$, we introduce a convenient way of splitting the indices of the coordinates $X^I$ that is extensively employed in Section~\ref{sec:ClassB_Jl>1}.

\subsection{Elementary tensions in Class B limits with \texorpdfstring{$|\mathcal{J}_\lambda| = 1$}{|Jl|=1}}   \label{app:ClassB1}

First, let us consider the case in which $|\mathcal{J}_\lambda| = 1$, and let us denote with $0$ the sole index residing in $\mathcal{J}_\lambda$. Below we establish lower bounds on the
elementary tensions $\mathcal{T}_{{\bf p}^0} = \mathcal{F}_0$ and $\mathcal{T}_{{\bf p}^{(\mu)}} = \mathcal{F}_\mu$ along these limits.

\noindent\textbf{Asymptotic behavior of $\mathcal{F}_0$.} As in \eqref{eq:N2_Sugra_JB_F}, by isolating the $0$-th index, we can rewrite the prepotential as
\begin{equation}
    \label{eq:Pos_JB_J1_F01}
    \mathcal{F}=X^0\mathcal{F}_0+\frac{1}{6} \mathcal{F}_{\mu\nu\rho}X^\mu X^\nu X^\rho\,, \qquad \text{with} \quad \mathcal{F}_0=\frac{1}{2} \mathcal{F}_{0\mu\nu}X^\mu X^\nu\,.
\end{equation}
Since $X^0 \sim \lambda$ as $\lambda \to \infty$, by the prepotential constraint \eqref{eq:N2_Fconstr} it has to hold $\mathcal{F}_0\precsim\lambda^{-1}$.
However, we will show that $\mathcal{F}_0\sim\lambda^{-1}$ necessarily along Class B limits characterized by $|\mathcal{J}_\lambda| = 1$.

To this end, it is enough to prove that the second term in the prepotential expression given in \eqref{eq:Pos_JB_J1_F01} falls off to zero as $\lambda \to \infty$. 
Indeed, let us assume that, on the contrary, $\mathcal{F}_{\mu\nu\rho}X^\mu X^\nu X^\rho$ approaches a constant along such a limit.
Then, we can choose three indices $\mu,\nu,\rho\in\mathcal{J}_2$, such that $X^\mu X^\nu X^\rho\sim 1$, and $\mathcal{F}_{\mu\nu\rho} \neq 0$.
If all Chern-Simons couplings $\mathcal{F}_{0\mu\nu}$, $\mathcal{F}_{0\nu\rho}$, $\mathcal{F}_{0\rho\mu}$ were non-zero, the
prepotential constraint \eqref{eq:N2_Fconstr} would imply
\begin{equation}
    \label{eq:Pos_JB_J1_F0munu}
    \mathcal{F}_{0\mu\nu} X^\mu X^\nu\precsim\lambda^{-1},\quad \mathcal{F}_{0\nu\rho} X^\nu X^\rho\precsim\lambda^{-1},\quad
    \mathcal{F}_{0\rho\mu} X^\rho X^\mu\precsim\lambda^{-1},
\end{equation}
yielding, in turn, $X^\mu X^\nu X^\rho\precsim\lambda^{-3/2}$. 
Hence, at least one of the couplings $\mathcal{F}_{0\mu\nu}$, $\mathcal{F}_{0\nu\rho}$, $\mathcal{F}_{0\rho\mu}$
has to vanish in order not to violate $X^\mu X^\nu X^\rho\sim 1$. 
Thus, we may assume $\mathcal{F}_{0\mu\nu}=0$ in the following.

Now, consider a supergravity string with elementary charges
${\bf p}^{(\nu)} = {\bm \delta}^\nu$, with associated ’t Hooft matrix $k^{({\bm\delta}^\nu)}_{IJ}=k_{IJ} = \mathcal{F}_{\nu IJ}$.
Expanding the generators $K_0$ and $K_\mu$ in the eigenbasis $\{\Tilde{K}_0,\Tilde{K}_\alpha,\Tilde{K}_\zeta\}$ where the ’t Hooft matrix acquires the form \eqref{eq:Pos_tHmat_diag} as
\begin{equation}
    K_0 =a_0\Tilde{K}_0+a_\alpha\Tilde{K}_\alpha+a_\zeta\Tilde{K}_\zeta,\quad
    K_\mu=b_0\Tilde{K}_0+b_\alpha\Tilde{K}_\alpha+b_\zeta\Tilde{K}_\zeta,
\end{equation}
and recalling that $\mathcal{F}_{000} = 0$, and using the condition $\mathcal{F}_{\nu\mu\mu} \geq 0$, we arrive at
\begin{equation}
    \label{eq:Pos_JB_J1_krel}
    k(K_0,K_0)=a_0^2-a_\alpha a_\alpha=0,\qquad
    k(K_\mu,K_\mu)=b_0^2-b_\alpha b_\alpha\geq0.
\end{equation}
Moreover, from the previously found relation \eqref{eq:Pos_JA_I_e}, we also have $b_0 = \frac{a_\alpha b_\alpha}{a_0}$. 
Then, using the first relation in \eqref{eq:Pos_JB_J1_krel} and employing the Cauchy-Schwarz inequality, we obtain
\begin{equation}
    |b_0|=\frac{|a_\alpha b_\alpha|}{\sqrt{a_\alpha a_\alpha}}\leq\sqrt{b_\alpha b_\alpha}.
\end{equation}
Comparing this with the second relation in \eqref{eq:Pos_JB_J1_krel}, we recognize that $|b_0| = \sqrt{b_\alpha b_\alpha}$.
Hence, on the proper subspace $\mathcal{B}^\nu\subset\mathcal{A}$, on which $k$ is non-degenerate, $K_\mu\vert_{\mathcal{B}^\nu}= \beta K_0\vert_{\mathcal{B}^\nu}$, for some $\beta\in\mathbb{R}$. 
This implies
\begin{equation}
    \beta \mathcal{F}_{0\rho\nu}=\beta k(K_0,K_\rho)=k(K_\mu,K_\rho)= \mathcal{F}_{\mu\nu\rho}>0\,,
\end{equation}
which leads to $\mathcal{F}_{0\nu\rho}>0$, due to the assumption \eqref{eq:Pos_III}. 
Thus, the prepotential constraint \eqref{eq:N2_Fconstr} implies $X^\nu X^\rho\precsim\lambda^{-1}$.
However, by our 
original assumption, according to which $X^\mu X^\nu X^\rho\sim 1$, it has to hold that $X^\mu\succsim\lambda$, but this contradicts the hypothesis that $\mathcal{J}_\lambda=\{0\}$. 
Hence, the prepotential constraint \eqref{eq:N2_Fconstr} can only be satisfied if 
\begin{equation}
    \label{eq:Pos_JB_Asym_I}
    \boxed{\mathcal{F}_0\sim\lambda^{-1}} \,.
\end{equation}
%

\noindent\textbf{Asymptotic behavior of $\mathcal{F}_\mu$.} 
We will now show that $\mathcal{F}_\mu\succ\lambda^{-1}$, with $\mu\in\mathcal{J}_2$, along any Class B limits characterized by $|\mathcal{J}_\lambda| = 1$. 
We proceed in two steps: first, we prove that if $\mathcal{F}_\mu\precsim\lambda^{-1}$, then $\mathcal{F}_\mu \sim \lambda^{-1}$ necessarily; second, we show that $\mathcal{F}_{\mu}\sim\lambda^{-1}$ leads to inconsistencies
in the structure of the Chern-Simons couplings.

\noindent\textit{Proof that, if $\mathcal{F}_\mu\precsim\lambda^{-1}$, then $\mathcal{F}_{\mu}\sim\lambda^{-1}$.} 
Preliminarily, notice that we can focus on those $\mu\in\mathcal{J}_2$
with $\mathcal{F}_{0\mu\mu}=0$. Indeed, it follows readily that, if $\mathcal{F}_{0\mu\mu}\neq0$, the physical charge \eqref{eq:N2_Qstr} of a BPS string with elementary charge $\mathbf{p}^{(\mu)}={\bm\delta}^\mu$ would be positive only if $\mathcal{F}_{\mu}\succsim\lambda^{1/2}$, a case which we are presently not considering.

Since $\mu \in \mathcal{J}_2$, there exists some $\nu\in\mathcal{J}_2$, with $\mu\neq\nu$, such that  $\mathcal{F}_{0\mu\nu}\neq0$. If $\mathcal{F}_\mu\sim\lambda^{-1}$ the proof is trivially over; hence, we assume
the strict inequality and write
\begin{equation}
    \label{eq:Pos_JB_Fmu}
      \mathcal{F}_\mu=\mathcal{F}_{0\mu\nu}X^0 X^\nu+\frac{1}{2}\mathcal{F}_{\mu\rho\sigma}X^\rho X^\sigma
     \prec\lambda^{-1},
\end{equation}
from which $X^\nu\prec\lambda^{-2}$ follows straightforwardly. 

Now, let us notice that, if $\mathcal{F}_{0\mu\rho_0} \neq 0$, the tension $\mathcal{F}_\mu$ cannot scale as we initially assumed. 
To see this, let us recall that, as shown in Appendix~\ref{app:ClassB1}, it holds that $\mathcal{F}_0\sim\lambda^{-1}$, and consequently, as illustrated in the proof therein, there are 
$\rho_0,\sigma_0\in\mathcal{J}_2$ such that
\begin{equation}
    \label{eq:Pos_JB_F0_cons}
    X^{\rho_0}X^{\sigma_0}\sim\lambda^{-1}\quad\mathrm{and}\quad
    \mathcal{F}_{0\rho_0\sigma_0}\neq0\,.
\end{equation}
To accommodate for \eqref{eq:Pos_JB_Fmu}, it would have to hold that $X^{\rho_0}\prec\lambda^{-2}$ and hence
$X^{\sigma_0}\succ\lambda$. This cannot be and so the tension $\mathcal{F}_\mu$ cannot fall off as in 
\eqref{eq:Pos_JB_Fmu}. 
Hence, in order for \eqref{eq:Pos_JB_Fmu} to hold, we need that $\mathcal{F}_{0\mu\rho_0}=0$, and similarly $\mathcal{F}_{0\mu\sigma_0}=0$.
However, we will show that $\mathcal{F}_{0\mu\rho_0}\neq0$ necessarily, which in turn implies that the behavior \eqref{eq:Pos_JB_Fmu} cannot be realized, thus proving the statement. 

Let us indeed assume, for instance, that $\mathcal{F}_{0\mu\sigma_0}=0$. First, we remark that either 
$\mu\neq\rho_0$, or $\mu\neq\sigma_0$. 
This obviously holds if $\rho_0\neq\sigma_0$; for $\rho_0 =\sigma_0$
we have $\mathcal{F}_{0\rho_0\rho_0}\neq0$ by \eqref{eq:Pos_JB_F0_cons}, but
$\mathcal{F}_{0\mu\mu}=0$ as remarked above. Hence, from now on, we will take $\mu\neq\rho_0$. 

Consider now a supergravity string with $\mathbf{p}^{(\sigma_0)}={\bm\delta}^{\sigma_0}$. The ’t Hooft matrix
$k=k^{({\bm\delta}^{\sigma_0})}$
on its worldsheet displays the elements
\begin{equation}
      k(K_\mu,K_\mu)=\mathcal{F}_{\mu\mu\sigma_0}\geq0,\quad k(K_{\rho_0},K_{\rho_0})=\mathcal{F}_{\rho_0\rho_0\sigma_0}\geq0 \,.
      \label{eq:Btensions1}
\end{equation}
We can now expand $K_\mu$ and $K_{\rho_0}$ as
\begin{equation}
    \label{eq:Pos_JB_Fl_Amurho_b}
    K_\mu=a_0\Tilde{K}_0+a_\alpha\Tilde{K}_\alpha+a_\zeta\Tilde{K}_\zeta,\quad
    K_{\rho_0}=b_0\Tilde{K}_0+b_\alpha\Tilde{K}_\alpha+b_\zeta\Tilde{K}_\zeta,
\end{equation}
where $\Tilde{K}_I$ is the eigenbasis of $k$, see \eqref{eq:Pos_tHmat_sig}. 
The second relation in 
\eqref{eq:Btensions1} implies $|b_0|\geq\sqrt{b_\alpha b_\alpha}$; then, 
$\mathcal{F}_{0\mu\sigma_0}=k(K_0,K_\mu)=0$ yields
\begin{equation}
      |b_0|=\frac{|a_\alpha b_\alpha|}{|a_0|}\leq\frac{|a_\alpha b_\alpha|}{\sqrt{a_\alpha a_\alpha}}
       \leq \sqrt{b_\alpha b_\alpha},
\end{equation}
where in the last step we have employed the Cauchy-Schwarz inequality. Hence, we need to have $|b_0|= \sqrt{b_\alpha b_\alpha}$, which is possible only if $K_{\rho_0}\vert_{\mathcal{B}^{\sigma_0}}
 =\beta K_\mu\vert_{\mathcal{B}^{\sigma_0}}$, for some real $\beta$, where $\mathcal{B}^{\sigma_0}$ is the subspace spanned by the generators $\Tilde{K}_0$ and $\Tilde{K}_\alpha$, see \eqref{eq:Pos_tHmat_diag}.
However, such a relation implies that
\begin{equation}
      0\neq \mathcal{F}_{0\rho_0\sigma_0}=k(K_0,K_{\rho_0})=\beta k(K_0,K_\mu)=\beta \mathcal{F}_{0\mu\sigma_0},
\end{equation}
in contradiction with the starting hypothesis $\mathcal{F}_{0\mu\sigma_0}=0$. 
Therefore, we conclude that $\mathcal{F}_{0\mu\sigma_0}>0$, and 
$\mathcal{F}_\mu\sim\lambda^{-1}$.

\noindent\textit{Inconsistency of $\mathcal{F}_{\mu}\sim\lambda^{-1}$.}  
We will now show that $\mathcal{F}_\mu\sim\lambda^{-1}$ leads to a further contradiction, and we can therefore conclude $\mathcal{F}_\mu\succ\lambda^{-1}$. Recall that
there are $\rho_0,\sigma_0\in\mathcal{J}_2$ with $\mathcal{F}_{0\rho_0\sigma_0} \neq 0$ 
and $X^0 X^{\rho_0} X^{\sigma_0} \sim 1$. Also for this scaling of $\mathcal{F}_\mu$,
it follows that $\mathcal{F}_{0\mu\rho_0}=0=\mathcal{F}_{0\mu\sigma_0}$. Indeed,
if it were $\mathcal{F}_{0\mu\rho_0} \neq 0$, then $X^0X^{\rho_0}\precsim\lambda^{-1}$ since
$\mathcal{F}_\mu\sim\lambda^{-1}$. 
But this contradicts $X^0X^{\rho_0}X^{\sigma_0}\sim1$, since $X^{\sigma_0}\prec\lambda$.
We then conclude that $\mathcal{F}_{0\mu\rho_0}=0$, and analogously $\mathcal{F}_{0\mu\sigma_0}=0$. 

Let us now consider the supergravity string, with elementary charges $\mathbf{p}^{(0)}={\bm\delta}^0$. 
The associated ’t Hooft matrix $k=k^{({\bm\delta}^0)}$ displays the null elements $k(K_\mu,K_{\rho_0}) = \mathcal{F}_{00\mu} =0$ and $k(K_\mu,K_{\rho_0})  = \mathcal{F}_{0\mu\rho_0} =0$, $k(K_\mu,K_{\sigma_0})  = \mathcal{F}_{0\mu\sigma_0} =0$. 
Switching to the eigenbasis $\{\Tilde{K}_I\}$, and rewriting $K_\mu$, $K_{\rho_0}$ and $K_{\sigma_0}$ as
\begin{equation}
    \label{eq:Pos_JB_Fl_Amurho}
    \begin{aligned}
        &K_\mu=a_0\Tilde{K}_0+a_\alpha\Tilde{K}_\alpha+a_\zeta\Tilde{K}_\zeta,
        \\
        &K_{\rho_0}=b_0\Tilde{K}_0+b_\alpha\Tilde{K}_\alpha+b_\zeta\Tilde{K}_\zeta, \quad K_{\sigma_0}=c_0\Tilde{K}_0+c_\alpha\Tilde{K}_\alpha+c_\zeta\Tilde{K}_\zeta,
    \end{aligned}    
\end{equation}
the conditions $k(K_\mu,K_{\rho_0})=0$, $k(K_\mu,K_{\sigma_0}) =0$ can be recast as
\begin{equation}
   a_0b_0-a_\alpha b_\alpha=0,\quad a_0c_0-a_\alpha c_\alpha=0 \,,
\end{equation}
Then, using the fact that $k(K_\mu,K_\mu) = a_0^2 - a_\alpha a_\alpha \geq0$, together with the Cauchy-Schwarz inequality, we get
\begin{equation}
    \label{eq:Pos_JB_Fl_b0}
  |b_0|=\frac{\sqrt{a_\alpha b_\alpha}}{|a_0|}
  \leq \frac{\sqrt{a_\alpha b_\alpha}}{\sqrt{a_\alpha a_\alpha}} \leq \sqrt{b_\alpha b_\alpha}\,,
\end{equation}
and similarly $|c_0|  \leq \sqrt{c_\alpha c_\alpha}$, $|a_0|  \leq \sqrt{a_\alpha a_\alpha}$.
This implies that $k(K_{\rho_0},K_{\rho_0}) \leq 0$, $k(K_{\sigma_0},K_{\sigma_0}) \leq 0$ as well as $k(K_\mu,K_\mu) \leq 0$, and compatibility with \eqref{eq:Pos_II} implies that all of them have to be zero. 
Moreover, the fact that the Cauchy-Schwarz inequality is saturated in \eqref{eq:Pos_JB_Fl_b0} implies that there exists some 
$\beta\in\mathbb{R}_{\geq0}$ such that $b_\alpha=\beta c_\alpha$ and $b_0=\beta c_0$. 
In particular, it follows that
\begin{equation*}
    k(K_{\rho_0},K_{\sigma_0})=b_0c_0-b_\alpha c_\alpha=\beta k(K_{\sigma_0},K_{\sigma_0})=0,
\end{equation*}
which is in contradiction with the fact that $\mathcal{F}_{0\rho_0\sigma_0} \neq 0$. 
Hence, we conclude that $\mathcal{F}_\mu$ cannot fall off as $\lambda^{-1}$, i.e. 
\begin{equation}
    \label{eq:Pos_JB_Asym_II}
    \boxed{\mathcal{F}_\mu\succ\lambda^{-1}} \,.
\end{equation}
%


\subsection{Index sets in Class B limits with \texorpdfstring{$|\mathcal{J}_\lambda| > 1$}{|Jl|>1}}
\label{app:B-Jl>1}

As argued in the second paragraph of Section~\ref{sec:N2_Sugra_JB}, for limits of Class B with
$|\mathcal{J}_\lambda|>1$ it is more convenient to employ a splitting of the index set $\mathcal{J}$
that is different from the original one in \eqref{eq:N2_Sugra_J_split}. Indeed, in this appendix we prove
that the set $\mathcal{J}$ can also be written as
\begin{equation}
    \mathcal{J}=\mathcal{J}_\lambda\sqcup\mathcal{J}_2'\sqcup\mathcal{J}_2'',
    \label{eq:AppB_Jl>1split}
\end{equation}
where
\begin{equation}
    \begin{gathered}
        \mathcal{J}_2'=\{\mu'\in\mathcal{J}_2\,:\,\mathcal{F}_{ab\mu'}=0\,\,\forall\,a,b\in\mathcal{J}_\lambda\},\\
        \mathcal{J}_2''=\{\mu''\in\mathcal{J}_2\,:\,\mathcal{F}_{ab\mu''}\neq0\,\,\forall a\neq b\in\mathcal{J}_\lambda\}.
    \end{gathered}
\end{equation}
In particular, this most readily implies $X^{\mu''}\precsim\lambda^{-2}$ for all $\mu''\in\mathcal{J}_2''$.

In order to show \eqref{eq:AppB_Jl>1split}, let us consider $\mu\in\mathcal{J}\setminus\mathcal{J}_\lambda$. 
It is then enough to prove the following statement: \textit{if there are $a_0\neq b_0\in\mathcal{J}_\lambda$ with $\mathcal{F}_{\mu a_0b_0}=0$, then $\mathcal{F}_{\mu ab}=0$ for all $a,b\in\mathcal{J}_\lambda$}. 
In order to show this statement, we proceed in two steps.

As a first step, we show that $\mathcal{F}_{a_0b\mu}=0$ for all $b\in\mathcal{J}_\lambda$.
To this end, consider a supergravity string with $\mathbf{p}^{(a_0)}={\bm\delta}^{a_0}$. 
The corresponding 't Hooft matrix does not have full rank, since $k^{({\bm\delta}^{a_0})}_{a_0I}=\mathcal{F}_{a_0a_0I}=0$ for all 
$I\in\mathcal{J}_\lambda$. As before, let us denote by $\mathcal{B}^{a_0}$ the proper subspace of $\mathcal{A}$, where $k^{({\bm\delta}^{a_0})}$ has no zero eigenvalues. 
Since $k^{({\bm\delta}^{a_0})}(K_{b_0},K_{b_0})= \mathcal{F}_{a_0b_0b_0} = 0$ and $k^{({\bm\delta}^{a_0})}(K_\mu,K_\mu)=
\mathcal{F}_{a_0\mu\mu}\geq0$ due to \eqref{eq:Pos_II}, 
it can be shown that $k^{({\bm\delta}^{a_0})}(K_{b_0},K_\mu)=\mathcal{F}_{a_0b_0\mu}=0$ can only be satisfied if 
$k^{({\bm\delta}^{a_0})}(K_\mu,K_\mu)=0$, which in turn implies that $K_\mu\vert_{\mathcal{B}^{a_0}}=\beta K_{b_0}\vert_{\mathcal{B}^{a_0}}$ for some real $\beta$. Hence,
\begin{equation}
    \mathcal{F}_{a_0b\mu}=k^{({\bm\delta}^{a_0})}(K_b,K_\mu)=\beta k^{({\bm\delta}^{a_0})}(K_b,K_{b_0})=\beta \mathcal{F}_{a_0b_0b}=0
\end{equation}
for all $b\in\mathcal{J}_\lambda$ since the subspace $\mathcal{A}\mysetminus\mathcal{B}^{a_0}$, where $K_\mu$ and
$K_{b_0}$ necessarily differ, does not enter the 't Hooft matrix computation.

In the second step, we show that $\mathcal{F}_{ab\mu}=0$ holds for all $a,b\in\mathcal{J}_\lambda$.
Let us pick some arbitrary $a_0\neq b\in\mathcal{J}_\lambda$. On the supergravity string characterized by the elementary charge
$\mathbf{p}^{(b)}={\bm\delta}^b$, the 't Hooft matrix has elements $k^{({\bm\delta}^b)}(K_{a_0},K_{a_0})=0$ and $k^{({\bm\delta}^b)}(K_\mu,K_\mu)\geq0$. 
As we have shown above, $k^{({\bm\delta}^b)}(K_{a_0},K_\mu)=\mathcal{F}_{a_0b\mu}=0$. 
But this can hold if and only if $k^{({\bm\delta}^b)}(K_\mu,K_\mu)=0$ and $K_\mu\vert_{\mathcal{B}^b}=\beta' K_{a_0}\vert_{\mathcal{B}^b}$, where $\mathcal{B}^b$ is a proper subspace of $\mathcal{A}$. 
Thus, $K_\mu$ and $K_{a_0}$ necessarily differ on $\mathcal{A}\mysetminus\mathcal{B}^b$,
and also
\begin{equation}
    \mathcal{F}_{ab\mu}=k^{({\bm\delta}^b)}(K_a,K_\mu)=\beta' k^{({\bm\delta}^b)}(K_a,K_{a_0})=\beta' \mathcal{F}_{aa_0b}=0,
\end{equation}
for all $a\in\mathcal{J}_\lambda$. Since the choice of $b\in\mathcal{J}_\lambda$ is arbitrary, this shows the equality
\begin{equation}
    \boxed{
        \mathcal{J}=\mathcal{J}_\lambda\sqcup\mathcal{J}_2'\sqcup\mathcal{J}_2''
    }\,.
\end{equation}

\bibliographystyle{jhep}
\bibliography{references.bib}

\providecommand{\href}[2]{#2}\begingroup\raggedright\begin{thebibliography}{10}

\bibitem{Vafa:2005ui}
C.~Vafa, \emph{{The String landscape and the swampland}},
  \href{https://arxiv.org/abs/hep-th/0509212}{{\ttfamily hep-th/0509212}}.

\bibitem{Palti:2019pca}
E.~Palti, \emph{{The Swampland: Introduction and Review}},
  \href{https://doi.org/10.1002/prop.201900037}{\emph{Fortsch. Phys.}
  {\bfseries 67} (2019) 1900037}
  [\href{https://arxiv.org/abs/1903.06239}{{\ttfamily 1903.06239}}].

\bibitem{vanBeest:2021lhn}
M.~van Beest, J.~Calder\'on-Infante, D.~Mirfendereski and I.~Valenzuela,
  \emph{{Lectures on the Swampland Program in String Compactifications}},
  \href{https://doi.org/10.1016/j.physrep.2022.09.002}{\emph{Phys. Rept.}
  {\bfseries 989} (2022) 1} [\href{https://arxiv.org/abs/2102.01111}{{\ttfamily
  2102.01111}}].

\bibitem{Grana:2021zvf}
M.~Gra\~na and A.~Herr\'aez, \emph{{The Swampland Conjectures: A Bridge from
  Quantum Gravity to Particle Physics}},
  \href{https://doi.org/10.3390/universe7080273}{\emph{Universe} {\bfseries 7}
  (2021) 273} [\href{https://arxiv.org/abs/2107.00087}{{\ttfamily
  2107.00087}}].

\bibitem{Agmon:2022thq}
N.~B. Agmon, A.~Bedroya, M.~J. Kang and C.~Vafa, \emph{{Lectures on the string
  landscape and the Swampland}},
  \href{https://arxiv.org/abs/2212.06187}{{\ttfamily 2212.06187}}.

\bibitem{Ooguri:2006in}
H.~Ooguri and C.~Vafa, \emph{{On the Geometry of the String Landscape and the
  Swampland}},
  \href{https://doi.org/10.1016/j.nuclphysb.2006.10.033}{\emph{Nucl. Phys.}
  {\bfseries B766} (2007) 21}
  [\href{https://arxiv.org/abs/hep-th/0605264}{{\ttfamily hep-th/0605264}}].

\bibitem{Lee:2019wij}
S.-J. Lee, W.~Lerche and T.~Weigand, \emph{{Emergent strings from infinite
  distance limits}}, \href{https://doi.org/10.1007/JHEP02(2022)190}{\emph{JHEP}
  {\bfseries 02} (2022) 190}
  [\href{https://arxiv.org/abs/1910.01135}{{\ttfamily 1910.01135}}].

\bibitem{Baume:2020dqd}
F.~Baume and J.~Calder\'on~Infante, \emph{{Tackling the SDC in AdS with CFTs}},
  \href{https://doi.org/10.1007/JHEP08(2021)057}{\emph{JHEP} {\bfseries 08}
  (2021) 057} [\href{https://arxiv.org/abs/2011.03583}{{\ttfamily
  2011.03583}}].

\bibitem{Perlmutter:2020buo}
E.~Perlmutter, L.~Rastelli, C.~Vafa and I.~Valenzuela, \emph{{A CFT distance
  conjecture}}, \href{https://doi.org/10.1007/JHEP10(2021)070}{\emph{JHEP}
  {\bfseries 10} (2021) 070}
  [\href{https://arxiv.org/abs/2011.10040}{{\ttfamily 2011.10040}}].

\bibitem{Baume:2023msm}
F.~Baume and J.~Calder\'on-Infante, \emph{{On higher-spin points and infinite
  distances in conformal manifolds}},
  \href{https://doi.org/10.1007/JHEP12(2023)163}{\emph{JHEP} {\bfseries 12}
  (2023) 163} [\href{https://arxiv.org/abs/2305.05693}{{\ttfamily
  2305.05693}}].

\bibitem{Ooguri:2024ofs}
H.~Ooguri and Y.~Wang, \emph{{Universal Bounds on CFT Distance Conjecture}},
  \href{https://arxiv.org/abs/2405.00674}{{\ttfamily 2405.00674}}.

\bibitem{Calderon-Infante:2024oed}
J.~Calder\'on-Infante and I.~Valenzuela, \emph{{Tensionless String Limits in 4d
  Conformal Manifolds}},  \href{https://arxiv.org/abs/2410.07309}{{\ttfamily
  2410.07309}}.

\bibitem{Etheredge:2022opl}
M.~Etheredge, B.~Heidenreich, S.~Kaya, Y.~Qiu and T.~Rudelius,
  \emph{{Sharpening the Distance Conjecture in diverse dimensions}},
  \href{https://doi.org/10.1007/JHEP12(2022)114}{\emph{JHEP} {\bfseries 12}
  (2022) 114} [\href{https://arxiv.org/abs/2206.04063}{{\ttfamily
  2206.04063}}].

\bibitem{Etheredge:2023odp}
M.~Etheredge, B.~Heidenreich, J.~McNamara, T.~Rudelius, I.~Ruiz and
  I.~Valenzuela, \emph{{Running Decompactification, Sliding Towers, and the
  Distance Conjecture}},  \href{https://arxiv.org/abs/2306.16440}{{\ttfamily
  2306.16440}}.

\bibitem{Etheredge:2024tok}
M.~Etheredge, B.~Heidenreich, T.~Rudelius, I.~Ruiz and I.~Valenzuela,
  \emph{{Taxonomy of Infinite Distance Limits}},
  \href{https://arxiv.org/abs/2405.20332}{{\ttfamily 2405.20332}}.

\bibitem{Castellano:2022bvr}
A.~Castellano, A.~Herr\'aez and L.~E. Ib\'a\~nez, \emph{{The emergence proposal
  in quantum gravity and the species scale}},
  \href{https://doi.org/10.1007/JHEP06(2023)047}{\emph{JHEP} {\bfseries 06}
  (2023) 047} [\href{https://arxiv.org/abs/2212.03908}{{\ttfamily
  2212.03908}}].

\bibitem{vandeHeisteeg:2023ubh}
D.~van~de Heisteeg, C.~Vafa and M.~Wiesner, \emph{{Bounds on Species Scale and
  the Distance Conjecture}},
  \href{https://doi.org/10.1002/prop.202300143}{\emph{Fortsch. Phys.}
  {\bfseries 71} (2023) 2300143}
  [\href{https://arxiv.org/abs/2303.13580}{{\ttfamily 2303.13580}}].

\bibitem{Blumenhagen:2023yws}
R.~Blumenhagen, A.~Gligovic and A.~Paraskevopoulou, \emph{{The emergence
  proposal and the emergent string}},
  \href{https://doi.org/10.1007/JHEP10(2023)145}{\emph{JHEP} {\bfseries 10}
  (2023) 145} [\href{https://arxiv.org/abs/2305.10490}{{\ttfamily
  2305.10490}}].

\bibitem{vandeHeisteeg:2023dlw}
D.~van~de Heisteeg, C.~Vafa, M.~Wiesner and D.~H. Wu, \emph{{Species scale in
  diverse dimensions}},
  \href{https://doi.org/10.1007/JHEP05(2024)112}{\emph{JHEP} {\bfseries 05}
  (2024) 112} [\href{https://arxiv.org/abs/2310.07213}{{\ttfamily
  2310.07213}}].

\bibitem{Herraez:2024kux}
A.~Herr\'aez, D.~L\"ust, J.~Masias and M.~Scalisi, \emph{{On the Origin of
  Species Thermodynamics and the Black Hole - Tower Correspondence}},
  \href{https://arxiv.org/abs/2406.17851}{{\ttfamily 2406.17851}}.

\bibitem{Dvali:2007hz}
G.~Dvali, \emph{{Black Holes and Large N Species Solution to the Hierarchy
  Problem}}, \href{https://doi.org/10.1002/prop.201000009}{\emph{Fortsch.
  Phys.} {\bfseries 58} (2010) 528}
  [\href{https://arxiv.org/abs/0706.2050}{{\ttfamily 0706.2050}}].

\bibitem{Dvali:2007wp}
G.~Dvali and M.~Redi, \emph{{Black Hole Bound on the Number of Species and
  Quantum Gravity at LHC}},
  \href{https://doi.org/10.1103/PhysRevD.77.045027}{\emph{Phys. Rev. D}
  {\bfseries 77} (2008) 045027}
  [\href{https://arxiv.org/abs/0710.4344}{{\ttfamily 0710.4344}}].

\bibitem{Dvali:2009ks}
G.~Dvali and D.~Lust, \emph{{Evaporation of Microscopic Black Holes in String
  Theory and the Bound on Species}},
  \href{https://doi.org/10.1002/prop.201000008}{\emph{Fortsch. Phys.}
  {\bfseries 58} (2010) 505} [\href{https://arxiv.org/abs/0912.3167}{{\ttfamily
  0912.3167}}].

\bibitem{Dvali:2010vm}
G.~Dvali and C.~Gomez, \emph{{Species and Strings}},
  \href{https://arxiv.org/abs/1004.3744}{{\ttfamily 1004.3744}}.

\bibitem{Montero:2022prj}
M.~Montero, C.~Vafa and I.~Valenzuela, \emph{{The dark dimension and the
  Swampland}}, \href{https://doi.org/10.1007/JHEP02(2023)022}{\emph{JHEP}
  {\bfseries 02} (2023) 022}
  [\href{https://arxiv.org/abs/2205.12293}{{\ttfamily 2205.12293}}].

\bibitem{FierroCota:2023bsp}
C.~Fierro~Cota, A.~Mininno, T.~Weigand and M.~Wiesner, \emph{{The minimal weak
  gravity conjecture}},
  \href{https://doi.org/10.1007/JHEP05(2024)285}{\emph{JHEP} {\bfseries 05}
  (2024) 285} [\href{https://arxiv.org/abs/2312.04619}{{\ttfamily
  2312.04619}}].

\bibitem{Heidenreich:2015nta}
B.~Heidenreich, M.~Reece and T.~Rudelius, \emph{{Sharpening the Weak Gravity
  Conjecture with Dimensional Reduction}},
  \href{https://doi.org/10.1007/JHEP02(2016)140}{\emph{JHEP} {\bfseries 02}
  (2016) 140} [\href{https://arxiv.org/abs/1509.06374}{{\ttfamily
  1509.06374}}].

\bibitem{Montero:2016tif}
M.~Montero, G.~Shiu and P.~Soler, \emph{{The Weak Gravity Conjecture in three
  dimensions}}, \href{https://doi.org/10.1007/JHEP10(2016)159}{\emph{JHEP}
  {\bfseries 10} (2016) 159}
  [\href{https://arxiv.org/abs/1606.08438}{{\ttfamily 1606.08438}}].

\bibitem{Andriolo:2018lvp}
S.~Andriolo, D.~Junghans, T.~Noumi and G.~Shiu, \emph{{A Tower Weak Gravity
  Conjecture from Infrared Consistency}},
  \href{https://doi.org/10.1002/prop.201800020}{\emph{Fortsch. Phys.}
  {\bfseries 66} (2018) 1800020}
  [\href{https://arxiv.org/abs/1802.04287}{{\ttfamily 1802.04287}}].

\bibitem{Lee:2018urn}
S.-J. Lee, W.~Lerche and T.~Weigand, \emph{{Tensionless Strings and the Weak
  Gravity Conjecture}},
  \href{https://doi.org/10.1007/JHEP10(2018)164}{\emph{JHEP} {\bfseries 10}
  (2018) 164} [\href{https://arxiv.org/abs/1808.05958}{{\ttfamily
  1808.05958}}].

\bibitem{Lee:2018spm}
S.-J. Lee, W.~Lerche and T.~Weigand, \emph{{A Stringy Test of the Scalar Weak
  Gravity Conjecture}},
  \href{https://doi.org/10.1016/j.nuclphysb.2018.11.001}{\emph{Nucl. Phys. B}
  {\bfseries 938} (2019) 321}
  [\href{https://arxiv.org/abs/1810.05169}{{\ttfamily 1810.05169}}].

\bibitem{Lee:2019tst}
S.-J. Lee, W.~Lerche and T.~Weigand, \emph{{Modular Fluxes, Elliptic Genera,
  and Weak Gravity Conjectures in Four Dimensions}},
  \href{https://doi.org/10.1007/JHEP08(2019)104}{\emph{JHEP} {\bfseries 08}
  (2019) 104} [\href{https://arxiv.org/abs/1901.08065}{{\ttfamily
  1901.08065}}].

\bibitem{Klaewer:2020lfg}
D.~Klaewer, S.-J. Lee, T.~Weigand and M.~Wiesner, \emph{{Quantum corrections in
  4d $N$ = 1 infinite distance limits and the weak gravity conjecture}},
  \href{https://doi.org/10.1007/JHEP03(2021)252}{\emph{JHEP} {\bfseries 03}
  (2021) 252} [\href{https://arxiv.org/abs/2011.00024}{{\ttfamily
  2011.00024}}].

\bibitem{Lee:2021qkx}
S.-J. Lee and T.~Weigand, \emph{{Elliptic K3 surfaces at infinite complex
  structure and their refined Kulikov models}},
  \href{https://doi.org/10.1007/JHEP09(2022)143}{\emph{JHEP} {\bfseries 09}
  (2022) 143} [\href{https://arxiv.org/abs/2112.07682}{{\ttfamily
  2112.07682}}].

\bibitem{Lee:2021usk}
S.-J. Lee, W.~Lerche and T.~Weigand, \emph{{Physics of infinite complex
  structure limits in eight dimensions}},
  \href{https://doi.org/10.1007/JHEP06(2022)042}{\emph{JHEP} {\bfseries 06}
  (2022) 042} [\href{https://arxiv.org/abs/2112.08385}{{\ttfamily
  2112.08385}}].

\bibitem{Alvarez-Garcia:2023gdd}
R.~\'Alvarez-Garc\'\i{}a, S.-J. Lee and T.~Weigand, \emph{{Non-minimal elliptic
  threefolds at infinite distance. Part I. Log Calabi-Yau resolutions}},
  \href{https://doi.org/10.1007/JHEP08(2024)240}{\emph{JHEP} {\bfseries 08}
  (2024) 240} [\href{https://arxiv.org/abs/2310.07761}{{\ttfamily
  2310.07761}}].

\bibitem{Alvarez-Garcia:2023qqj}
R.~\'Alvarez-Garc\'\i{}a, S.-J. Lee and T.~Weigand, \emph{{Non-minimal Elliptic
  Threefolds at Infinite Distance II: Asymptotic Physics}},
  \href{https://arxiv.org/abs/2312.11611}{{\ttfamily 2312.11611}}.

\bibitem{Collazuol:2022jiy}
V.~Collazuol, M.~Gra\~na and A.~Herr\'aez, \emph{{E$_{9}$ symmetry in the
  heterotic string on S$^{1}$ and the weak gravity conjecture}},
  \href{https://doi.org/10.1007/JHEP06(2022)083}{\emph{JHEP} {\bfseries 06}
  (2022) 083} [\href{https://arxiv.org/abs/2203.01341}{{\ttfamily
  2203.01341}}].

\bibitem{Collazuol:2022oey}
V.~Collazuol, M.~Gra\~na, A.~Herr\'aez and H.~Parra De~Freitas, \emph{{Affine
  algebras at infinite distance limits in the Heterotic String}},
  \href{https://doi.org/10.1007/JHEP07(2023)036}{\emph{JHEP} {\bfseries 07}
  (2023) 036} [\href{https://arxiv.org/abs/2210.13471}{{\ttfamily
  2210.13471}}].

\bibitem{Alvarez-Garcia:2021pxo}
R.~\'Alvarez-Garc\'\i{}a, D.~Kl\"awer and T.~Weigand, \emph{{Membrane limits in
  quantum gravity}},
  \href{https://doi.org/10.1103/PhysRevD.105.066024}{\emph{Phys. Rev. D}
  {\bfseries 105} (2022) 066024}
  [\href{https://arxiv.org/abs/2112.09136}{{\ttfamily 2112.09136}}].

\bibitem{Rudelius:2023odg}
T.~Rudelius, \emph{{Gopakumar-Vafa invariants and the Emergent String
  Conjecture}}, \href{https://doi.org/10.1007/JHEP03(2024)061}{\emph{JHEP}
  {\bfseries 03} (2024) 061}
  [\href{https://arxiv.org/abs/2309.10024}{{\ttfamily 2309.10024}}].

\bibitem{Xu:2020nlh}
F.~Xu, \emph{{On TCS G$_{2}$ manifolds and 4D emergent strings}},
  \href{https://doi.org/10.1007/JHEP10(2020)045}{\emph{JHEP} {\bfseries 10}
  (2020) 045} [\href{https://arxiv.org/abs/2006.02350}{{\ttfamily
  2006.02350}}].

\bibitem{Baume:2019sry}
F.~Baume, F.~Marchesano and M.~Wiesner, \emph{{Instanton Corrections and
  Emergent Strings}},
  \href{https://doi.org/10.1007/JHEP04(2020)174}{\emph{JHEP} {\bfseries 04}
  (2020) 174} [\href{https://arxiv.org/abs/1912.02218}{{\ttfamily
  1912.02218}}].

\bibitem{Grimm:2018ohb}
T.~W. Grimm, E.~Palti and I.~Valenzuela, \emph{{Infinite Distances in Field
  Space and Massless Towers of States}},
  \href{https://doi.org/10.1007/JHEP08(2018)143}{\emph{JHEP} {\bfseries 08}
  (2018) 143} [\href{https://arxiv.org/abs/1802.08264}{{\ttfamily
  1802.08264}}].

\bibitem{Grimm:2018cpv}
T.~W. Grimm, C.~Li and E.~Palti, \emph{{Infinite Distance Networks in Field
  Space and Charge Orbits}},
  \href{https://doi.org/10.1007/JHEP03(2019)016}{\emph{JHEP} {\bfseries 03}
  (2019) 016} [\href{https://arxiv.org/abs/1811.02571}{{\ttfamily
  1811.02571}}].

\bibitem{Corvilain:2018lgw}
P.~Corvilain, T.~W. Grimm and I.~Valenzuela, \emph{{The Swampland Distance
  Conjecture for K\"ahler moduli}},
  \href{https://doi.org/10.1007/JHEP08(2019)075}{\emph{JHEP} {\bfseries 08}
  (2019) 075} [\href{https://arxiv.org/abs/1812.07548}{{\ttfamily
  1812.07548}}].

\bibitem{Grimm:2019bey}
T.~W. Grimm, F.~Ruehle and D.~van~de Heisteeg, \emph{{Classifying
  Calabi\textendash{}Yau Threefolds Using Infinite Distance Limits}},
  \href{https://doi.org/10.1007/s00220-021-03972-9}{\emph{Commun. Math. Phys.}
  {\bfseries 382} (2021) 239}
  [\href{https://arxiv.org/abs/1910.02963}{{\ttfamily 1910.02963}}].

\bibitem{Basile:2022zee}
I.~Basile, \emph{{Emergent Strings at an Infinite Distance with Broken
  Supersymmetry}},
  \href{https://doi.org/10.3390/astronomy2030015}{\emph{Astronomy} {\bfseries
  2} (2023) 206} [\href{https://arxiv.org/abs/2201.08851}{{\ttfamily
  2201.08851}}].

\bibitem{Aoufia:2024awo}
C.~Aoufia, I.~Basile and G.~Leone, \emph{{Species scale, worldsheet CFTs and
  emergent geometry}},  \href{https://arxiv.org/abs/2405.03683}{{\ttfamily
  2405.03683}}.

\bibitem{Basile:2023blg}
I.~Basile, D.~L\"ust and C.~Montella, \emph{{Shedding black hole light on the
  emergent string conjecture}},
  \href{https://doi.org/10.1007/JHEP07(2024)208}{\emph{JHEP} {\bfseries 07}
  (2024) 208} [\href{https://arxiv.org/abs/2311.12113}{{\ttfamily
  2311.12113}}].

\bibitem{Basile:2024dqq}
I.~Basile, N.~Cribiori, D.~Lust and C.~Montella, \emph{{Minimal black holes and
  species thermodynamics}},
  \href{https://doi.org/10.1007/JHEP06(2024)127}{\emph{JHEP} {\bfseries 06}
  (2024) 127} [\href{https://arxiv.org/abs/2401.06851}{{\ttfamily
  2401.06851}}].

\bibitem{Bedroya:2024ubj}
A.~Bedroya, R.~K. Mishra and M.~Wiesner, \emph{{Density of States, Black Holes
  and the Emergent String Conjecture}},
  \href{https://arxiv.org/abs/2405.00083}{{\ttfamily 2405.00083}}.

\bibitem{Cribiori:2023ffn}
N.~Cribiori, D.~Lust and C.~Montella, \emph{{Species entropy and
  thermodynamics}}, \href{https://doi.org/10.1007/JHEP10(2023)059}{\emph{JHEP}
  {\bfseries 10} (2023) 059}
  [\href{https://arxiv.org/abs/2305.10489}{{\ttfamily 2305.10489}}].

\bibitem{Heidenreich:2020ptx}
B.~Heidenreich and T.~Rudelius, \emph{{Infinite distance and zero gauge
  coupling in 5D supergravity}},
  \href{https://doi.org/10.1103/PhysRevD.104.106016}{\emph{Phys. Rev. D}
  {\bfseries 104} (2021) 106016}
  [\href{https://arxiv.org/abs/2007.07892}{{\ttfamily 2007.07892}}].

\bibitem{Katz:2020ewz}
S.~Katz, H.-C. Kim, H.-C. Tarazi and C.~Vafa, \emph{{Swampland Constraints on
  5d $\mathcal{N}=1$ Supergravity}},
  \href{https://doi.org/10.1007/JHEP07(2020)080}{\emph{JHEP} {\bfseries 07}
  (2020) 080} [\href{https://arxiv.org/abs/2004.14401}{{\ttfamily
  2004.14401}}].

\bibitem{Lanza:2020qmt}
S.~Lanza, F.~Marchesano, L.~Martucci and I.~Valenzuela, \emph{{Swampland
  Conjectures for Strings and Membranes}},
  \href{https://doi.org/10.1007/JHEP02(2021)006}{\emph{JHEP} {\bfseries 02}
  (2021) 006} [\href{https://arxiv.org/abs/2006.15154}{{\ttfamily
  2006.15154}}].

\bibitem{Lanza:2021udy}
S.~Lanza, F.~Marchesano, L.~Martucci and I.~Valenzuela, \emph{{The EFT stringy
  viewpoint on large distances}},
  \href{https://doi.org/10.1007/JHEP09(2021)197}{\emph{JHEP} {\bfseries 09}
  (2021) 197} [\href{https://arxiv.org/abs/2104.05726}{{\ttfamily
  2104.05726}}].

\bibitem{Marchesano:2022axe}
F.~Marchesano and L.~Melotti, \emph{{EFT strings and emergence}},
  \href{https://doi.org/10.1007/JHEP02(2023)112}{\emph{JHEP} {\bfseries 02}
  (2023) 112} [\href{https://arxiv.org/abs/2211.01409}{{\ttfamily
  2211.01409}}].

\bibitem{Grimm:2022sbl}
T.~W. Grimm, S.~Lanza and C.~Li, \emph{{Tameness, Strings, and the Distance
  Conjecture}}, \href{https://doi.org/10.1007/JHEP09(2022)149}{\emph{JHEP}
  {\bfseries 09} (2022) 149}
  [\href{https://arxiv.org/abs/2206.00697}{{\ttfamily 2206.00697}}].

\bibitem{Martucci:2024trp}
L.~Martucci, N.~Risso, A.~Valenti and L.~Vecchi, \emph{{Wormholes in the
  axiverse, and the species scale}},
  \href{https://doi.org/10.1007/JHEP07(2024)240}{\emph{JHEP} {\bfseries 07}
  (2024) 240} [\href{https://arxiv.org/abs/2404.14489}{{\ttfamily
  2404.14489}}].

\bibitem{Marchesano:2022avb}
F.~Marchesano and M.~Wiesner, \emph{{4d strings at strong coupling}},
  \href{https://doi.org/10.1007/JHEP08(2022)004}{\emph{JHEP} {\bfseries 08}
  (2022) 004} [\href{https://arxiv.org/abs/2202.10466}{{\ttfamily
  2202.10466}}].

\bibitem{Wiesner:2022qys}
M.~Wiesner, \emph{{Light strings and strong coupling in F-theory}},
  \href{https://doi.org/10.1007/JHEP04(2023)088}{\emph{JHEP} {\bfseries 04}
  (2023) 088} [\href{https://arxiv.org/abs/2210.14238}{{\ttfamily
  2210.14238}}].

\bibitem{Marchesano:2023thx}
F.~Marchesano, L.~Melotti and L.~Paoloni, \emph{{On the moduli space curvature
  at infinity}}, \href{https://doi.org/10.1007/JHEP02(2024)103}{\emph{JHEP}
  {\bfseries 02} (2024) 103}
  [\href{https://arxiv.org/abs/2311.07979}{{\ttfamily 2311.07979}}].

\bibitem{Marchesano:2024tod}
F.~Marchesano, L.~Melotti and M.~Wiesner, \emph{{Asymptotic curvature
  divergences and non-gravitational theories}},
  \href{https://arxiv.org/abs/2409.02991}{{\ttfamily 2409.02991}}.

\bibitem{Castellano:2024gwi}
A.~Castellano, F.~Marchesano, L.~Melotti and L.~Paoloni, \emph{{The Moduli
  Space Curvature and the Weak Gravity Conjecture}},
  \href{https://arxiv.org/abs/2410.10966}{{\ttfamily 2410.10966}}.

\bibitem{Kim:2019vuc}
H.-C. Kim, G.~Shiu and C.~Vafa, \emph{{Branes and the Swampland}},
  \href{https://doi.org/10.1103/PhysRevD.100.066006}{\emph{Phys. Rev. D}
  {\bfseries 100} (2019) 066006}
  [\href{https://arxiv.org/abs/1905.08261}{{\ttfamily 1905.08261}}].

\bibitem{Lee:2019skh}
S.-J. Lee and T.~Weigand, \emph{{Swampland Bounds on the Abelian Gauge
  Sector}}, \href{https://doi.org/10.1103/PhysRevD.100.026015}{\emph{Phys. Rev.
  D} {\bfseries 100} (2019) 026015}
  [\href{https://arxiv.org/abs/1905.13213}{{\ttfamily 1905.13213}}].

\bibitem{Kim:2019ths}
H.-C. Kim, H.-C. Tarazi and C.~Vafa, \emph{{Four-dimensional
  $\mathbf{\mathcal{N}=4}$ SYM theory and the swampland}},
  \href{https://doi.org/10.1103/PhysRevD.102.026003}{\emph{Phys. Rev. D}
  {\bfseries 102} (2020) 026003}
  [\href{https://arxiv.org/abs/1912.06144}{{\ttfamily 1912.06144}}].

\bibitem{Tarazi:2021duw}
H.-C. Tarazi and C.~Vafa, \emph{{On The Finiteness of 6d Supergravity
  Landscape}},  \href{https://arxiv.org/abs/2106.10839}{{\ttfamily
  2106.10839}}.

\bibitem{Martucci:2022krl}
L.~Martucci, N.~Risso and T.~Weigand, \emph{{Quantum gravity bounds on $
  \mathcal{N} $ = 1 effective theories in four dimensions}},
  \href{https://doi.org/10.1007/JHEP03(2023)197}{\emph{JHEP} {\bfseries 03}
  (2023) 197} [\href{https://arxiv.org/abs/2210.10797}{{\ttfamily
  2210.10797}}].

\bibitem{Lee:2022swr}
S.-J. Lee and P.-K. Oehlmann, \emph{{Geometric bounds on the 1-form gauge
  sector}}, \href{https://doi.org/10.1103/PhysRevD.108.086021}{\emph{Phys. Rev.
  D} {\bfseries 108} (2023) 086021}
  [\href{https://arxiv.org/abs/2212.11915}{{\ttfamily 2212.11915}}].

\bibitem{Kim:2024tdh}
H.-C. Kim and C.~Vafa, \emph{{Exploring new constraints on Kahler moduli space
  of 6d N = 1 Supergravity}},
  \href{https://arxiv.org/abs/2406.06704}{{\ttfamily 2406.06704}}.

\bibitem{Kim:2024hxe}
H.-C. Kim, C.~Vafa and K.~Xu, \emph{{Finite Landscape of 6d N=(1,0)
  Supergravity}},  \href{https://arxiv.org/abs/2411.19155}{{\ttfamily
  2411.19155}}.

\bibitem{Bergshoeff:2004kh}
E.~Bergshoeff, S.~Cucu, T.~de~Wit, J.~Gheerardyn, S.~Vandoren and
  A.~Van~Proeyen, \emph{{N = 2 supergravity in five-dimensions revisited}},
  \href{https://doi.org/10.1088/0264-9381/23/23/C01}{\emph{Class. Quant. Grav.}
  {\bfseries 21} (2004) 3015}
  [\href{https://arxiv.org/abs/hep-th/0403045}{{\ttfamily hep-th/0403045}}].

\bibitem{Lauria:2020rhc}
E.~Lauria and A.~Van~Proeyen, \emph{{${\cal N}=2$ Supergravity in $D=4,5,6$
  Dimensions}}, vol.~966. Springer, 3, 2020,
  \href{https://doi.org/10.1007/978-3-030-33757-5}{10.1007/978-3-030-33757-5},
  [\href{https://arxiv.org/abs/2004.11433}{{\ttfamily 2004.11433}}].

\bibitem{Ferrara:1996hh}
S.~Ferrara, R.~R. Khuri and R.~Minasian, \emph{{M theory on a Calabi-Yau
  manifold}}, \href{https://doi.org/10.1016/0370-2693(96)00270-5}{\emph{Phys.
  Lett. B} {\bfseries 375} (1996) 81}
  [\href{https://arxiv.org/abs/hep-th/9602102}{{\ttfamily hep-th/9602102}}].

\bibitem{Mizoguchi:1998wv}
S.~Mizoguchi and N.~Ohta, \emph{{More on the similarity between D = 5 simple
  supergravity and M theory}},
  \href{https://doi.org/10.1016/S0370-2693(98)01122-8}{\emph{Phys. Lett. B}
  {\bfseries 441} (1998) 123}
  [\href{https://arxiv.org/abs/hep-th/9807111}{{\ttfamily hep-th/9807111}}].

\bibitem{Boyarsky:2002ck}
A.~Boyarsky, J.~A. Harvey and O.~Ruchayskiy, \emph{{A Toy model of the
  M5-brane: Anomalies of monopole strings in five dimensions}},
  \href{https://doi.org/10.1006/aphy.2002.6294}{\emph{Annals Phys.} {\bfseries
  301} (2002) 1} [\href{https://arxiv.org/abs/hep-th/0203154}{{\ttfamily
  hep-th/0203154}}].

\bibitem{Polchinski:2003bq}
J.~Polchinski, \emph{{Monopoles, duality, and string theory}},
  \href{https://doi.org/10.1142/S0217751X0401866X}{\emph{Int. J. Mod. Phys.}
  {\bfseries A19S1} (2004) 145}
  [\href{https://arxiv.org/abs/hep-th/0304042}{{\ttfamily hep-th/0304042}}].

\bibitem{Banks:2010zn}
T.~Banks and N.~Seiberg, \emph{{Symmetries and Strings in Field Theory and
  Gravity}}, \href{https://doi.org/10.1103/PhysRevD.83.084019}{\emph{Phys.
  Rev.} {\bfseries D83} (2011) 084019}
  [\href{https://arxiv.org/abs/1011.5120}{{\ttfamily 1011.5120}}].

\bibitem{Montero:2022vva}
M.~Montero and H.~Parra~de Freitas, \emph{{New supersymmetric string theories
  from discrete theta angles}},
  \href{https://doi.org/10.1007/JHEP01(2023)091}{\emph{JHEP} {\bfseries 01}
  (2023) 091} [\href{https://arxiv.org/abs/2209.03361}{{\ttfamily
  2209.03361}}].

\bibitem{Heidenreich:2021yda}
B.~Heidenreich, M.~Reece and T.~Rudelius, \emph{{The Weak Gravity Conjecture
  and axion strings}},
  \href{https://doi.org/10.1007/JHEP11(2021)004}{\emph{JHEP} {\bfseries 11}
  (2021) 004} [\href{https://arxiv.org/abs/2108.11383}{{\ttfamily
  2108.11383}}].

\bibitem{Cota:2022yjw}
C.~F. Cota, A.~Mininno, T.~Weigand and M.~Wiesner, \emph{{The asymptotic Weak
  Gravity Conjecture for open strings}},
  \href{https://doi.org/10.1007/JHEP11(2022)058}{\emph{JHEP} {\bfseries 11}
  (2022) 058} [\href{https://arxiv.org/abs/2208.00009}{{\ttfamily
  2208.00009}}].

\bibitem{Cota:2022maf}
C.~F. Cota, A.~Mininno, T.~Weigand and M.~Wiesner, \emph{{The asymptotic weak
  gravity conjecture in M-theory}},
  \href{https://doi.org/10.1007/JHEP08(2023)057}{\emph{JHEP} {\bfseries 08}
  (2023) 057} [\href{https://arxiv.org/abs/2212.09758}{{\ttfamily
  2212.09758}}].

\bibitem{Rudelius:2023mjy}
T.~Rudelius, \emph{{Revisiting the refined Distance Conjecture}},
  \href{https://doi.org/10.1007/JHEP09(2023)130}{\emph{JHEP} {\bfseries 09}
  (2023) 130} [\href{https://arxiv.org/abs/2303.12103}{{\ttfamily
  2303.12103}}].

\bibitem{Antoniadis:1997eg}
I.~Antoniadis, S.~Ferrara, R.~Minasian and K.~S. Narain, \emph{{R**4 couplings
  in M and type II theories on Calabi-Yau spaces}},
  \href{https://doi.org/10.1016/S0550-3213(97)00572-5}{\emph{Nucl. Phys. B}
  {\bfseries 507} (1997) 571}
  [\href{https://arxiv.org/abs/hep-th/9707013}{{\ttfamily hep-th/9707013}}].

\bibitem{Grimm:2017okk}
T.~W. Grimm, K.~Mayer and M.~Weissenbacher, \emph{{Higher derivatives in Type
  II and M-theory on Calabi-Yau threefolds}},
  \href{https://doi.org/10.1007/JHEP02(2018)127}{\emph{JHEP} {\bfseries 02}
  (2018) 127} [\href{https://arxiv.org/abs/1702.08404}{{\ttfamily
  1702.08404}}].

\end{thebibliography}\endgroup

\end{document}